\documentstyle[prl,aps,preprint]{revtex}

\let\ssection=\section
\renewcommand{\section}{\setcounter{equation}{0}\ssection}

\begin{document}
\draft
\title{Regular solutions in Abelian gauge model}
\author{Yuri N. Obukhov\footnote{On leave from: Department of 
Theoretical Physics, Physics Faculty, Moscow State University, 
117234 Moscow, Russia}
\footnote{e-mail: yo@thp.uni-koeln.de}
and Franz E. Schunck\footnote{Present address: Astronomy Centre, 
School of Chemistry, Physics and Environmental Science,
University of Sussex, Falmer, Brighton BN1 9QJ, UK}
\footnote{e-mail: fs@astr.maps.susx.ac.uk}
}
\address{Institute for Theoretical Physics, University of Cologne,
 D--50923 K\"oln, Germany}
\maketitle

\begin{abstract}
\noindent
The regular solutions for the Ginzburg-Landau (-Nielsen-Olesen) Abelian
gauge model are studied numerically. We consider the static isolated 
cylindrically symmetric configurations. The well known (Abrikosov) vortices,
which present a particular example of such solutions, play an important
role in the theory of type II superconductors and in the models of 
structure formation in the early universe. We find new regular static 
isolated cylindrically symmetric solutions which we call the type B and
the flux tube solutions. In contrast to the pure vortex configurations
which have finite energy, the new regular solutions possess a finite 
Gibbs free energy. The flux tubes appear to be energetically the most 
preferable configurations in the interval of external magnetic fields 
between the thermodynamic critical value $H_{c}$ and the upper critical 
field $H_{c_2}$, while the pure vortex dominate only between the lower
critical field $H_{c_1}$ and $H_{c}$. Our conclusion is thus that type
B and flux tube solutions are important new elements necessary for
the correct understanding of a transition from the vortex state to the
completely normal state. 
\end{abstract}
\bigskip\bigskip
\pacs{PACS no.: 03.50.-z; 11.15.Kc; 74.20.De; 74.60.-w}
\section{Introduction}

The Ginzburg-Landau theory of superconductivity \cite{gl} is mathematically 
equivalent to the Abelian theory of coupled gauge and Higgs fields \cite{no}. 
The existence of vortex (string-like) solutions in it was predicted by
Abrikosov in 1952 (and published 5 years later \cite{abr}) in the context
of the phenomenological model of superconductors and was discussed in 
\cite{no} in the framework of the dual string approach (see also a recent 
generalization to the case of nontrivial helicity in \cite{owc}). The aim of 
this work is to clarify, with the help of the careful numerical analysis, 
the properties of such solutions and to construct new solutions. We confine 
ourselves to the case of cylindrical symmetry, thus considering an isolated 
vortex and related field configurations. In the literature devoted to this 
subject (see, e.g., \cite{no,abr,gen,parks,saint,hueb,tin,ufn}) main 
attention was paid to approximate solutions and qualitative methods, but 
there were very few attempts to study exact solutions by numerical methods. 
Moreover, as recently was shown in \cite{peri}, the old qualitative results 
may contain mistakes and are incomplete. Approximate results are normally
confined to the domains of very small ($\lambda\ll 1$) or extremely large
($\log\lambda\gg 1$) values of the coupling constant $\lambda$ (or the 
characteristic Ginzburg-Landau parameter $\kappa =\sqrt{\lambda}$, see
Appendix). As for numerical studies, one can mention the earlier reports
\cite{hard,meis,fink,neu,doll} and more recent (variational) analysis in
\cite{jac}. In our paper we present new numerical results for the isolated
regular structures in the Ginzburg-Landau (-Nielsen-Olesen) model. 
Although here we confine ourselves to the case of Abelian gauge field,
we will consider elsewhere the non-Abelian generalizations (cf. previous
approaches in \cite{jpw,conf}). 
Our results demonstrate a rich structure of the space of exact 
solutions for the classical Ginzburg-Landau model. Besides the well 
known vortex solutions, for which we compute a variety of parameters in an
interval of values of $\lambda$ close to the critical value $\lambda=
{1\over 2}$ ($\kappa=\sqrt{1\over 2}$), we describe some new exact regular 
solutions. In our opinion, of particular interest are the configurations
which we call the {\it flux-tube} and the {\it oscillating} solutions below. 
The former appear to comprise a new nontrivial structure in ideal type II 
superconductors which is important for understanding the transition between 
a pure vortex state and a normal state. 

As it is well known, the vortex (or the mixed) state of a type II 
superconductor exists between the critical magnetic fields $H_{c_1}$ and 
$H_{c_2}$. Their values, as is usually claimed in the literature 
\cite{abr,gen,parks,saint,hueb,ufn}, can be determined from the analysis 
of isolated vortex solutions. However, from our results we conclude that 
such a claim is correct only partly. Indeed, $H_{c_1}$ is determined, for 
each value of $\lambda=\kappa^2$, by a relevant isolated vortex solution 
from  a comparison of the Gibbs free energy of the vortex configuration with 
that of the pure Meissner configuration. But the upper critical magnetic 
field $H_{c_2}$ appears to be completely unrelated to the vortex
configurations (cf. \cite{tin}). Instead, one can only determine $H_{c_2}$ 
from a similar comparison of the Gibbs free energy for a normal state with 
that of a different type of exact regular solutions of the Ginzburg-Landau 
equations. We call these the type B solutions. Actually, {\it approximate} 
type B solutions are usually described in the literature 
\cite{abr,gen,parks,saint,hueb,tin,ufn} in the framework of analysis of the 
{\it linearized} Ginzburg-Landau equations. We construct numerically exact 
self-consistent type B solutions for all possible values of external magnetic 
field. Although interesting enough themselves as mathematical structures, the 
type B solutions have a limited physical value as compared to the new flux 
tube solutions. The latter configurations correctly describe the transition 
from a pure vortex to a normal state, as will be demonstrated below. 

It seems worthwhile to notice that the Ginzburg-Landau equations describe
common extremals for two different (action type) functionals: for the usual 
{\it energy} and for the {\it Gibbs free energy} integrals. The principal 
difference of the vortex and the new solutions is that the former are the 
{\it finite energy} configurations, while the latter are the {\it finite 
Gibbs free energy} configurations. Previously, attention in the literature 
was paid only to the finite energy regular solutions of the Ginzburg-Landau 
equations. In the context of cylindrical symmetry, these are the famous 
Abrikosov(-Nielsen-Olesen) vortices. It is our aim to demonstrate the 
existence and the physical relevance of two large families of {\it finite 
Gibbs free energy} regular solutions which are described in our paper as  
the flux tube (or type A) and the type B solutions. In our opinion, the 
correct understanding of the mixed state of type II superconductor can only 
be achieved after taking into account these new solutions. In particular, 
it turns out that the class of the flux tube solutions is divided into an 
infinite number of families labeled by a number of nodes $n$ for the scalar 
field configuration. Each family exists on a finite interval of magnetic 
field (definition of limiting points see below). In this way, one finds a 
certain {\it fine structure of the mixed state} for an ideal 
type II superconductor. 

The plan of the paper is as follows: Section 2 contains a general 
introduction to the model, we discuss the two energy functionals and 
formulate general regularity conditions at the origin. A brief account of
numerical analysis of the pure vortex solutions is given in Section 3. 
Sections 4 and 5 contain the description of new solutions, the type B
and the flux tubes, respectively. In Section 6 we perform the linearization
analysis of the Ginzburg-Landau equations, while Section 7 is devoted to
the regular oscillating solutions. The latter appear to be certain unstable 
``relatives'' of the vortices. Finally, Section 8 contains a short discussion 
and a summary of the results obtained. In the Appendix we compare our 
notations and conventions with the old ones used in the literature. A general 
remark is necessary for the tables and figures: We find it convenient to put
all the available numerical data into the separate addendum \cite{tab},
available in electronic form. In the present paper, the Tables contain only 
some selected reasonable minimum of data, while the Figures 
present additional information.

\section{Energy functionals and regularity conditions}

\subsection{Nielsen-Olesen Lagrangian}

The Lagrangian of the Abelian (``Nielsen-Olesen'' \cite{no}) gauge model 
describing interacting electromagnetic $F=d A$ and complex scalar field 
$\Phi$ reads
\begin{equation}
L_{\rm NO}=-{1\over 2}(F\wedge{}^{\ast}F + D\Phi\wedge
{}^{\ast}\overline{D\Phi}) - {\cal V}(|\Phi |)\eta,\label{lagrNO}
\end{equation}
with the potential 
\begin{equation}
{\cal V}(|\Phi|)={\lambda\over 4}\left(|\Phi |^2 - 
{\mu^2\over\lambda}\right)^2 ,\label{pot}
\end{equation}
where the overbar denotes complex conjugation and $D=d +iA$, and star $^\ast$
is the Hodge dual operator, $\eta:=^\ast\!\! 1$ being the volume form.

Noticing that $\mu$ has a dimension of inverse length, we can introduce
for the cylindrical system $(\rho, \theta, z)$ a new (dimensionless) radial
coordinate $r$ via $\rho={\sqrt{\lambda}\over\mu}r$. We are looking for
{\it static} configurations, and use the following
cylindrically symmetric ansatz (cf. \cite{no})
\begin{equation}
A=f(r)d\theta,\quad \Phi={\mu\over\sqrt{\lambda}}\varphi(r),\label{ansatz}
\end{equation}
where $f, \varphi$ are two {\it real} functions. 

The magnetic field 1-form, defined as the (three-dimensional) Hodge dual
$^\ast\! F$, has only one component in $z$-direction, $^\ast\! F= H dz$, where 
the latter is given by the expression
\begin{equation}
H ={\mu^2\over\lambda}\; {1\over r}{df\over dr} .\label{mag}
\end{equation}
We will denote the {\it dimensionless} magnetic field by $h:=H\lambda/\mu^2 
=f'/r$.

Magnetic field is conveniently characterized by the flux it produces through
a two-dimensional surface. In a cylindrically symmetric case the total flux
over a surface orthogonal to the $z$-axis is
\begin{equation}
F=\int \rho\; d\rho\; d\theta\; H = 2\pi\int_{0}^{\infty}dr {df\over dr}=
2\pi [f(\infty)-f(0)].\label{flux}
\end{equation}

\subsection{Energy (line density) functional and equations of motion}

For the Lagrangian (\ref{lagrNO}) one can immediately 
write the energy per unit length in the form
\begin{equation}
{\cal E}={2\pi\mu^2\over\lambda}\int_{0}^{\infty}dr\;r
{1\over 2}\left[\left({1\over r}{df\over dr}\right)^2
+ \left({d\varphi\over dr}\right)^2 +
{1\over r^2}f^2\varphi^2 + {\lambda\over 2}
\left(\varphi^2 -1\right)^2\right].\label{energyNO}
\end{equation}
Notice that it is not necessary to include a phase factor 
$e^{in\theta}$ for the scalar field in the ansatz (\ref{ansatz}), as it is
done in a number of different approaches. The field 
$\varphi$ is always defined up to a gauge transformation and we find it 
more convenient to work in the gauge (\ref{ansatz}). 

Now, it is straightforward to see that the equations of motion of the 
Nielsen-Olesen model \cite{no} read:
\begin{eqnarray}
r^2 f'' - r f' &=&r^2\varphi^2 f,\label{1a}\\
r^2 \varphi'' + r\varphi' &=&\varphi(f^2 + 
\lambda r^2(\varphi^2 - 1)).\label{3a}
\end{eqnarray}
The value of the constant 
$\lambda$ plays an important role. If it equals to the critical value
$\lambda={1\over2}$, the Nielsen-Olesen equations are consequences of the
{\it first order} (Bogomolny) system
\begin{eqnarray}
{1\over r}f' + {1\over 2}(\varphi^2 -1)&=&0,\label{1st1a}\\
\varphi' + {1\over r}f\varphi &=&0\label{1st2a}.
\end{eqnarray}
In the theory of superconductivity $\lambda={1\over2}$ separates two phases:
for $\lambda>{1\over2}$ (resp., $\lambda<{1\over2}$) one has a type II (resp.,
type I) superconductor. The critical subcase was extensively studied in the 
literature \cite{vega,wein,taubes,jac}. The general noncritical case for
$\lambda\neq{1\over2}$ is much less investigated.

Using (\ref{mag}), we can write the equation (\ref{1a}) in the form
\begin{equation}
h'={1\over r}f\varphi^2,\label{hprime}
\end{equation}
and hence the system (\ref{1a})-(\ref{3a}) can be transformed into
\begin{eqnarray}
h'' + {1\over r}h' &=& h\varphi^2 + 2h'{\varphi'\over\varphi},\label{1b}\\
\varphi'' + {1\over r}\varphi' &=&\varphi\left({1\over\varphi^4}(h')^2 + 
\lambda (\varphi^2 - 1)\right),\label{3b}
\end{eqnarray}
explicitly for the coupled magnetic field $h$ and scalar field $\varphi$ 
variables. This system is however not particularly useful for numerical
investigation because of explicit $1/\varphi$ terms.

\subsection{Gibbs free energy functional}

The line energy density functional (\ref{energyNO}) is minimal 
(${\cal E}_{min}=0$) for the Meissner state, i.e., when $\varphi=1$ 
(superconducting order in all points of a sample) and $h=0$ (no magnetic
field inside a superconductor). This corresponds to a trivial solution of
(\ref{1a})-(\ref{3a}), $f=0, \varphi=1$. The normal state is described by
another simple solution, $f=C + {1\over 2}h_0 r^2, \varphi=0$, ($C$ and 
$h_0$ are constants), which always (with or without magnetic field $h_0$) has 
formally infinite energy ${\cal E}$. However physically, of interest is a 
difference of energies, not an energy itself.

In particular, let us turn our attention to the Gibbs free energy 
(per unit length) which is defined by 
\begin{equation}
{\cal G} = {\cal E} - 
\int d^2 x\, ({\hbox{\boldmath $H H$}}_{\rm ext}).\label{gibbs}
\end{equation}
We assume that both magnetic fields, the internal one
${\hbox{\boldmath $H$}}$ and an external ${\hbox{\boldmath $H$}}_{\rm ext}$
are directed along the $z$-axis, and use the dimensionless values
of fields defined, in accordance with (\ref{mag}), by $H=(\mu^2/\lambda)h$
and $H_{\rm ext}=(\mu^2/\lambda)h_{\rm ext}$. The Gibbs free energy of the
normal state with a magnetic field $h_{0}=h_{\rm ext}$ is given by the
integral
\begin{equation}
{\cal G}_{nh}={\cal E}[\varphi=0,h=h_{0}] -
{2\pi\mu^2\over\lambda}\int_{0}^{\infty}dr\,r\, h_{0}h_{\rm ext}
 ={2\pi\mu^2\over\lambda}\int_{0}^{\infty}dr\,r\,
{1\over 2}\left[ - h_{0}^2 + {\lambda\over 2}\right].\label{gibbsNH}
\end{equation}
For a sample without boundary, this is an infinite constant, while for a
cylinder of a radius $R$ this is a finite positive (negative) constant
${\pi R^2\mu^2}(1/2 - h_{0}^2/\lambda)/2$ for a magnetic field
$h_{0} < h_{c}$ ($h_{0} > h_{c}$), and zero for $h_{0}= h_{c}:=
\sqrt{\lambda/2}$. This observation underlies the physical interpretation
of $h_{c}$ as a critical (so called thermodynamic) value of the magnetic
field which distinguishes normal and superconducting phases.

Now, let us consider the difference,
\begin{eqnarray}
\Delta{\cal G}&:=&{\cal G}-{\cal G}_{nh}\nonumber\\
&=&{2\pi\mu^2\over\lambda}\int_{0}^{\infty}dr\,r\,
{1\over 2}\left[\left({1\over r}{df\over dr} - h_{0}\right)^2
+ \left({d\varphi\over dr}\right)^2 +{1\over r^2}f^2\varphi^2 + 
{\lambda\over 2}\left(\varphi^4 -2\varphi^2\right)\right].\label{gibbsdiff}
\end{eqnarray}
It is very important to notice that the Gibbs functional (\ref{gibbsdiff}) 
has the {\it same equations for extremals} as the energy functional 
(\ref{energyNO}), namely (\ref{1a})-(\ref{3a}). However, unlike the
strictly positive (\ref{energyNO}), the functional (\ref{gibbsdiff}) can
have any sign. 

\subsection{Regularity at the symmetry axis}

Looking for solutions which are regular at the origin, we substitute
the series expansions
\begin{equation}
f=\sum_{k=0}a_{k}r^k,\quad \varphi=\sum_{k=0}b_{k}r^k,\label{ser}
\end{equation}
into (\ref{1a})-(\ref{3a}). We then find two types of conditions: 

(A) Potential $f$ is non-zero while scalar field $\varphi$ vanishes at 
the origin,
\begin{eqnarray}
f&=&N + ar^2 + {1\over 4N(N+1)}b^2\;r^{2N+2} + 
O(r^{2N+4}) ,\label{ser1}\\
\varphi&=&b\,r^N\left(1+ {N\over 2(N+1)}\left(a-
{{\lambda\over 2N}}\right)r^2 + O(r^4)\right),
\label{ser3}
\end{eqnarray}
where $N=\pm 1,\pm 2,...$ is a nonzero integer and parameters $a,b$ are 
arbitrary.

(B) Potential $f$ vanishes while scalar field $\varphi$ is nontrivial at the 
origin,
\begin{eqnarray}
f&=& ar^2\left(1 + {1\over 8}b^2\;r^2 + O(r^4)\right) ,\label{0ser1}\\
\varphi&=&b\,\left(1+ {\lambda\over 4}(b^2 - 1)r^2 + 
O(r^4)\right),\label{0ser3}
\end{eqnarray}
with some parameters $a,b$.

In both cases, parameter $a$ determines the value of the magnetic field
at the origin, $h(0)=2a$. When $b=0$, both cases reduce to the solution
which describes a normal superconductor $\varphi=0$ filled by the constant
homogeneous magnetic field $h(r)=h(0)=2a$.

\section{Vortex solutions}

Vortex solutions of the Ginzburg-Landau equations (\ref{1a})-(\ref{3a})
are distinguished among others by the special conditions at infinity which
read:
\begin{eqnarray}
f(r)\vert_{r\rightarrow\infty}&&\longrightarrow 0,\label{infpot1}\\
\varphi(r)\vert_{r\rightarrow\infty}&&\longrightarrow 1.\label{infscal1}
\end{eqnarray}
When combined with the type A regularity conditions at the origin 
(\ref{ser1})-(\ref{ser3}), these asymptotic conditions define uniquely 
parameters $a, b$ for any value of $\lambda$ and $N$. The meaning of the 
constant $N$ is clear: this is the value of the magnetic flux 
described by such solutions. Indeed, we substitute $f(0)=N$ and $f(\infty)=0$ 
into (\ref{flux}) to obtain $F=-2\pi N$. This is the well known flux 
quantization result for superconductors. Notice that the asymptotics
(\ref{infpot1})-(\ref{infscal1}) cannot be realized for any solution
with type B regularity conditions (\ref{0ser1})-(\ref{0ser3}), see the
discussion in the next section.

The results of the numerical integration are given in Figure \ref{fig1}
and Table \ref{vorAll}, where left and right columns describe solutions for 
$N=1$ and $N=2$, respectively. For convenience of comparison with the old 
calculations \cite{hard,meis,fink,neu,doll} we write the coupling constant 
as $\lambda=\kappa^2$ [see Appendix for definitions, in particular 
(\ref{kappa})].

Energy (\ref{energyNO}) is always positive on all vortex configurations. 
It decreases with growing $\kappa$ and increases when 
$\kappa\rightarrow 0$. A useful physical variable, which allows to find 
necessary conditions for the existence of vortices, is the Gibbs free energy 
(line density) (\ref{gibbs}). This quantity is zero for the Meissner state 
$(h=0, \varphi=1)$, while for a vortex state it is always positive when an 
external magnetic field ${\hbox{\boldmath $H$}}_{\rm ext}$ is oriented 
oppositely to the magnetic field ${\hbox{\boldmath $H$}}$ inside a sample 
(hence such vortices are ruled out), and it can become negative for a certain 
value of external field with the same orientation as in the sample. In the 
latter case, for a constant external magnetic field 
${\hbox{\boldmath $H$}}_{\rm ext}=(0,0,H_{\rm ext})$ one finds
\begin{equation}
{\cal G} = {\cal E} - 2\pi H_{\rm ext}
\left|\int_{0}^{\infty} dr\,r\,h\right|= {\cal E} - 
2\pi\,|N|\,H_{\rm ext}.
\end{equation}
Hence the lower critical value of an external magnetic field $H_{\rm ext}=
H_{c_1}$, determined by the condition ${\cal G}=0$, is equal to 
\begin{equation}
H_{c_1}={{\cal E}\over 2\pi|N|}={\kappa\epsilon\over 4\pi|N|}\sqrt{2}H_{c},
\end{equation}
where in the last equality we switched to the old conventions summarized in 
Appendix. 

It seems worthwhile to make the following short remark: The analysis of 
exact vortex configurations shows that the {\it actual} (computed for 
vortices) magnetic field penetration length for different values of 
$\kappa$ is always greater than the domain of an essential change of the 
scalar field, even for small $\kappa$. In fact, this observation was also 
confirmed in earlier numerical studies \cite{meis,hard}. This casts doubts
on a possible qualitative and quantitative understanding (cf., 
\cite{gen,parks,saint,ufn}) of these solutions on the basis of so called 
penetration lengths $\delta$ and $\xi$ (see the Appendix) for the magnetic and 
the scalar fields, and stresses the need of exact numerical investigations
(see details in addendum \cite{tab} where the numerous vortex solutions for
a wide range of $\kappa$ are described). 

\section{Regular solutions for type B conditions}

The type B conditions (\ref{0ser1})-(\ref{0ser3}), like the type A conditions
(\ref{ser1})-(\ref{ser3}), guarantee a regular behavior of solutions at
the origin. (It seems worthwhile to notice that type B conditions are {\it
not} a particular case of type A for $N=0$). It is straightforward to 
analyze a qualitative behavior of such solutions for finite values of $r$.
For example, equation (\ref{1a}) immediately yields that $f(r)$ is a 
monotonous increasing for $a>0$ (decreasing for $a<0$) function. Indeed, 
if we assume an extremum for a finite $r=r_{0}$, one finds from (\ref{1a})
$f''(r_{0})=\varphi^2 f(r_{0})$ which means that such an extremum is a 
minimum for positive $f(r_{0})$ and a maximum for negative $f(r_{0})$.
For a function starting from a zero (\ref{0ser1}) both possibilities are
excluded and thus $f(r)$ is a monotonous function. In particular, this
means that $f(r)\rightarrow\infty$ for $r\rightarrow\infty$, and one thus
concludes that {\it no finite energy} solutions for type B regularity
conditions exist: the functional (\ref{energyNO}) is infinite.

Let us however look at (\ref{1a})-(\ref{3a}) as the equations for extremals
of the Gibbs free energy functional (\ref{gibbsdiff}). It is immediately clear 
that there exist {\it finite Gibbs free energy} regular type B solutions, 
provided they satisfy at infinity
\begin{eqnarray}
h(r)={1\over r}f'(r)\,\vline\,{\hbox{\raisebox{-1.5ex}{\scriptsize 
$r\rightarrow\infty$}}}
&&\longrightarrow h_{0},\label{infpotB}\\
\varphi(r)\vert_{r\rightarrow\infty}&&\longrightarrow 0.\label{infscalB}
\end{eqnarray}
Notice that unlike in a vortex configuration (\ref{infscal1}), the scalar 
field cannot approach at infinity any finite value except 0. 

The results of numerical integration are presented in Tables 
\ref{B1}-\ref{B225} and Figure \ref{fig2}. 
A principal difference of these solutions (which we will call ``type B'' 
solutions for brevity) from the vortex solutions lies in the fact that 
the magnetic field is asymptotically 
constant, and hence the flux integral defined by (\ref{flux}) is formally 
infinite. However a reasonable replacement is provided by the quantity
\begin{equation}
M:=\Delta F = \int \rho\; d\rho\; d\theta\;(H-H_{0}) = 
2\pi\int_{0}^{\infty}dr\, r\, (h - h_{0}).\label{diflux}
\end{equation}
Defined formally as a difference of fluxes, this variable is usually
interpreted as a {\it magnetization} per unit volume \cite{saint,fink}.
Unlike the quantized flux for the vortices, $M$ can have an arbitrary value. 

{}From Tables \ref{B1}-\ref{B225} we see, for the type B solutions without 
node, that when $a\rightarrow\lambda/2$ the external magnetic field is 
approaching $h_{\rm ext}=h_{c_2}=\lambda$, while  $M/(2\pi)=\Delta{\cal G}=0$. 
For $\lambda >1/2$, we find a higher external magnetic field for larger values 
of magnetic field at the center $h(0)=2a$. When $\lambda =1/2$, for all 
solutions $h_{\rm ext}=1/2$ (see \cite{tab}). Finally, for $\lambda <1/2$ 
\cite{tab} the external magnetic field is decreasing when $h(0)=2a$ grows 
(the reverse order as compared to the $\lambda >1/2$ case).

Besides the simple type B solutions without nodes, there exist more nontrivial 
solutions with nodes. Both are displayed on Figure \ref{fig2}, while the
configurations with one node are represented by the right columns in the 
Tables \ref{B1}-\ref{B225}. 

For $\lambda >1/2$ all type B solutions without node have a 
negative Gibbs free energy $\Delta{\cal G}$, for $\lambda =1/2$ always 
$\Delta{\cal G}=0$, and for $\lambda <1/2$ the Gibbs free energy is positive.
One can find also negative $\Delta{\cal G}$ for type B solutions
with one node when $h_{c}< \lambda/3$ (i.e., $\kappa > 3/\sqrt{2}$), 
see the case of $\kappa=2.25$ (Table \ref{B225}). This will be
clarified later in Section 6. 

The magnetization curves for the type B solutions are given in Figure 
\ref{fig4}. For any $\lambda=\kappa^2$, $M$ diverges at $h_{c}$. 

In the present paper, we display the results for $\kappa =1.0, 2.25$, the
detailed tables and figures for $\kappa$ ranging from 0.5 to 5.0 are
collected in the addendum \cite{tab}. 

\section{Flux tube solutions}

Let us look for the other solutions which yield finite values for the 
Gibbs functional (\ref{gibbsdiff}). At the origin $r=0$ we take the  
type A regularity conditions (\ref{ser1})-(\ref{ser3}), while at infinity 
we consider the asymptotics (\ref{infpotB})-(\ref{infscalB}). Such a 
combination of two type conditions at zero and at infinity suggests a possible 
physical interpretation of such solutions which appear as a result of a
certain ``gluing'' of a vortex configuration at the origin with a type B
solutions at large radial values. 

The results of numerical integration are given in Figures 
\ref{fig3}-\ref{fig6} and Tables \ref{ft1}-\ref{ft225}. As in the previous 
section, we present results only for the values of $\kappa =1.0, 2.25$, see
\cite{tab} for more solutions. 

Figure \ref{fig3} explains why we call these 
solutions the {\it flux tubes}: There is a core where matter is in a state 
close to the normal one filled by the magnetic field (this is in fact a 
vortex), surrounded by a superconducting tube (almost completely free of a 
magnetic field). Outside such a tube the sample quickly reduces to a normal 
state with the external field penetrated in it. For the solutions with
one node we have a ``sandwich-like'' structure: a tube of normal state between 
two superconducting tubes. 

Each family of flux tube solutions has two 
branches. E.g., for $\kappa=1$, one of these branches is characterized by the 
positive Gibbs free energy, and another has negative Gibbs free energy. 
However, for sufficiently large $\kappa$ both branches describe the 
negative Gibbs free energy configurations. We find it convenient to depict 
these branches in the form of magnetization curves, Figure \ref{fig4}. For 
$\lambda\leq {1\over 2}$, two branches with positive Gibbs free energy exist. 
As in the case of type B solutions, $M$ diverges at $h_{c}$. 

It is interesting to notice that among the flux tube 
configurations there are solutions with quite unusual behavior of the 
magnetic field, which in the center is oppositely oriented with respect to the 
direction of external magnetic field. See the data with negative $a$ in the 
Tables \ref{ft1}-\ref{ft225} (notice that for $\kappa=2.25$ 
there exist negative Gibbs free energy solutions with such property).

Like for the type B configurations, we also find the flux tube solutions 
with one node, relevant data is displayed in the right half of Tables 
\ref{ft1}-\ref{ft225}. Notice that for $\kappa=1$ all the flux tube solutions
with one node  have positive Gibbs free energy. However, with increasing
$\kappa$ this changes. The important thing is the position of the
thermodynamic critical value $h_{c}$ relative to the ``limiting points''
for $h_{\rm ext}$ of the flux tube families, at which the magnetization and 
the Gibbs free energy vanishes. We have determined numerically the values
of such limiting points which depend on the value of $\kappa$
and are located, on the $h_{\rm ext}$ axis, at $\lambda, {1\over 3}\lambda,
{1\over 5}\lambda, {1\over 7}\lambda, \dots$. In the next section we 
explain the values of these limiting points with the help of linearization 
analysis. 

The reader should compare the magnetization curve and the 
$(\Delta{\cal G}/h_{\rm ext})$ energy/field plots for $\kappa=1.0, 2.25$, 
Figures \ref{fig4}-\ref{fig5}, which demonstrate that the ``motion'' of the 
limiting points to the right of $h_{c}$ is accompanied by creation of flux 
tubes with one node which have negative Gibbs free energy.

Both, the type B solutions and the flux tube solutions, have a well defined
Gibbs free energy for an infinite sample. It is natural to compare them. Of
course, this must be done in a correct way: one should compare energies
of configurations with the same values of external magnetic field 
$h_{\rm ext}$. Using our data, we can display the Gibbs free energy 
$\Delta{\cal G}$ as a function of $h_{\rm ext}$. These functions for the
type B and the flux tubes are given on the Figure \ref{fig5},
showing that the flux tube configurations are energetically more preferable. 

\section{Linearized system and critical magnetic fields}

The best way to understand the structure of type B and flux tube solutions 
in the limit of vanishing magnetization $M\rightarrow 0$ and the Gibbs free
energy $\Delta{\cal G}\rightarrow 0$ is to study the linearized 
Ginzburg-Landau equations. 

Let us consider, in the spirit of \cite{abr,gen,parks,saint,hueb,tin,ufn}, 
the system (\ref{1a})-(\ref{3a}) in the situation when the square of the 
scalar field $\varphi^2$ is negligibly small. Mathematically this means that, 
in the lowest order, one drops out the terms containing $\varphi^2$ in 
(\ref{1a}) and (\ref{3a}). We then immediately notice that such a linearized 
system
\begin{eqnarray}
r^2 f'' - r f' &=&0,\label{1c}\\
r^2 \varphi'' + r\varphi' &=&\varphi(f^2  - \lambda r^2),\label{3c}
\end{eqnarray}
is a consequence of any of the {\it first order} systems
\begin{eqnarray}
{1\over r}f' - s\lambda &=&0,\label{1st1b}\\
\varphi' + s{1\over r}f\varphi &=&0\label{1st2b},
\end{eqnarray}
where $s=\pm 1$ is the sign factor. 

This system is straightforwardly integrated and yields
\begin{eqnarray}
f(r)&=&N + s{\lambda\over 2}r^2,\label{potlin}\\
\varphi(r)&=& 
\varphi_{0}r^{-sN}\exp\left(-{\lambda\over 4}r^2\right),\label{sclin}
\end{eqnarray}
where $N$ and $\varphi_{0}$ are integration constants. In the linear 
approximation, the constant $N$ is arbitrary, but as we can see from 
the analysis of the complete {\it self-consistent} system at the origin 
(\ref{ser1})-(\ref{0ser3}), this constant should be either $0$ or $\pm 1, 
\pm 2, ...\,$. In order to have a regular behavior of (\ref{sclin}) at $r=0$, 
one should choose the sign of $N$ in such a way that $sN<0$. Notice that the 
potential (\ref{potlin}) describes an homogeneous constant magnetic field 
$h=s\lambda$, and thus the value of $s$ shows its direction (up or down 
along the $z$ axis).

It is easy to check that all the solutions (\ref{potlin}) and (\ref{sclin}),
for arbitrary values of integration constants, have the same (zero) energy 
integral computed for the linearized system (\ref{1c})-(\ref{3c}). For
$N=0$ the field (\ref{sclin}) evidently describes the linearized type B
solution, while for $N=1$ this is a linearized flux tube solution. As we
see, the linearized solutions are energetically equivalent. However, the
numerical results (see Figure \ref{fig5}) definitely show 
that self-consistent flux tubes are energetically more preferable than the 
type B configurations.

The first correction to the magnetic field is easily computed. One must
now take the complete system, and consider the first equation (\ref{1a})
in the form (\ref{hprime}) where the right hand side is constructed from
the lowest order configurations (\ref{potlin}) and (\ref{sclin}). Since these
satisfy (\ref{1st2b}), we find from it ${1\over r}f\varphi^2 = - {1\over 2}s
(\varphi^2)'$, and hence (\ref{hprime}) is immediately integrated, yielding
for the magnetic field
\begin{equation}
h=s\left(\lambda - {1\over 2}\varphi^2\right).
\end{equation}
As we see, an important role is played here by the normalization of the
linearized solution, i.e. by the constant $\varphi_{0}$. 

Further insight can be obtained also directly
from the analysis of the second order linearized system (\ref{1c})-(\ref{3c}).
Indeed, integration of (\ref{1c}) is straightforward, giving
\begin{equation}
f=N + {h\over 2}r^2,\label{magH}
\end{equation}
where $N$ and $h$ are integration constants with the latter representing 
the value of an homogeneous constant magnetic field. After substituting 
(\ref{magH}) into (\ref{3c}) we find a Schr\"odinger type of equation for 
$\varphi$ with the potential of a circular oscillator. Regular solutions 
exist only when 
\begin{equation}
h = {s\lambda\over {1 + 2n + sN + |N|}},\label{eigenvalue}
\end{equation}
where $n=0,1,2,...\,$. Corresponding eigenfunctions $\varphi_{n,N}$ are given
in terms of the Laguerre polynomials, with $n$ equal to the number of zeros
(nodes). Let us introduce the notation
\begin{equation}
{h}_{k}:={\lambda\over 2k+1},\quad k=0,1,2,\dots\, .
\end{equation}
It is easy to see that the maximal eigenvalue (\ref{eigenvalue}), 
$h=s\lambda=s{h}_{0}$, is achieved for $n=0$ and $sN=-|N|$, and the 
scalar field is then described exactly by (\ref{sclin}). This maximal
eigenvalue is precisely the second critical field $h_{c_{2}}=\lambda=
\kappa^2$. The rest of eigenvalues also have clear physical meaning: these 
define the values of the external magnetic field at which the exact type B and
flux tube solutions become ``linearizable'' and thus disappear. Looking
at the Tables \ref{ft1}-\ref{ft225}, \ref{B1}-\ref{B225}, we find the 
complete agreement with the 
above linearization analysis. Indeed, the flux tube (without nodes) and the 
type B (without nodes) configurations have the limit magnetic field values 
${h}_{1}=\lambda/3$ and ${h}_{0}=h_{c_2}$, while the flux tubes and type B 
solutions with one-node ``live'' between ${h}_{2}=\lambda/5$ and ${h}_{1}=
\lambda/3$. In general, the family of solutions with $k$ nodes have the 
limiting points $h_{k+1}$ and $h_{k}$. If $h_{c}$ belongs to the interval 
$[h_{k+1}, h_{k}]$, then $h_{k+1}<h_{\rm ext}<h_{k}$ for all solutions in
this family. However when $h_{c}$ does not belong to this interval,
then $[h_{k+1}, h_{k}]$ is extended up to $h_{c}$. 

This linearization analysis clearly supports the existence of the flux tube
type solutions. 

\section{Oscillating solutions}

Let us now consider weaker conditions at infinity: potential  
still satisfy (\ref{infpot1}), however instead of (\ref{infscal1}) we 
require regularity and finiteness of the scalar field. From the physical 
interpretation of $\varphi$ as the ``density of superconducting electrons''
one concludes that $|\varphi|<1$ for all values of the radial coordinate $r$.

Qualitatively, one can understand the behavior of a scalar field for
large $r$ as follows. When in (\ref{3a}) the potential $f^2$ and
the scalar field $\varphi^2$ nonlinear terms become small enough, one is left 
with the linearized equation
\begin{equation}
\varphi'' + {1\over r}\varphi' + \lambda\varphi =0,\label{linscal}
\end{equation}
which has the Bessel function as a solution $\varphi=J_{0}(\sqrt{\lambda}r)=
J_{0}(\kappa r)$. Such an asymptotic behavior is confirmed by direct
numerical integration, see Figure \ref{fig7} and Table \ref{tabos}.

To the best of our knowledge, this type of solution was never reported
in the literature. It is interesting to find out, what physics corresponds
to it. One interpretation is that these new oscillating solutions are 
unstable configurations preceding to the completely formed Abrikosov vortex 
state, they appear when the external magnetic field is switched on and 
reaches $H_{c_1}$. We may draw attention to the following remarks in an 
experimental research paper: ``Consistent values of the magnetization were 
obtained for fields just above $H_{c1}$ only after the sample had been moved 
between the coils a number of times. The change in the magnetization of a 
sample upon a slight increase (or decrease) of the field is very dependent 
on the fact that the sample has been jarred as it is pulled between the 
measuring coils, and a final, steady-state value of the magnetization is 
sometimes obtained only after 10 or 20 sample translations. It is as if 
vibration assists the flux movement into or out of the sample. Hence, all of 
the data reported below refer to the final steady state of the magnetization;
that is, further sample motion would produce no further change'' \cite{fin}.

For the {\it type A} conditions at the origin, the (selected) results of 
numerical integration are given in Table \ref{tabos}. In general, oscillating 
solutions exist only for initial values $(a,b)$ below the ones $(a^{*},b^{*})$
of the vortex solution (first line of Table \ref{tabos}). The flux 
for the displayed solutions is always $F/(2\pi)=-1$. The values here are
given for $\kappa =1.0$. From Figure \ref{fig7}, one recognizes that for 
$b$ close to $b^*$ the scalar field reaches almost the value $1$, i.e.~a 
complete superconducting state. 

Due to a not so quick decay of the scalar field at infinity, approximately
$\varphi\sim \cos(\kappa r)/\sqrt{r}$, the energy of an oscillating solution 
is infinite for an infinite sample, but for a real finite cylindrical
sample it is finite, although larger than the energy of a vortex
configuration. Notice, that an oscillating character of such
solutions may resemble an ``intermediate'' superconducting state with
coexisting normal and superconducting regions \cite{parks}. However a 
considerable difference is that the magnetic field penetrates only at 
the center, exactly like in a vortex case. Moreover, the
magnetic flux is quantized in a precisely same manner as for vortices,
which is immediately seen after using (\ref{ser1}) and $f(\infty)=0$ in 
(\ref{flux}), flux is $F/(2\pi)=-N$. Notice also that the magnetic energy 
of oscillating solutions is always finite even for an infinite sample.

Oscillating solutions exist also for the type B conditions. However, the
numerical analysis revealed that in this case the parameter $a$ must 
vanish and hence the magnetic field is  completely absent. Nevertheless,
the scalar field configuration is nontrivial. In fact, one is left then
with the nonlinear scalar field equation
\begin{equation}
\varphi'' + {1\over r}\varphi' + 
\lambda\left(\varphi - \varphi^3\right)=0,\label{0scal}
\end{equation}
which in the limit of $r\rightarrow\infty$, when $\varphi$ approaches $0$,
reduces to the linearized equation (\ref{linscal}). 

It is evident, that it is enough to find explicitly only a solution 
$\varphi_{1}(r)$ for the case $\lambda=\kappa=1$. For an arbitrary value of
the coupling constant the solution is then given by $\varphi=\varphi_{1}
(\sqrt{\lambda}r)$. In particular, this defines positions of extrema
and zeros of $\varphi(r)$ for all values of $\lambda$ from that of 
$\varphi_{1}(r)$. The latter evidently depends only on the value of the 
parameter $b$ in (\ref{0ser3}). Numerical results are displayed in Figure
\ref{fig7}.

As we already mentioned, for an infinite sample both energy and the Gibbs
free energy are divergent for oscillating solutions. However, for a finite
cylinder we discover convergent results. As it is well known, the boundary
conditions in the Ginzburg-Landau theory require vanishing of derivative
$\varphi'|_{r=R}=0$ on a cylinder's surface $r=R$. Using (\ref{0scal}), one
then finds from (\ref{gibbsdiff}) that for such zero magnetic field 
oscillating solutions the Gibbs free energy is always negative,
\begin{equation}
\Delta{\cal G}=-{\pi\mu^2\over 2}\int_{0}^{R} dr\, r\, \varphi^{4}(r).
\label{gibbsosc}
\end{equation}
A curious conclusion is thus that for a finite sample in absence of an
external magnetic field an oscillating state is energetically more 
preferable than a purely normal state. 

We have calculated the Gibbs free energy for the oscillating solutions in 
finite samples. A boundary can be placed at any of the positions of extrema 
of the solutions, and the numerical results are displayed in the lower part
of Table \ref{tabos}. We present explicitly only the case $\lambda=1$, while
for an arbitrary $\lambda$, the relevant data are easily obtained from
Table \ref{tabos} by replacing $(R, \Delta{\cal G})$ with 
$(R/\sqrt{\lambda}, \Delta{\cal G}/\lambda)$ (cf.~(\ref{gibbsosc})).

Oscillating scalar field solutions appear also if one takes instead of an
electromagnetic theory a general relativistic gravitational theory 
\cite{sch1}. The solutions for this Einstein-scalar-field theory
describe a dark halo of galaxies or galaxy clusters, respectively.
The oscillating behavior of the scalar field can be removed simply by
adding a mass term for the scalar field potential.
In this case one speaks about boson star solutions
\cite{jetzer} which have some characteristics similar to the neutron
stars but also decisive differences \cite{sch2}. These boson stars
could be formed in the very early universe from Higgs or axion particles.
That stable configurations of these boson stars can exist was investigated
with the help of the catastrophe theory \cite{kus}.

\section{Discussion and conclusion}

In each class of solutions the decisive role is played by the values of
the parameters $(a,b)$ which appear in the regularity conditions at the
origin (\ref{ser1})-(\ref{ser3}) and (\ref{0ser1})-(\ref{0ser3}). All the
solutions are obtained after a ``fine tuning'' of these parameters. It is
worthwhile to draw a kind of a ``phase diagram'' on the $(a,b)$ plane 
which shows explicitly the domains of existence for different solutions.
Since the vortices, flux tubes and the oscillating solutions all belong
to the type A regularity conditions (\ref{ser1})-(\ref{ser3}), we can 
display them on the same $(a,b)$-plane, see Figure \ref{fig6} for
different values of $\kappa$. The encircled dots denote the ``position''
of a vortex solution while each curve represents a complete family of a 
flux tube or oscillating solutions for a fixed $\kappa$. Notice that all 
curves end on the $a$-axis ($b=0$) at the points which correspond to the 
half of the relevant limit magnetic field values, i.e., $a={1\over 2}
{h}_{k}, k=0,1,\dots\,$. Curves which represent flux tubes with 
increasing number of nodes are concentrating in the close neighborhood of 
the ``oscillating'' curve which seem to indicate that the oscillating 
solutions are unstable and a small perturbation may cause their decay into 
a nearby flux tube with a finite number of nodes. When moving along any flux
tube curve away from the $a$-axis, one inevitably hits the vortex dot, where
magnetization diverges. 

Technically, it is impossible (because of the limitations on numerical 
precision) to make integration for
the parameters $(a,b)$ in the close vicinity of a vortex. Thus, from the
data which we obtained, it is not clear what is the limiting value of an
external magnetic field to which all the flux tube configurations approach
when $(a,b)$ are coming closer and closer to the vortex parameters 
$(a^{*},b^{*})$. We can see however (cf.~Tables \ref{ft1}-\ref{ft225}), 
that such a limit is close to the thermodynamic
critical field $h_{c}$ for each $\kappa$. The following simple argument
demonstrates that in fact such a limit is equal to $h_{c}$. Let us formally
compare the values of the Gibbs free energy for a vortex and for a flux
tube solution. We find that these are equal when
\begin{equation}
{\cal E}_{\rm V} - {\cal E}_{\rm FT} = 2\pi H_{\rm ext}\left(1 - 
\int_{0}^{\infty}dr\,r\, h_{\rm FT}\right),\label{delg}
\end{equation}
where the subscripts $\scriptstyle{\rm V}$ and $\scriptstyle{\rm FT}$ 
denote the vortex and flux tube variables, respectively.
The right and the left hand sides are both formally divergent, but 
comparing the leading terms, one can use (\ref{delg}) and (\ref{energyNO}) 
to find [noticing that after a certain finite value of $r$ one has 
$h_{\rm FT}=h_{\rm ext}$ and $\varphi_{\rm FT}=0$] $h_{\rm ext}=\left(
h_{\rm ext}^2 + {\lambda\over 2}\right)/(2 h_{\rm ext})$, from which
\begin{equation}
h_{\rm ext}=\sqrt{\lambda\over 2}=h_{c}.
\end{equation}
Below (above) this value, the pure vortices (the flux tubes) are
energetically more preferable.

Summarizing, in this paper we present the numerical solutions of the
cylindrically symmetric Ginzburg-Landau equations. Besides the well known
vortex configurations with finite energy we find new solutions (we
call them type B and the flux tube solutions) which have finite Gibbs free 
energy. Direct numerical integration reveals many interesting properties 
of these solutions. One of the most important points is perhaps the 
clarification of the meaning and value of the upper critical field $h_{c_2}$. 
Contrary to what is usually claimed in the literature (with an exception of
\cite{tin}), $h_{c_2}$ by no means denotes the magnetic field below which the 
vortex becomes more energetically preferable than the normal state. Instead, 
as we demonstrated, $h_{c_2}=\lambda$ is the value of an external magnetic 
field at which the type B solutions and the flux tubes have zero Gibbs free 
energy. Below it, for $h_{\rm ext}<h_{c_2}$, $\Delta{\cal G}$ is negative for 
both flux tubes and type B configurations. The analysis of linearized 
Ginzburg-Landau equations near $h_{c_2}$ which usually (and incorrectly) is 
described in the literature (see, e.g., \cite{abr,gen,parks,saint,hueb,ufn}) 
as relevant to vortices, in fact is the linearization of flux tubes and 
type B solutions. Our results show that the flux tube solutions without node 
remain the most energetically preferable from $h_{c_2}$ down to the 
thermodynamic critical field $h_{c}$, after which the vortices become 
energetically more preferable and such a vortex state ends at the lower 
critical field $h_{c_1}$. We find it convenient to depict this 
observation on Figure \ref{fig8}. 

It is worthwhile to stress, that our results do not contradict the previous 
knowledge about the mixed state in the type II superconductors. On the 
contrary, they again support the significance of such a fundamental structure 
as a vortex: notice that, after all, one can interpret a flux tube solution 
as a vortex ``surrounded'' by a type B configuration. However, in our opinion, 
the flux tubes provide us with a new understanding that the mixed state 
reveals a rich structure in which a ``pure vortex'' is only part of the whole 
picture valid near $h_{c_1}$. We are convinced that a correct transition from 
such a pure vortex state to the normal state (starting at $h_{c}$ up to 
$h_{c_2}$) can only be correctly described with the help of the flux tube 
and the type B solutions. As is well known, the Ginzburg-Landau theory is 
only an approximation (valid near the critical temperature of superconducting 
phase transition) to the underlying microscopic Gorkov theory within the BCS 
scheme. The isolated vortices and the vortex lattice structures are discussed 
in the broader aspects in the recent review \cite{brandt}. It is worthwhile 
to mention the success in theoretical construction of a vortex lattice 
solution for the Gorkov equations \cite{gor}. The work is now in progress 
aiming at a generalizing our isolated flux tube solutions to the lattice 
type structures. Physically, it would be interesting to study a possibility 
of relating the new solutions to the problem of the origin of the so called 
irreversible line on the phase diagram for the high-temperature 
superconductors (currently discussed explanation of which is the phenomenon 
of the vortex lattice ``melting'' \cite{melt}).

\appendix\section*{Correspondence with the Ginzburg-Landau notation}

In the Ginzburg-Landau theory of superconductivity \cite{gl}, the scalar
field $\varphi$ is interpreted as the ``order parameter'' with the square
describing the ``density of superconducting electrons'', $n_{s}=|\varphi|^2$.
The potential is usually written in the form
\begin{equation}
{\cal V}_{GL}=\alpha |\varphi|^2 + {\beta\over 2}|\varphi|^4 ,\label{potgl}
\end{equation}
with constant parameters $\alpha <0, \beta>0$. Their physical meaning is
clarified by the following quantities they define: the thermodynamic 
critical magnetic field for a bulk superconductor $H_{c}^2:=\alpha^2/\beta$; 
the equilibrium density of superconducting electrons $|\varphi_{\infty}|:=
|\alpha|/\beta$; the order parameter coherence length $\xi:=1/
\sqrt{2|\alpha|}$; the magnetic field penetration length $\delta:=
\sqrt{\beta/|\alpha|}$. (We are using the units in which the mass and the
charge of the electron is equal one). Of particular importance is the ratio 
of two lengths
\begin{equation}
\kappa:={\delta\over\xi}.\label{ratio}
\end{equation}
Comparing (\ref{pot}) and (\ref{potgl}), we find the relation between our 
and the Ginzburg-Landau notations:
\begin{equation}
\alpha=-{\mu^2\over 2},\quad \beta={\lambda\over 2},
\end{equation}
hence, in our notation, we have
\begin{eqnarray}
\xi&=&{1\over\mu}, \quad \delta={\sqrt{\lambda}\over\mu},\label{lengths}\\
\kappa&=&\sqrt{\lambda},\quad 2H_{c}^2={\mu^4\over\lambda}.\label{kappa}
\end{eqnarray}
Technically, there are also other notational differences: in the literature 
on type II superconductors instead of $f$ one often uses 
$Q:=-f/\sqrt{\lambda}$, while the dimensionless line energy density is
defined \cite{gl,abr,meis,hard,gen,parks,saint,hueb,ufn} by 
\begin{equation}
\epsilon := {{\cal E}\over \delta^2 H_{c}^2}
={2\pi\over\lambda}\left\{[r\varphi'\varphi]_{0}^{\infty} + \int_{0}^{\infty}
dr\;r\left[h^2 + {\lambda\over 2}(1-\varphi^4)\right]\right\},\label{energy}
\end{equation}
where we used the field equation (\ref{3a}). 

\bigskip
{\bf Acknowledgments} 
\bigskip

We would like to thank Friedrich W. Hehl and Eckehard W. Mielke for very 
useful criticism and advice. The helpful comments of E.H.~Brandt, D.F~Brewer 
and Fjodor Kusmartsev are gratefully acknowledged. The work of YNO was 
supported by the Deutsche Forschungsgemeinschaft (Bonn) grant He 528/17-1, 
and for FES by the European Union.

\begin{figure}
\caption[]
{A typical vortex solution. At the center, the magnetic field $h$ is
maximal. The limit of the scalar field $\varphi $ is 1, which means
physically a completely superconducting state. With our gauge choice
the potential $f$ is vanishing at infinity.}
\label{fig1}
\end{figure}

\begin{figure}
\caption[]
{Type B solutions without node (above) and with one node (below) for 
$\kappa =1.0$. In both cases the scalar field $\varphi$ has initial values 
$\varphi(0)=0.9, 0.6, 0.3$ and magnetic potential $f(0)=0.0$.}
\label{fig2}
\end{figure}

\begin{figure}
\caption[]
{Flux tube solutions without node for $\kappa =0.5, 1.0, 1.5$ with the same 
magnetization $M/(2\pi)=-6.0$ and positive Gibbs free energy (above), and the
magnetization $M/(2\pi)=-4.0$ and negative Gibbs free energy (below). The 
scalar field of a flux tube reaches a maximum before going to zero at infinity 
(and not to 1, as in the case of vortices). The magnetic potential $f$ (not 
shown) starts at 1, is followed by a minimum and then grows with an asymptotic 
$r^2$ behavior, so that the magnetic field $h$ is asymptotically constant.
Note that for $\kappa=0.5$ all solutions have positive Gibbs free energy,
cf. Fig.~\ref{fig4}.}
\label{fig3}
\end{figure}

\begin{figure}
\caption[]
{Magnetization curves for type B and A solutions for different $\kappa$.
Above, the type B solutions without node (right curve for each $\kappa$), and 
with one node (left curve). For $\kappa=2.25$, we notice a change of the sign 
of the Gibbs free energy within a solution family, while for $\kappa 
=0.5, 1.0, 1.5$ we find only positive values of ${\cal G}$. Below, for the 
flux tubes without node for each value $\kappa=0.5, 1.0, 1.5$, the 
corresponding limit values $h_1$, $h_2$, $h_{c}$, and $h_{c_2}$ are drawn. 
Solid (broken) lines denote negative (positive) Gibbs free energy. In all 
cases the magnetization diverges at $h_{c}$. For $\kappa =0.5$ the point 
$h_c$ does not lie within the two limits $h_1$ and $h_{c_2}$ so that one 
can find also solutions up to $h_c$.}
\label{fig4}
\end{figure}

\begin{figure}
\caption[]
{$\Delta {\cal G}$ against the external magnetic field for
$\kappa =1.0$ and $\kappa =2.25$ for the flux tube and the type B solutions.
In each case the flux tubes have lower values of $\Delta{\cal G}$ than the 
type B solutions. Hence, flux tubes are energetically more preferable.}
\label{fig5}
\end{figure}
 
\begin{figure}
\caption[]
{$(a,b)$ diagrams for $\kappa =1.0, 2.25, 5.0$. The big dot describes
in each case the corresponding vortex solution. The drawn, broken and dotted 
lines represent the flux tubes without node, the flux tubes with one node, 
and the oscillating solutions, respectively. The type B solutions cannot be
compared in these diagrams because they have different initial values.}
\label{fig6}
\end{figure}

\begin{figure}
\caption[]
{The oscillating solutions for type A (above) and type B (below) initial 
conditions. For an initial value (here, $b=0.777$) very near to the one of 
the vortex, the scalar field closely approaches 1 before the oscillation 
starts. Near the origin, the vortex can be recognized. No type A oscillating 
solution with an initial value above that of the vortex can be found. For a 
type B oscillating solution, the magnetic field vanishes, while the square of 
the scalar field, the density of superconducting electrons, oscillates without 
any external magnetic field, and every finite sample has a negative Gibbs 
free energy.}
\label{fig7}
\end{figure}

\begin{figure}
\caption[]
{The general diagram for different solutions. The vortex state is 
energetically most preferable between $H_{c_1}$ and $H_c$, while above $H_c$, 
the flux tube configurations replace them. Hence, $H_c$ gains the following 
physical meaning for a type II superconductor: Above $H_c$, isolated vortices 
come into contact with the external magnetic field and the flux tube solutions 
are constructed. Energetically, the most preferable solution is a flux tube 
without node. But also the flux tubes with nodes do exist there, if $\kappa$ 
is large enough. Near $H_{c_2}$, only the flux tube without node exists. Of
course, the flux tubes can `live' also in the vortex and the Meissner state 
but they are energetically less preferable. In this way, one finds a rich 
fine structure of a superconductor's mixed state.}
\label{fig8}
\end{figure}
\bigskip\bigskip\bigskip
\vfill
Figures 1-8 are not enclosed. See instead the Addendum for complete set of
figures.

\pagebreak

\begin{table}
\caption[]{Vortex solutions}
\begin{tabular}{l|cccc|cccc}
&\multicolumn{4}{c|}{$N=1$}&\multicolumn{4}{c}{$N=2$}\\
\tableline
$\kappa $ & $a$ & $b$ & ${\cal E}$[$\pi\mu^2$]&${H_{c_1}\over\sqrt{2}H_{c}}$&
$a$ & $b$ & ${\cal E}$[$\pi\mu^2$]&${H_{c_1}\over\sqrt{2}H_{c}}$ \\
\tableline
2.00 & $-$0.46614 & 1.35218 & 0.38816 & 0.38816 &
$-$0.52764 & 1.01524 & 0.88928 & 0.44464 \\
1.75 & $-$0.43108 & 1.20961 & 0.47922 & 0.41932 &
$-$0.48000 & 0.82862 & 1.07872 & 0.47194 \\
1.50 & $-$0.39344 & 1.06659 & 0.61094 & 0.45821 &
$-$0.42993 & 0.65857 & 1.34754 & 0.50532 \\
1.25 & $-$0.35273 & 0.92273 & 0.81402 & 0.50876 &
$-$0.37708 & 0.50680 & 1.75284 & 0.54776 \\
1.00 & $-$0.30828 & 0.77735 & 1.15676 & 0.57838 &
$-$0.32096 & 0.37219 & 2.41915 & 0.60478 \\
0.75 & $-$0.25904 & 0.62916 & 1.82190 & 0.68321 &
$-$0.26080 & 0.25461 & 3.67126 & 0.68836 \\ 
$1/\sqrt{2}$ & $-$0.25000 & 0.60328 & 1.99999 & 0.70710 &
$-$0.25000 & 0.23614 & 3.99999 & 0.70710 \\
0.50& $-$0.20314 & 0.47525 & 3.47164 & 0.86791 &
$-$0.19527 & 0.15401 & 6.64856 & 0.83107 \\
\end{tabular}
\label{vorAll}
\end{table}

\begin{table}
\caption[]{Type B solutions for $\kappa =1.0$.}
\begin{tabular}{l|cccc|cccc}
&\multicolumn{4}{c|}{Without node}&\multicolumn{4}{c}{With one node}\\
\tableline
$b$ & $a$ & $M/(2\pi)$ & $\Delta{\cal G}$ [$\pi \mu^2$] & $h_{\rm ext}$
& $a$ & $M/(2\pi)$ & $\Delta{\cal G}$ [$\pi \mu^2$] & $h_{\rm ext}$ \\
\tableline
$1-10^{-16}$ & 2.62 $10^{-13}$ & $-$319.831 & $-$3.336 & 0.712 &
1.88 $10^{-14}$ & $-$399.137 & 22.180 & 0.672 \\
$1-10^{-12}$ & 2.80 $10^{-9}$ & $-$150.422 & $-$2.301 & 0.714 &
2.62 $10^{-10}$ & $-$201.853 & 14.548 & 0.659 \\ 
$1-10^{-8}$ & 1.79 $10^{-6}$ & $-$69.240 & $-$1.582 & 0.718 &
2.20 $10^{-7}$ &  $-$102.176 &  9.272 &   0.641 \\ 
$1-10^{-4}$ & 1.11 $10^{-3}$ & $-$18.774 & $-$0.846 & 0.728 &
2.11 $10^{-4}$ &   $-$34.388 &  4.108  &   0.597 \\
0.9 & 0.1466 & $-$1.412 & $-$0.169 & 0.800 &
0.0439 &  $-$5.299 &   0.610 &   0.477 \\
0.7 & 0.3030 & $-$0.430 & $-$0.042 & 0.878 &
0.0957 &  $-$2.376 &   0.188 &   0.420 \\
0.5 & 0.4033 & $-$0.160 & $-$0.009 & 0.937 &
0.1307 &  $-$1.126 &   0.051 &   0.380 \\
0.3 & 0.4659 & $-$0.048 & $-$0.001 & 0.977 &
0.1536 &  $-$0.401 &   0.007 &   0.351 \\
0.1 & 0.4962 & $-$0.005 & $-$1.25 $10^{-5}$ & 0.997 &
0.1652 &  $-$0.044 &   9.30 $10^{-5}$ &   0.335 \\
0.0 & 0.5000 & 0.000 & 0.000 & 1.000 &
0.1666 &  0.000 & 0.000 & 0.333 \\
\end{tabular}
\label{B1}
\end{table}

\begin{table}
\caption[]{Type B solutions for $\kappa =2.25$.}
\begin{tabular}{l|cccc|cccc}
&\multicolumn{4}{c|}{Without node}&\multicolumn{4}{c}{With one node}\\
\tableline
$b$ & $a$ & $M/(2\pi)$ & $\Delta{\cal G}$ [$\pi \mu^2$] & $h_{\rm ext}$
& $a$ & $M/(2\pi)$ & $\Delta{\cal G}$ [$\pi \mu^2$] & $h_{\rm ext}$ \\
\tableline
$1-10^{-8}$ &  3.96 $10^{-4}$ & $-$66.111 & $-$2.249 &  1.670 &
3.96 $10^{-4}$ & $-$70.821 & 0.526  &  1.534 \\
$1-10^{-4}$ &  0.0396   & $-$14.189 & $-$1.107 & 1.765 & 
0.0337 & $-$17.351 & $-$4.34 $10^{-2}$ &  1.495 \\
$1-10^{-2}$ &  0.3836   &  $-$2.443 & $-$0.428 & 2.069 &
0.2154 & $-$5.055 & $-$0.101 &  1.486 \\
0.9 &  1.1088  &  $-$0.402 & $-$9.99 $10^{-2}$ & 2.892 &
0.4756 & $-$1.673 & $-$3.78 $10^{-2}$ &  1.544 \\
0.7 &  1.7810  &  $-$9.67 $10^{-2}$ & $-$1.83 $10^{-2}$ & 3.865 & 
0.6640 & $-$0.621 & $-$8.38 $10^{-3}$ &  1.613 \\
0.5 &  2.1691  &  $-$3.33 $10^{-2}$ & $-$3.52 $10^{-3}$ & 4.475 &
0.7597 & $-$0.259 & $-$1.72 $10^{-3}$ &  1.652 \\
0.3 &  2.4044  &  $-$9.81 $10^{-3}$ & $-$3.90 $10^{-4}$ & 4.855 & 
0.8147 & $-$8.41 $10^{-2}$ & $-$1.97 $10^{-4}$ &  1.675 \\ 
0.1 &  2.5173  &  $-$9.98 $10^{-4}$ & $-$4.48 $10^{-6}$ & 5.039 &
0.8405 & $-$8.93 $10^{-3}$ & $-$2.30 $10^{-6}$ &  1.686 \\
0.0 &  2.53125  &  0.000 &  0.000 &  5.0625 &
0.84375 &  0.000 & 0.000 & 1.6875 \\
\end{tabular}
\label{B225}
\end{table}

\begin{table}
\caption[]{Flux-tube solutions for $\kappa=1.00$}
\begin{tabular}{ccccc|ccccc}
&\multicolumn{4}{c|}{Without node}&\multicolumn{5}{c}{With one node}\\
\tableline
$a$ & $b$ & $M/(2\pi)$ & $\Delta{\cal G}$ [$\pi \mu^2$] & $h_{\rm ext}$ &
$a$ & $b$ & $M/(2\pi)$ & $\Delta{\cal G}$ [$\pi \mu^2$] & $h_{\rm ext}$ \\
\tableline
   0.16666 &  0.00192 &   $-$0.0001 &    1.01 $10^{-9}$   &  0.3333 &
   0.09500 &  0.05752 &   $-$0.1982 &    0.0012           &  0.2061 \\ 
   0.09149 &  0.24083 &   $-$0.9999 &    0.1087           &  0.4408 &
   0.05000 &  0.19570 &   $-$1.6964 &    0.0968           &  0.2559 \\ 
   0.00770 &  0.38158 &   $-$1.9000 &    0.3749           &  0.5333 &
   0.00000 &  0.29595 &   $-$3.0598 &    0.3264           &  0.3041 \\ 
$-$0.00172 &  0.39592 &   $-$2.0000 &    0.4109           &  0.5426 &
$-$0.10000 &  0.46591 &   $-$5.5963 &    1.0575           &  0.3892 \\ 
$-$0.29562 &  0.76607 &  $-$12.0000 &    1.8288           &  0.7304 &
$-$0.28500 &  0.74590 &  $-$15.4375 &    3.8649           &  0.5416 \\ 
$-$0.30818 &  0.77728 &  $-$40.0000 &    1.3970           &  0.7216 &
$-$0.30825 &  0.77732 &  $-$56.9794 &    8.4137           &  0.6195 \\ 
$-$0.30828 &  0.77735 & $-$154.2204 &    0.2883           &  0.7145 &
   0.30840 &  0.77735 &  $-$39.2231 &    4.6106           &  0.6053 \\ 
   0.30828 &  0.77735 & $-$151.5651 & $-$3.0233           &  0.7130 &
   0.30000 &  0.74194 &  $-$10.8059 &    1.5583           &  0.5326 \\ 
   0.39637 &  0.72724 &   $-$1.5999 & $-$0.2007           &  0.8270 &
   0.18000 &  0.23311 &   $-$0.7157 &    0.0176           &  0.3582 \\ 
   0.49998 &  0.00999 &   $-$0.0001 & $-$2.49 $10^{-9}$   &  0.9999 &
   0.16670 &  0.01173 &   $-$0.0018 &    1.20 $10^{-7}$   &  0.3333 \\ 
\end{tabular}
\label{ft1}
\end{table}

\begin{table}
\caption[]{Flux-tube solutions for $\kappa=2.25$}
\begin{tabular}{ccccc|ccccc}
&\multicolumn{4}{c|}{Without node}&\multicolumn{5}{c}{With one node}\\
\tableline
$a$ & $b$ & $M/(2\pi)$ & $\Delta{\cal G}$ [$\pi \mu^2$] & $h_{\rm ext}$ &
$a$ & $b$ & $M/(2\pi)$ & $\Delta{\cal G}$ [$\pi \mu^2$] & $h_{\rm ext}$ \\
\tableline
   0.84325 & 0.03080 &  $-$0.0010 &    0.26 $10^{-7}$   & 1.6876 &
   0.50620 & 0.01167 &  $-$0.0003 &    0.22 $10^{-8}$   & 1.0125 \\ 
   0.43741 & 0.87389 &  $-$1.0000 &    0.0146           & 1.7754 &
   0.50000 & 0.13056 &  $-$0.0415 &    0.0000           & 1.0168 \\ 
   0.00000 & 1.23234 &  $-$3.0166 &    0.0421           & 1.8179 &
   0.00000 & 1.15113 &  $-$3.7504 &    0.1824           & 1.2955 \\ 
$-$0.01143 & 1.23977 &  $-$3.1000 &    0.0424           & 1.8181 &
$-$0.10000 & 1.24623 &  $-$4.8332 &    0.2573           & 1.3399 \\ 
$-$0.33764 & 1.42257 &  $-$8.0000 & $-$0.0170           & 1.7936 &
$-$0.45000 & 1.47366 & $-$17.7012 &    0.7550           & 1.4838 \\ 
$-$0.49780 & 1.49403 & $-$55.1538 & $-$1.0370           & 1.6803 &
$-$0.48750 & 1.48977 & $-$30.9911 &    0.9853           & 1.5089 \\ 
   0.50045 & 1.49512 & $-$49.9903 & $-$2.3225           & 1.6823 &
   0.50000 & 1.49494 & $-$59.2582 &    0.1456           & 1.5301 \\ 
   0.70035 & 1.56827 &  $-$5.0000 & $-$0.8568           & 1.9461 &
   0.55000 & 1.51459 & $-$14.7491 & $-$0.2253           & 1.5043 \\ 
   1.13578 & 1.65419 &  $-$1.2000 & $-$0.3092           & 2.5200 &
   0.70000 & 1.52843 &  $-$4.5843 & $-$0.0988           & 1.5491 \\ 
   2.53100 & 0.03330 &  $-$0.0000 & $-$4.27 $10^{-9}$   & 5.0620 &
   0.84373 & 0.03177 &  $-$0.0005 &    0.0000           & 1.6874 \\ 
\end{tabular}
\label{ft225}
\end{table}

\begin{table}
\caption[]{Oscillating solutions for $\kappa =1.0$ 
($R$ gives the positions of extrema).} 
\begin{tabular}{c|cc|cc}
Type A&$a$ & $b$ & $R$ (1st max) & ${\cal E}_{\rm mag}$ [$\pi \mu^2$]\\
\tableline
              &$-$0.30828 & 0.77735 &$\infty$ &0.24523 \\
              &$-$0.30700 & 0.77500 & 3.36848 &0.24328 \\
$F/(2\pi)=-1$ &$-$0.11438 & 0.40000 & 1.85499 &0.04429 \\
              &$-$0.03500 & 0.20000 & 1.83312 &0.00579 \\
              &$-$0.00246 & 0.05000 & 1.84246 &0.00005 \\
\tableline\tableline
Type B& $R$ (min)& ${\cal G}$[$\pi\mu^2$] &$R$ (max)& ${\cal G}$[$\pi\mu^2$]\\
\tableline\tableline
$\{a=0.0,$ &   3.84 & $-$2.484 $10^{-5}$ &   7.02 &  $-$2.978 $10^{-5}$ \\
$b=0.1\}$  &  98.2  & $-$5.009 $10^{-5}$ & 208.1  &  $-$5.583 $10^{-5}$ \\
\tableline
$\{a=0.0,$ &   5.13 & $-$0.54496         &   8.48 &  $-$0.62000 \\
$b=0.9\}$  & 295.5  & $-$1.08209         & 198.0  &  $-$1.03113 \\
\end{tabular}
\label{tabos}
\end{table}

\pagebreak
\centerline{\bf ADDENDUM: COMPLETE SET OF FIGURES AND TABLES}
\bigskip
\setcounter{figure}{0}

\begin{figure}
\caption[]
{A typical vortex solution. At the center, the magnetic field $h$ is
maximal. The limit of the scalar field $\varphi $ is 1, which means
physically a completely superconducting state. With our gauge choice
the potential is vanishing at infinity.}
\label{fig1c}
\end{figure}

\begin{figure}
\caption[]
{The scalar field $\varphi $ and the normalized magnetic field $h/(2a)$ for
several values of $\kappa $. From these curves, one can read
off the penetration lengths $\delta_{cal}$ of the scalar field and
$\xi_{cal}$ of the magnetic field which are determined at the $1/e$ level.}
\label{fig2c}
\end{figure}

\begin{figure}
\caption[]
{The energy per unit length for vortices with flux $F/(2\pi )=-1$ ($N=1$) 
and $F/(2\pi )=-2$ ($N=2$). For comparison, the double energy value of the
1-vortex is also given. One recognizes clearly that for $\kappa >1/\sqrt{2}$ 
(the vertical line) a 1-vortex is energetically more preferable than a 
2-vortex.}
\label{fig3c}
\end{figure}

\begin{figure}
\caption[]
{The comparison of the formal parameter $\kappa $, which appears
in the Ginzburg-Landau system, and of the ratio $\delta_{cal}/\xi_{cal}$
calculated from the solution of the Ginzburg-Landau system. Near 
$\kappa =1.0$, we find an essential deviation for $F/(2\pi )=-1,-2$.}
\label{fig4c}
\end{figure}

\begin{figure}
\caption[]
{Type B solutions without node for $\kappa =1.0$: The scalar field $\varphi$
with initial values $\varphi (0)=0.9, 0.6, 0.3$ and $f(0)=0.0$.}
\label{fig5c}
\end{figure}

\begin{figure}
\caption[]
{The same type B solutions without node for $\kappa =1.0$ as in
Fig.~\ref{fig5c}: The potential $f$ and the magnetic field $h$.}
\label{fig6c}
\end{figure}

\begin{figure}
\caption[]
{Type B solutions without node with very high magnetization. The scalar
field $\varphi $ has initial values which are almost 1 (see Table V of
Addendum). With increasing magnetization the scalar field has a wider central 
core where it is almost 1, i.e.~full superconductivity at the center. When 
the scalar field becomes zero, the magnetic field achieves a constant value.}
\label{fig7c}
\end{figure}

\begin{figure}
\caption[]
{Type B solutions with one node for $\kappa =1.0$: The scalar field
$\varphi $ with initial values $\varphi (0)=0.9, 0.6, 0.3$ and $f(0)=0.0$.}
\label{fig8c}
\end{figure}

\begin{figure}
\caption[]
{The same type B solutions with one node for $\kappa =1.0$  as in
Fig.~\ref{fig8c}: The potential $f$ and the magnetic field $h$.
At zeros of the scalar field, the magnetic field has
a step-like behavior. For the constant magnetic field, the potential
grows as $r^2$.}
\label{fig9c}
\end{figure}

\begin{figure}
\caption[]
{Flux tube solutions without node: The scalar field $\varphi $
for $\kappa =0.5, 1.0, 1.5$ with the same magnetization $M/(2\pi)=6.0$. The 
scalar field of a flux tube produces a maximum before going to zero (and not 
to 1, as in the case of vortices).}
\label{fig10c}
\end{figure}

\begin{figure}
\caption[]
{Flux tube solutions without node: The potential $f$ and the magnetic
field $h$ for $\kappa =0.5, 1.0, 1.5$ with the same magnetization
$M/(2\pi)=6.0$. The potential starts at 1, is followed by a minimum and
then grows with an asymptotic $r^2$ behavior, so that the magnetic
field is constant.}
\label{fig11c}
\end{figure}

\begin{figure}
\caption[]
{Flux tube solutions without node: Configurations with external magnetic
field near $h_c$. Both solutions are close to the vortex solution for 
$\kappa =1.0$ which `lies' between these two flux tube solutions;
cf.~Table~XIII of Addendum. The dotted curve has a negative Gibbs free energy,
while the drawn curve has positive Gibbs free energy. Coming closer and 
closer to the vortex, the magnetization increases without a limit, while 
the scalar field at the center approaches the vortex configuration, and 
simultaneously the interval grows where it is almost 1. Such a limit flux
tube consists then of a vortex in the center and a type B solution at
higher radial values. The initial values $(a,b)$ of this solution are 
close to that of the vortex; cf.~Table~I of Addendum.}
\label{fig12c}
\end{figure}

\begin{figure}
\caption[]
{The same flux tube solutions without node as in Fig. \ref{fig12c}:
potential and magnetic field configurations.}
\label{fig13c}
\end{figure}

\begin{figure}
\caption[]
{Flux tube solutions with one node: A solution near $h_c$
with negative Gibbs free energy. The scalar field has one node while 
the magnetic field has a `step' at the node's position. Again one can
describe this solution as a combination of a vortex at the center and
a type B solution outwards.}
\label{fig14c}
\end{figure}

\begin{figure}
\caption[]
{Magnetization curves for type B solutions without (right curve) and with 
one node (left curve) for different $\kappa$. For $\kappa =2.25$, we
notice a change of the sign of the Gibbs free energy within a solution
family. For $\kappa =0.5, 1.0, 1.5$ we find only positive values of 
${\cal G}$.}
\label{fig15c}
\end{figure}

\begin{figure}
\caption[]
{The magnetization curve for flux tube solutions with 
$\kappa =0.5, 1.0, 1.5$ for a scalar field without node.
For each $\kappa$, the corresponding limit values $h_1$, $h_2$, $h_{c}$, and 
$h_{c_2}$ is drawn. For instance, the curve for $\kappa =1.0$ has a branch 
with a negative Gibbs free energy in the range [$h_{c},h_{c_2}$]=[0.707,1.0] 
of the external magnetic field and a branch with positive values in the range 
of [$h_{1},h_{c}$]=[1/3,0.707]. The magnetization diverges at $h_{c}$.
For $\kappa =0.5$ the point $h_c$ does not lie within the two limits $h_1$ 
and $h_{c_2}$ so that one can find also solutions up to $h_c$.
The lower figure displays in detail how the branches of solutions with 
positive and negative $\Delta{\cal G}$ behave near the $h_c$ limit 
($\kappa=1$).}
\label{fig16c}
\end{figure}

\begin{figure}
\caption[]
{Magnetization curve for flux tube solutions with $\kappa =1.0$ for a scalar 
field without (limits [$h_1,h_{c_2}$]) and with one node (limits [$h_2,h_1$]).
In both cases the magnetization diverges at the external magnetic field 
$h_c$ which is the limiting point for the vortex.}
\label{fig17c}
\end{figure}

\begin{figure}
\caption[]
{$\Delta {\cal G}$ against the external magnetic field for
$\kappa =1.0$ for the flux tube and the type B solutions.}
\label{fig18c}
\end{figure}
 
\begin{figure}
\caption[]
{$\Delta {\cal G}$ against the external magnetic field for
$\kappa =2.25$ for the flux tubes and the type B solutions.
The lower figure gives an enlarged view of the flux tube curve with one node, 
while the broken line denotes type B solutions with one node. 
In each case the flux tubes have lower values of $\Delta{\cal G}$ than the 
type B solutions. Hence, flux tubes are energetically more preferable.} 
\label{fig19c}
\end{figure}

\begin{figure}
\caption[]
{$\Delta {\cal G}$ against the external magnetic field for $\kappa =5.0$ 
flux tubes without and with one node. Both curves limits into $h_c$.}
\label{fig20c}
\end{figure}

\begin{figure}
\caption[]
{$(a,b)$ diagrams for $\kappa =1.0, 2.25, 5.0$. The big dot describes
in each case the corresponding vortex solution. The drawn, broken and dotted 
lines represent the flux tubes without node, the flux tubes with one node, 
and the oscillating solutions, respectively. The type B solutions cannot be
compared in these diagrams because they have different initial values.}
\label{fig21c}
\end{figure}

\begin{figure}
\caption[]
{The oscillating solutions for type A initial conditions. For an
initial value (here, $b=0.777$) very near to the one of the vortex, the
scalar field has a part which is almost 1 before the oscillation
starts. At the center, the vortex can be recognized. No oscillating
solution with an initial value above that of the vortex can be found.}
\label{fig22c}
\end{figure}

\begin{figure}
\caption[]
{This figure shows what happens with the magnetic field for an
oscillating scalar field. At the zeros of the scalar field, the
magnetic field has a step-like behavior. As in the case of the
vortex, the magnetic field quickly vanishes, and the flux of the
oscillating solution is quantized. Here, we have $F/(2\pi )=-1$.}
\label{fig23c}
\end{figure}

\begin{figure}
\caption[]
{The oscillating solution for type B. Here, a solution
with a non-vanishing magnetic field does not exist. The square of 
the scalar field, the density of superconducting electrons, oscillates 
without any external
magnetic field, and every finite sample has a negative Gibbs free energy.}
\label{fig24c}
\end{figure}

\begin{figure}
\caption[]
{The general diagram for different solutions.
The vortex state is energetically most preferable between $H_{c_1}$ and 
$H_c$, while above $H_c$, the flux tube configurations replace them. Hence, 
$H_c$ gains the following physical meaning for a type II superconductor: 
Above $H_c$, isolated vortices come into contact with the external
magnetic field and the flux tube solutions are constructed. Energetically,
the most preferable solution is a flux tube without node. But
also the flux tubes with nodes do exist there, if $\kappa $ is large
enough. Near $H_{c_2}$, only the flux tube without node exists. Of
course, the flux tubes `live' also in the vortex and the Meissner
state but they are energetically less preferable. In this way, we find
a rich fine structure of a superconductor's mixed state.}
\label{fig25c}
\end{figure}   

\pagebreak

\input psfig

\centerline{Figure 1:}
\vskip3cm
\centerline{\psfig{figure=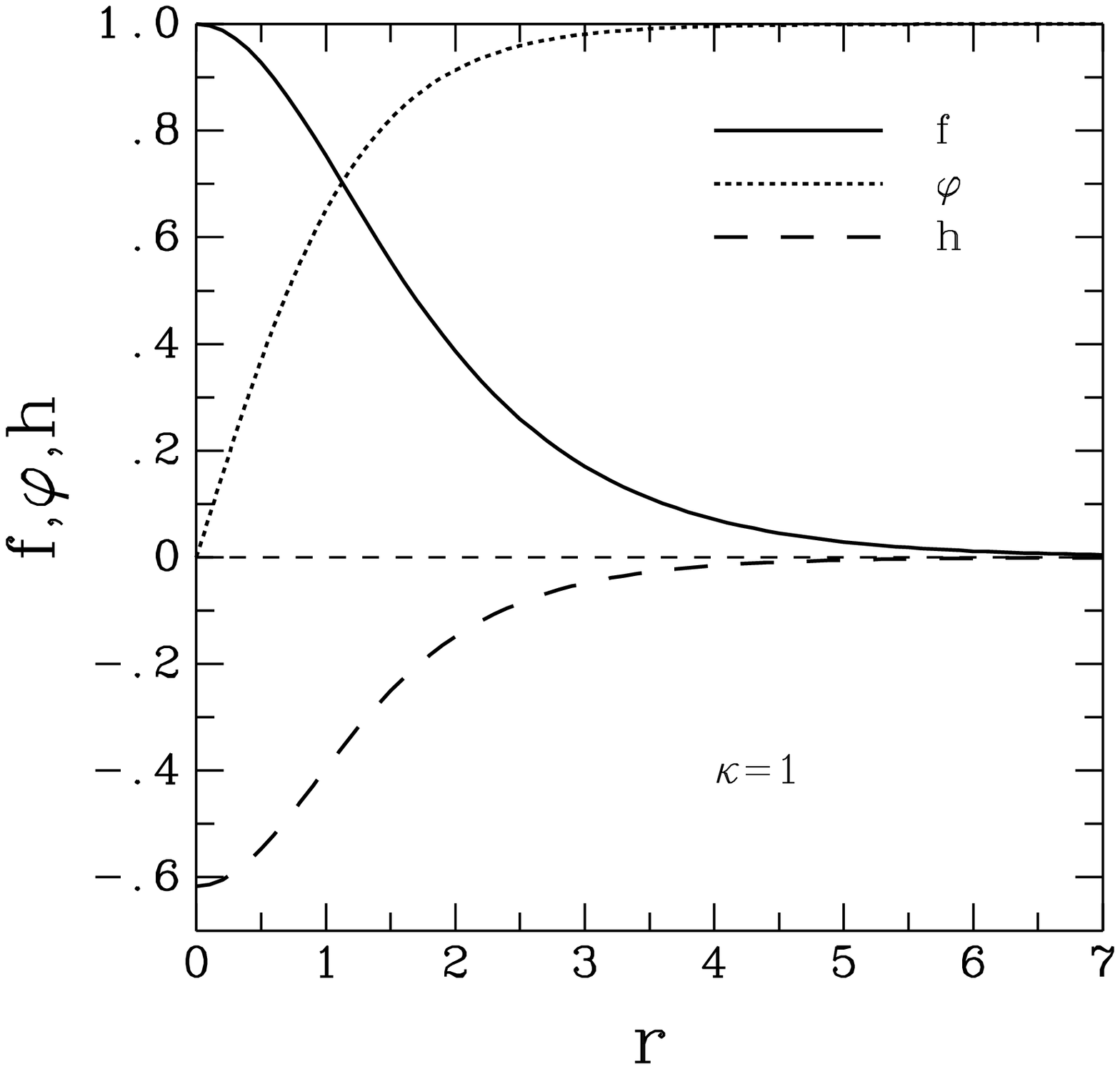,height=8cm} \hskip 2cm}
\pagebreak

\centerline{Figure 2:}
\centerline{\psfig{figure=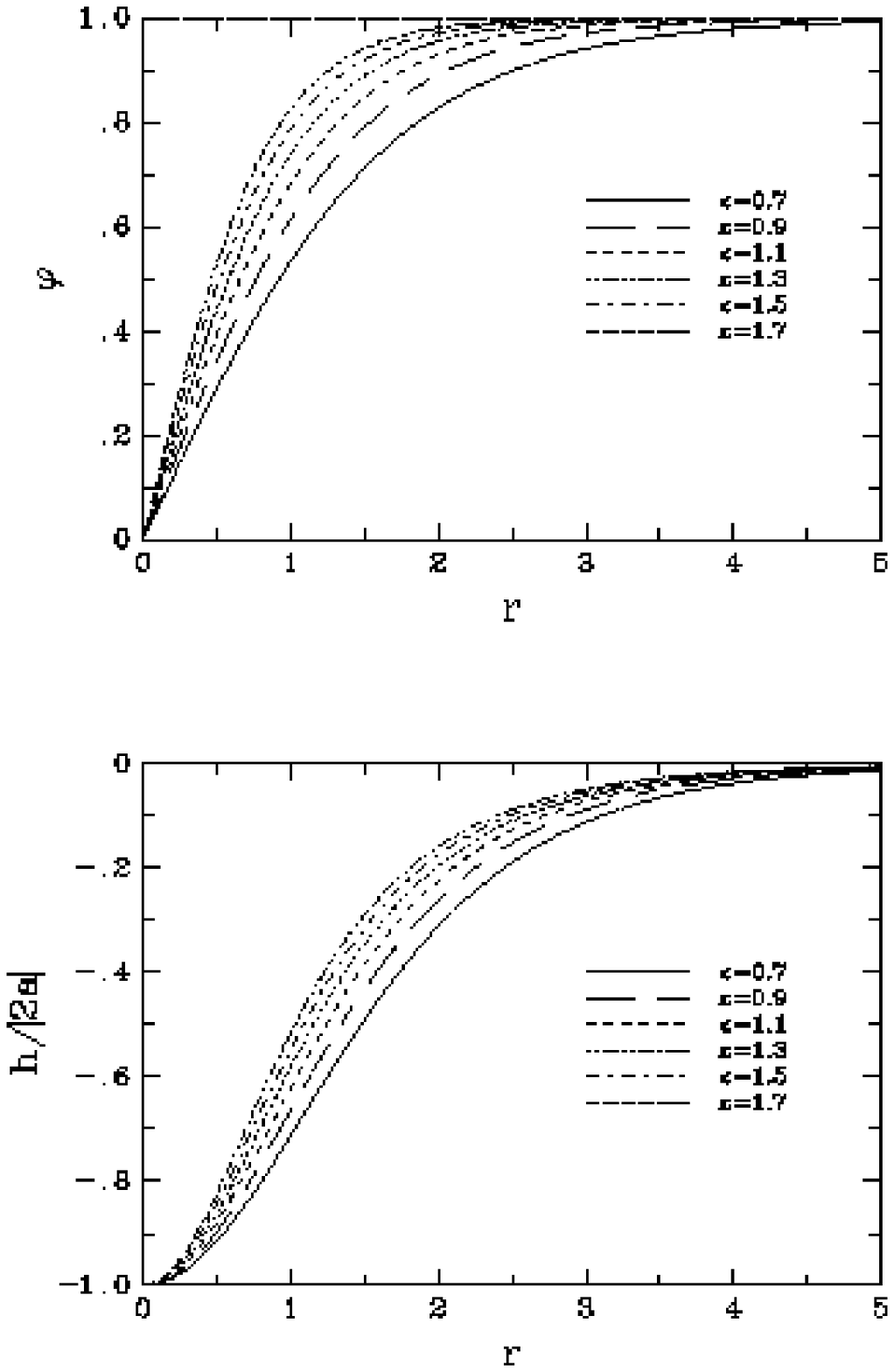,height=8cm} \hskip 2cm}
\pagebreak

\centerline{Figure 3:}
\vskip3cm
\centerline{\psfig{figure=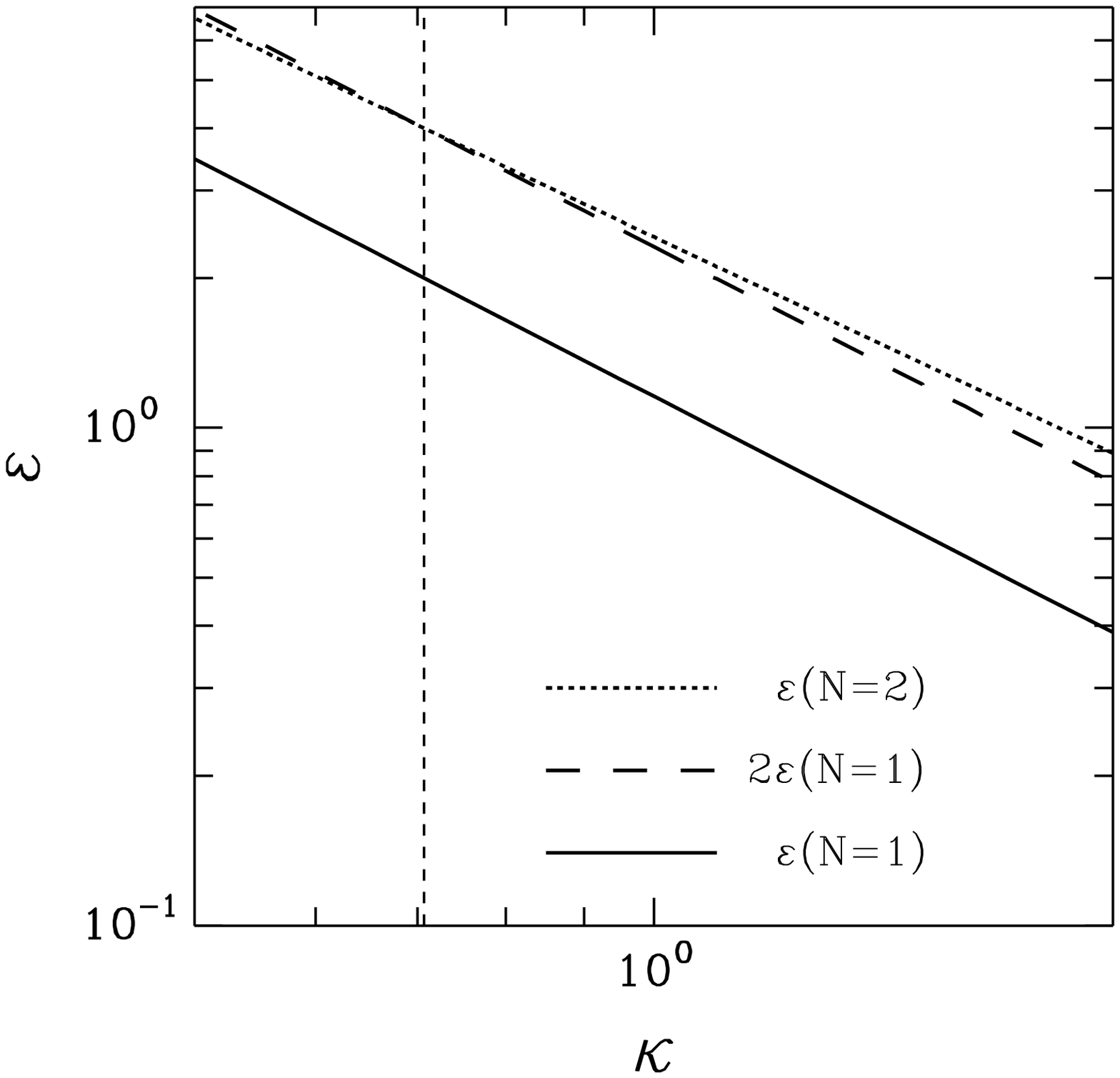,height=8cm} \hskip 2cm}
\pagebreak

\centerline{Figure 4:}
\vskip3cm
\centerline{\psfig{figure=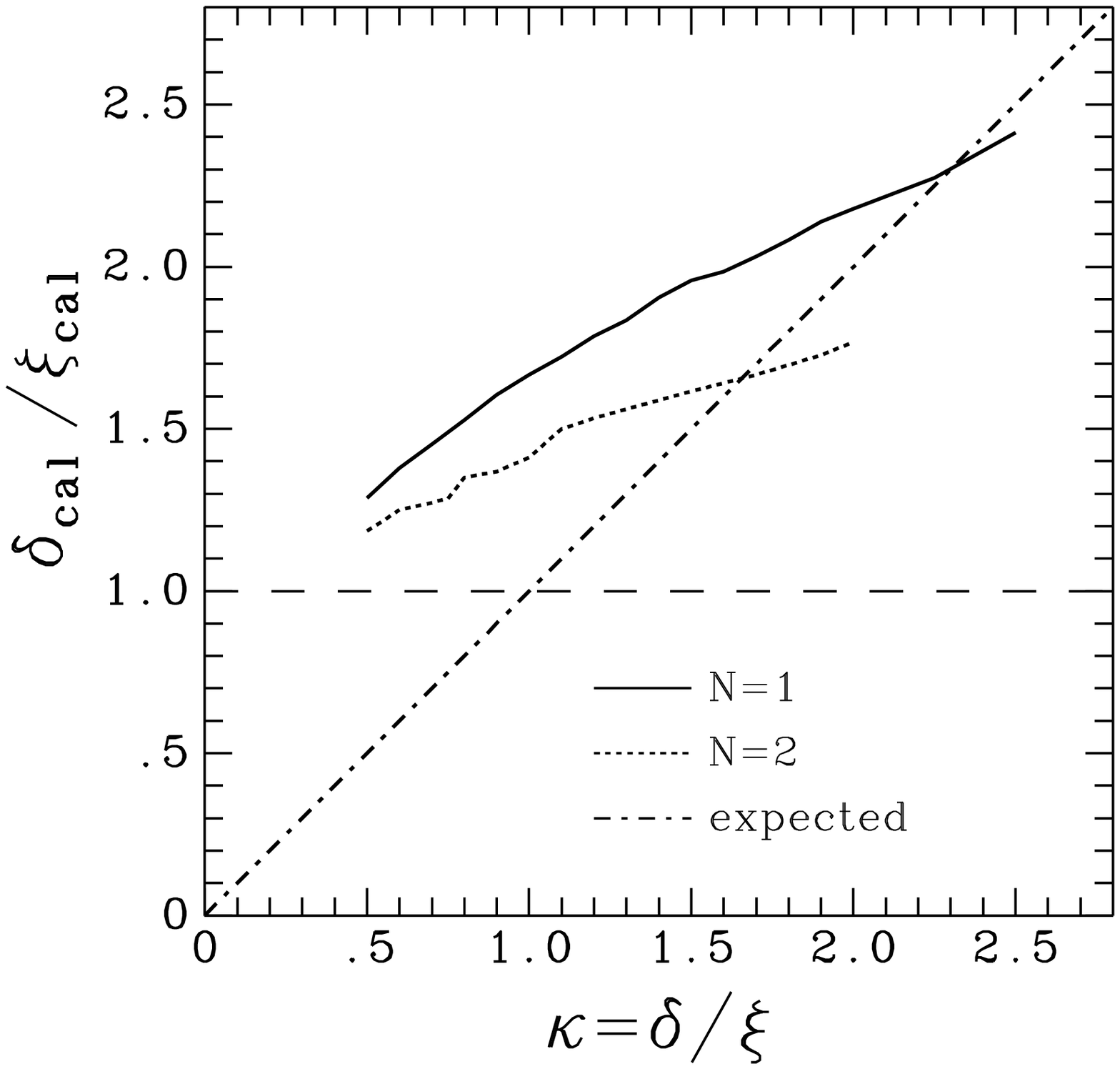,height=8cm} \hskip 2cm}
\pagebreak

\centerline{Figure 5:}
\vskip3cm
\centerline{\psfig{figure=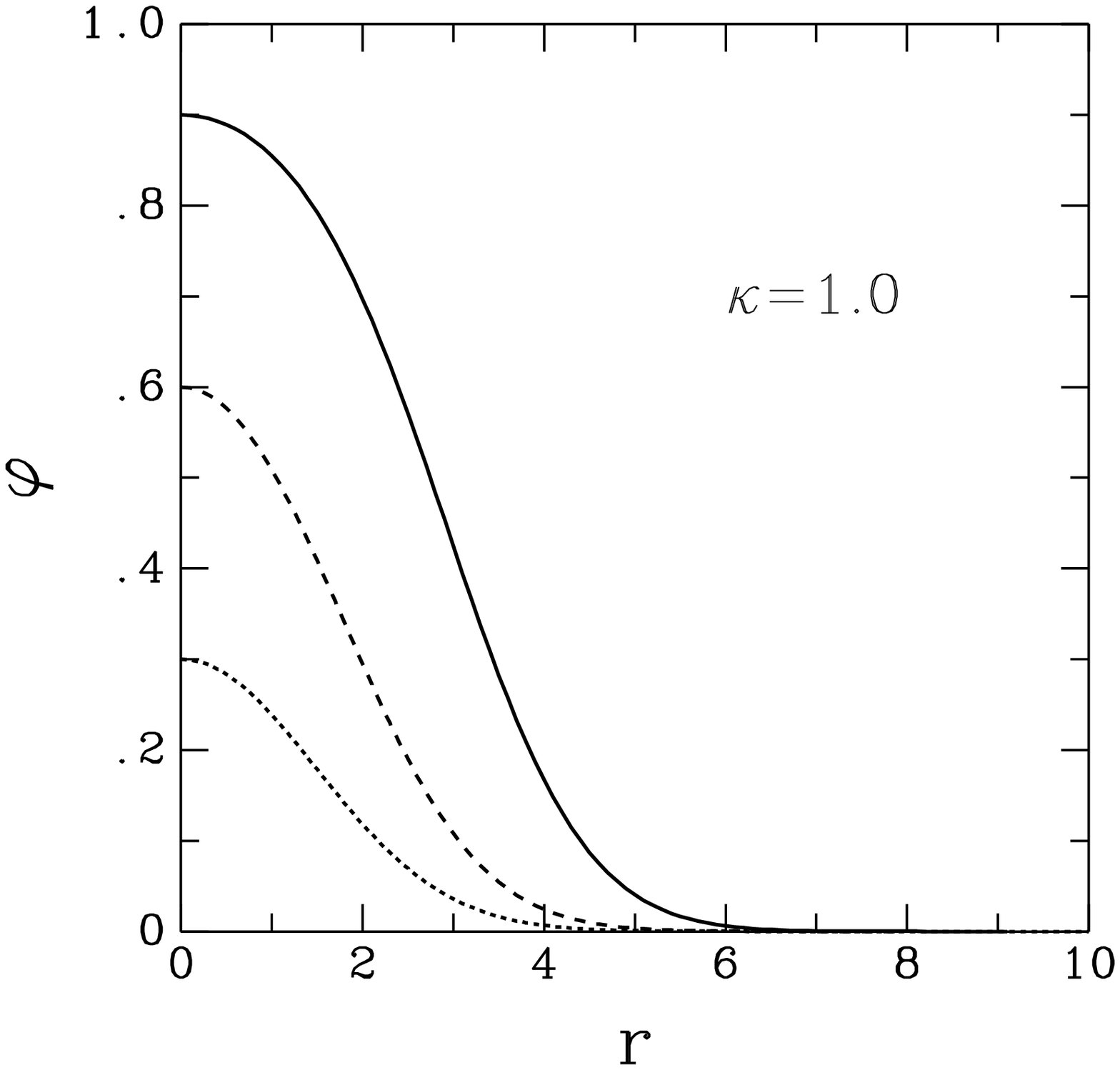,height=8cm} \hskip 2cm}
\pagebreak

\centerline{Figure 6:}
\vskip5cm
\centerline{\psfig{figure=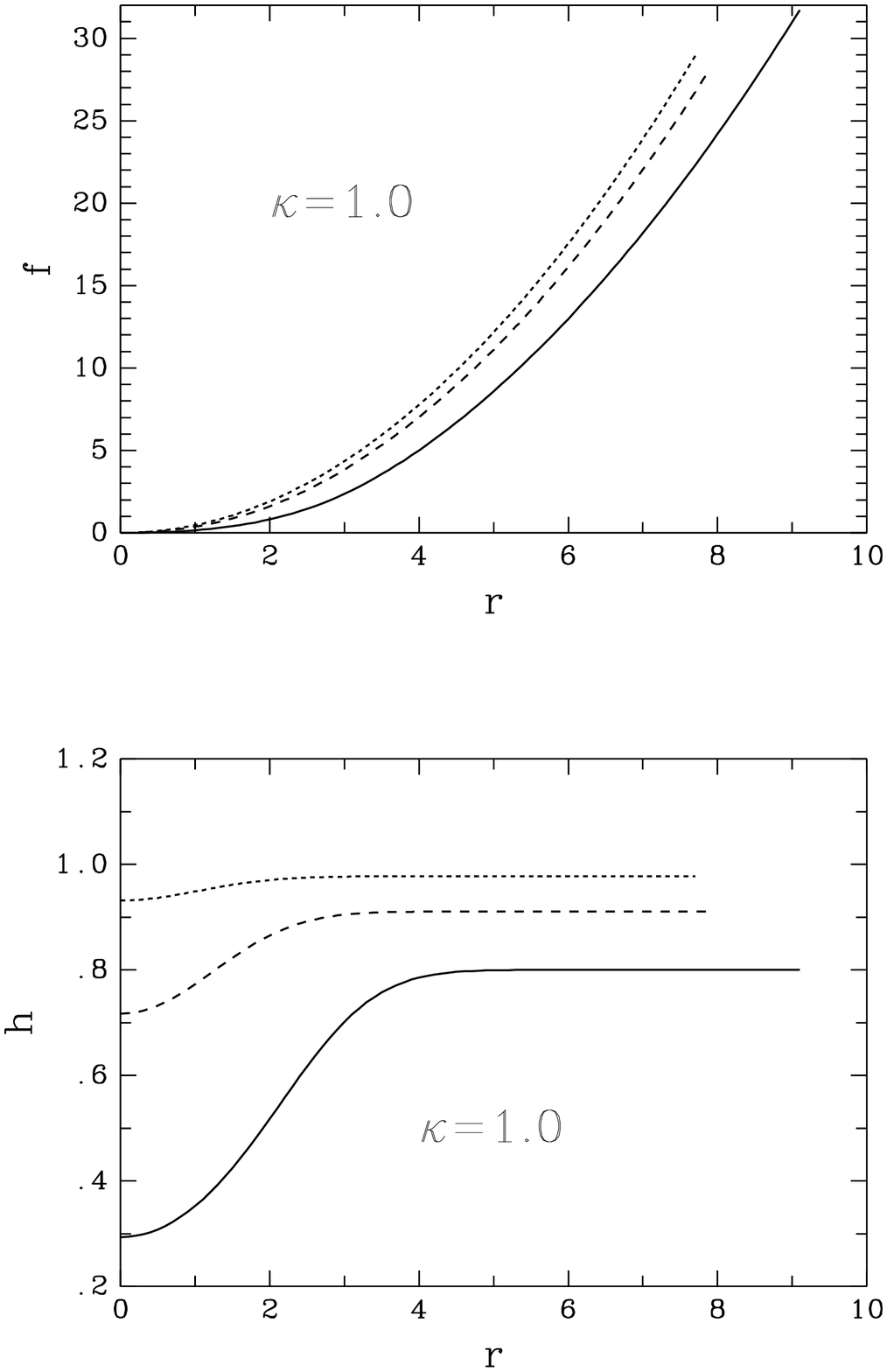,height=8cm} \hskip 2cm}
\pagebreak

\centerline{Figure 7:}
\centerline{\psfig{figure=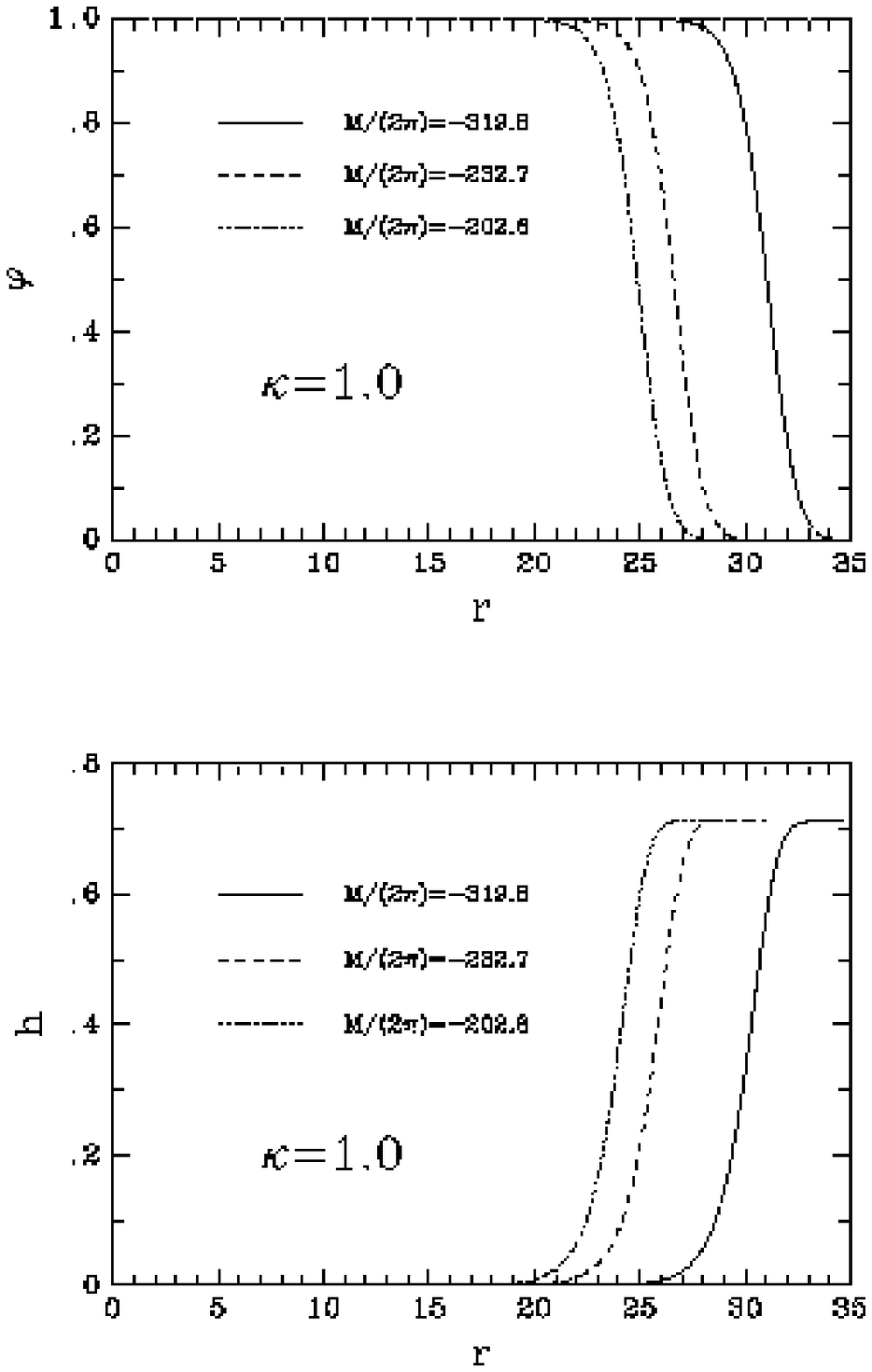,height=8cm} \hskip 2cm}
\pagebreak

\centerline{Figure 8:}
\vskip3cm
\centerline{\psfig{figure=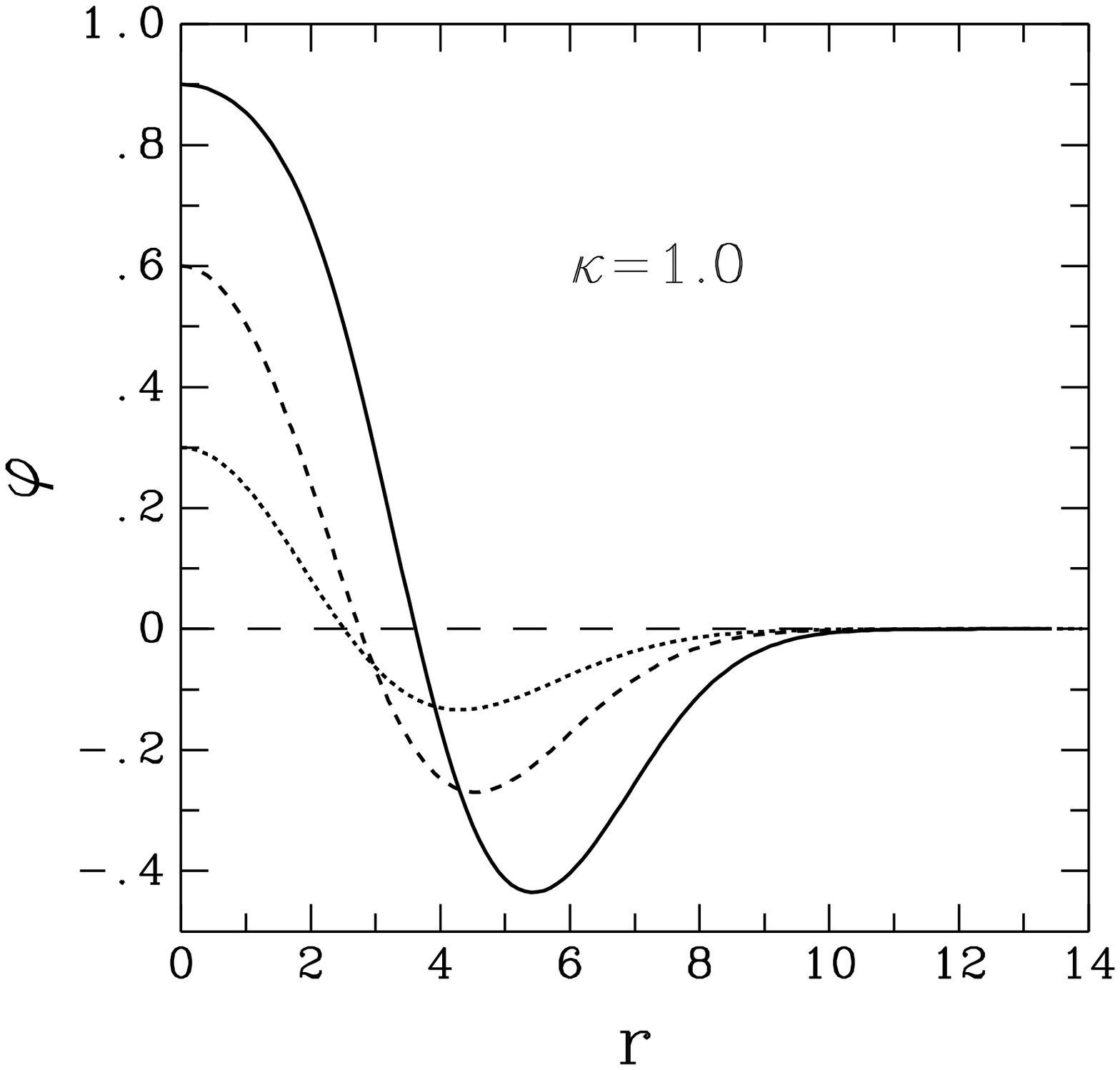,height=8cm} \hskip 2cm}
\pagebreak

\centerline{Figure 9:}
\vskip5cm
\centerline{\psfig{figure=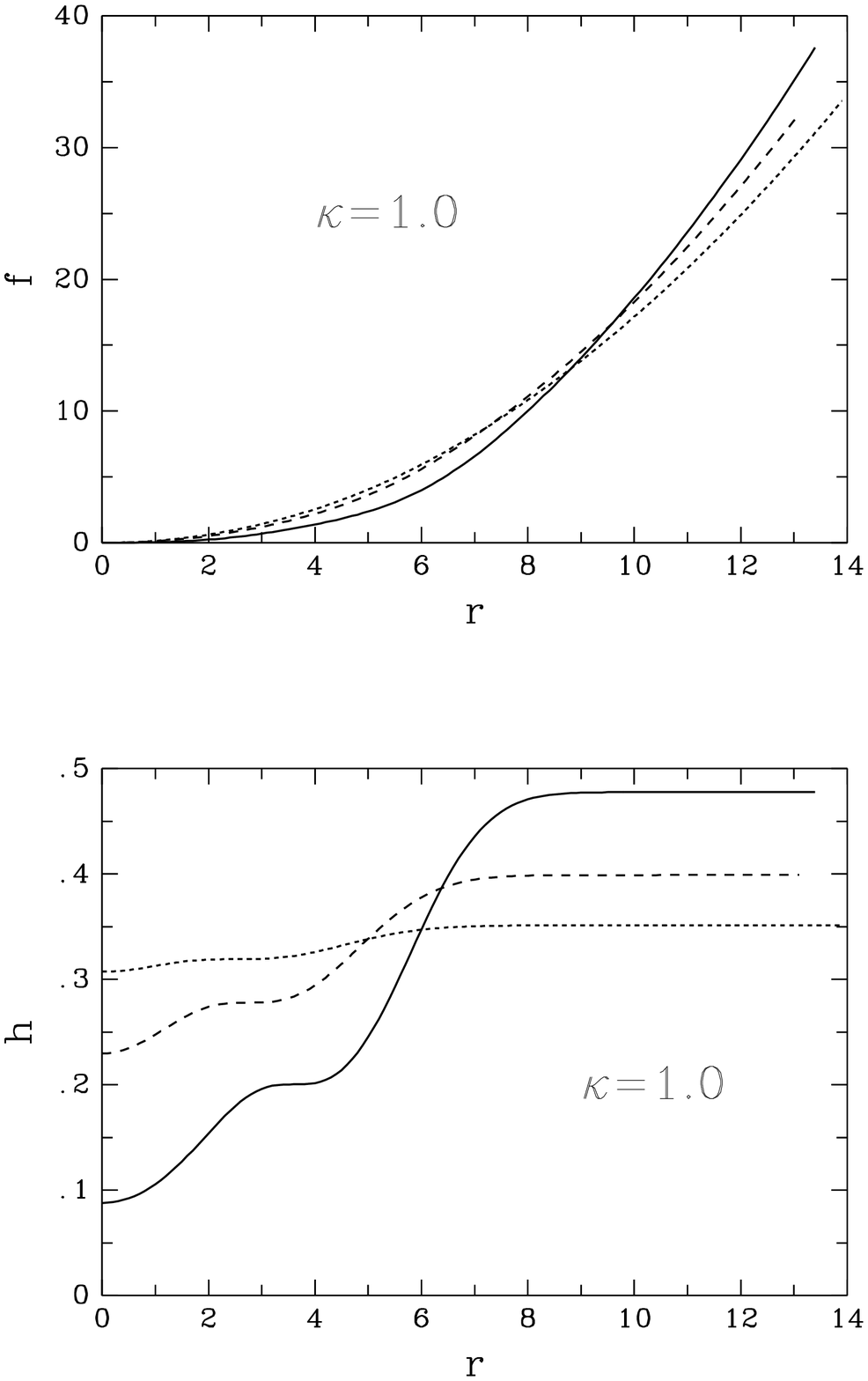,height=8cm} \hskip 2cm}
\pagebreak

\centerline{Figure 10:}
\vskip3cm
\centerline{\psfig{figure=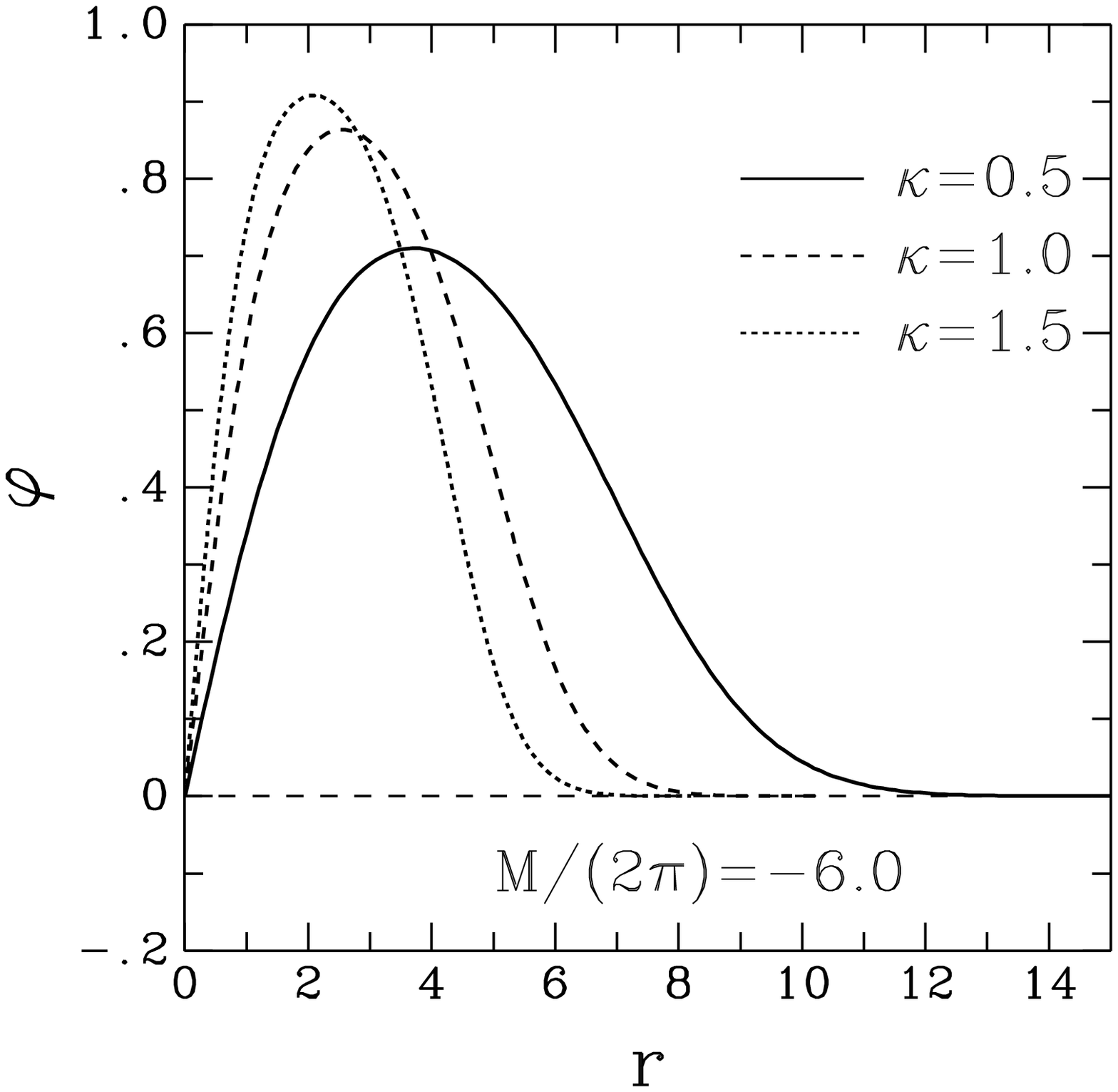,height=8cm} \hskip 2cm}
\pagebreak

\centerline{Figure 11:}
\centerline{\psfig{figure=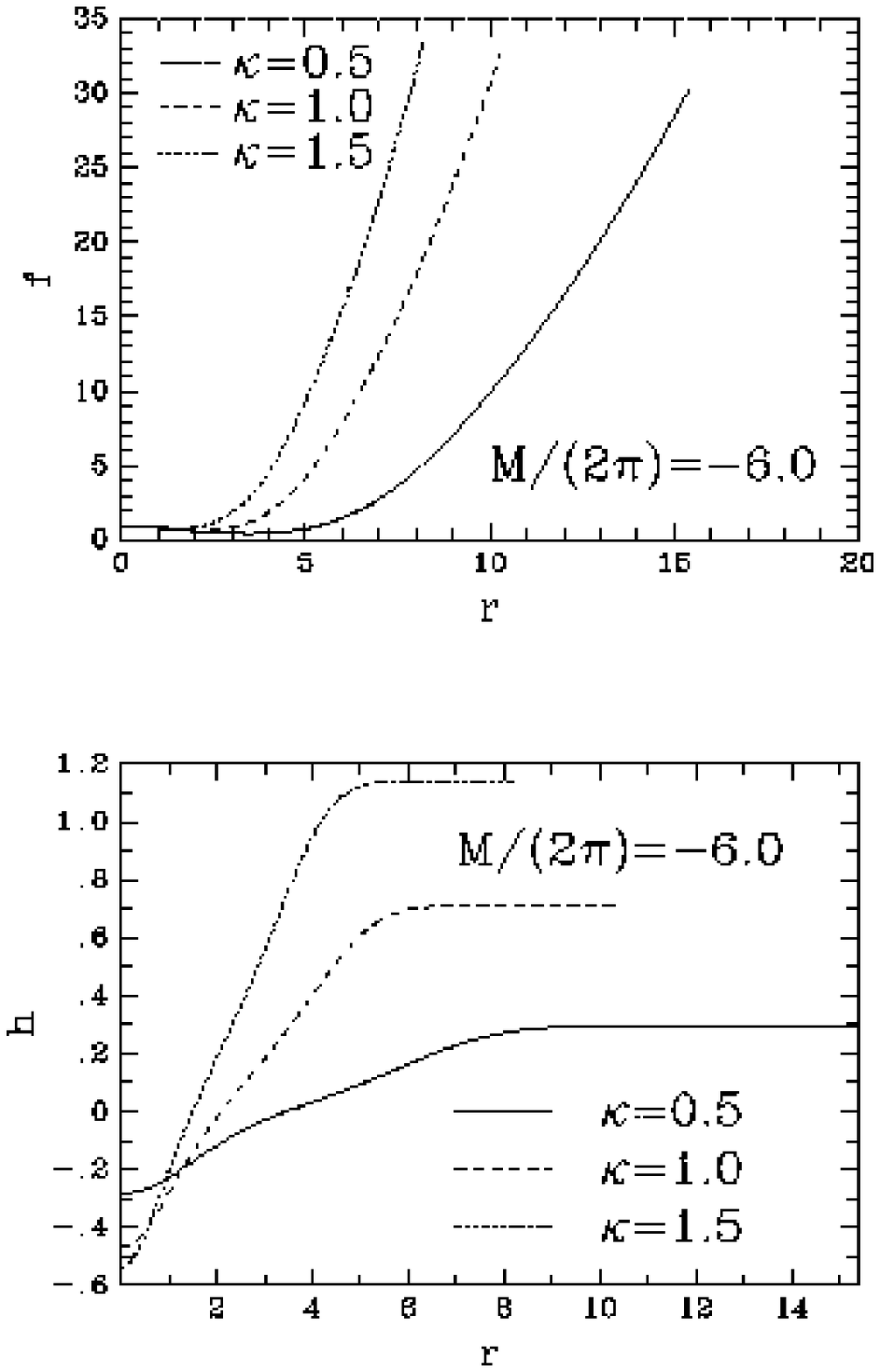,height=8cm} \hskip 2cm}
\pagebreak

\centerline{Figure 12:}
\vskip3cm
\centerline{\psfig{figure=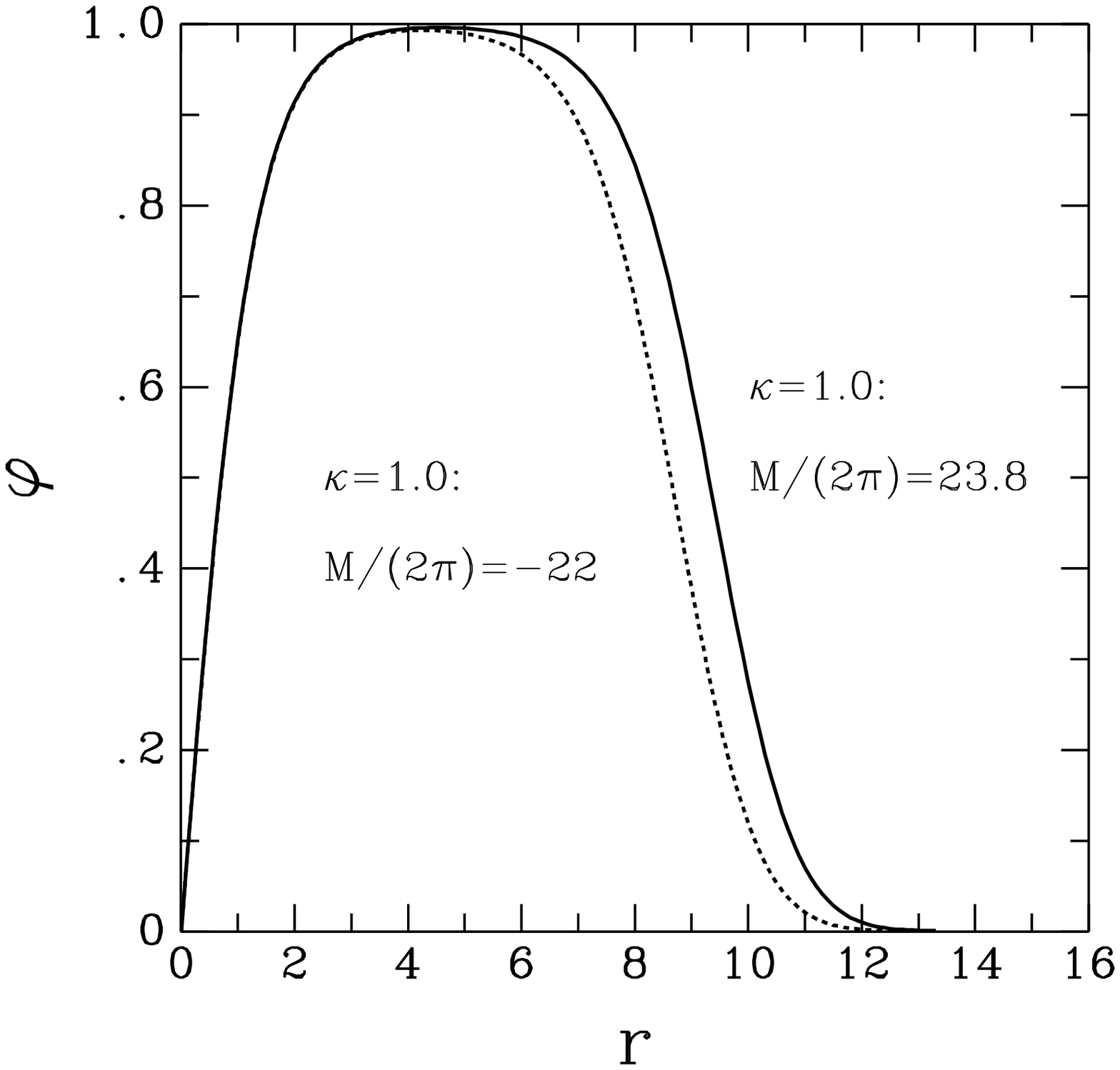,height=8cm} \hskip 2cm}
\pagebreak

\centerline{Figure 13:}
\centerline{\psfig{figure=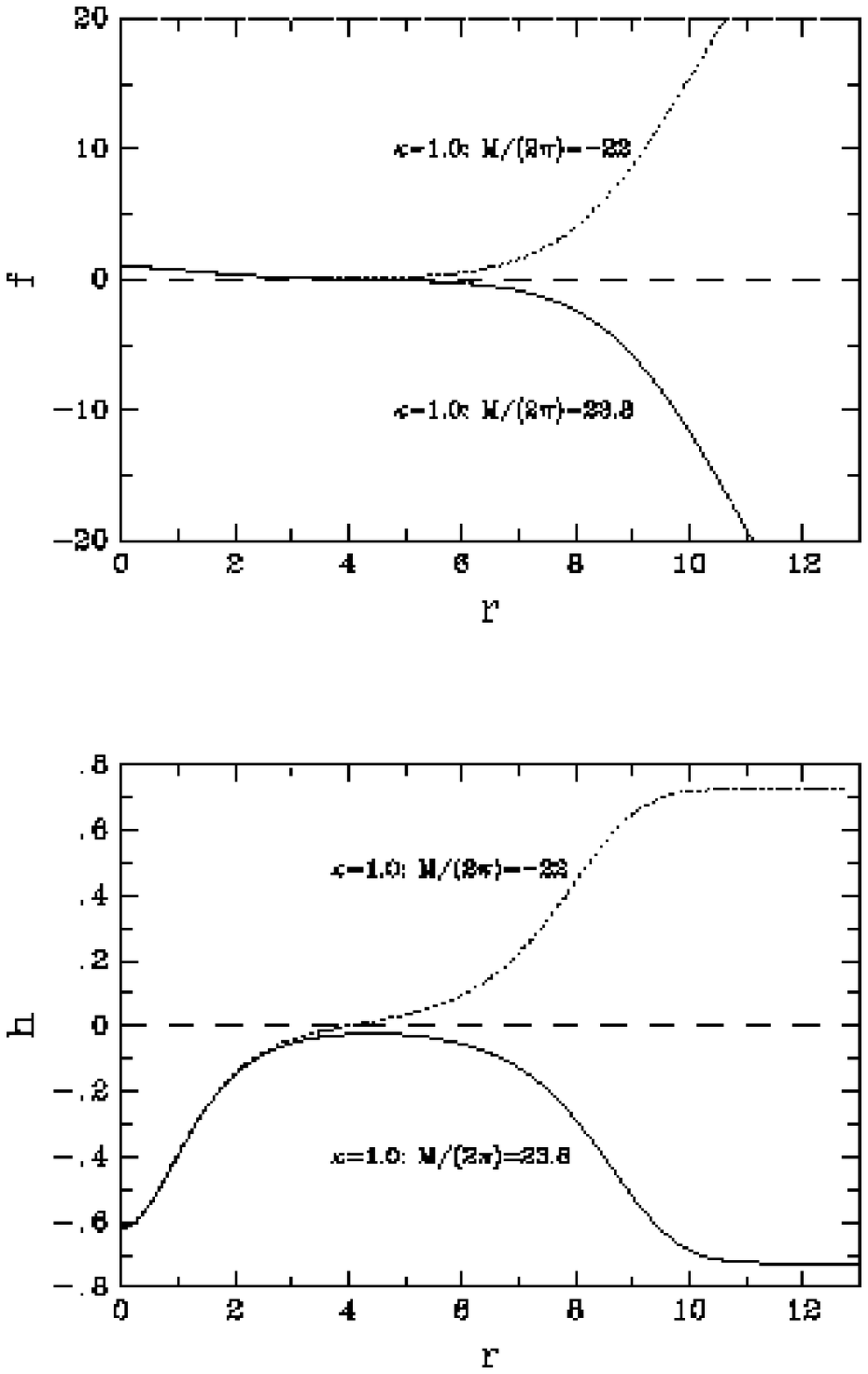,height=8cm} \hskip 2cm}
\pagebreak

\centerline{Figure 14:}
\vskip5cm
\centerline{\psfig{figure=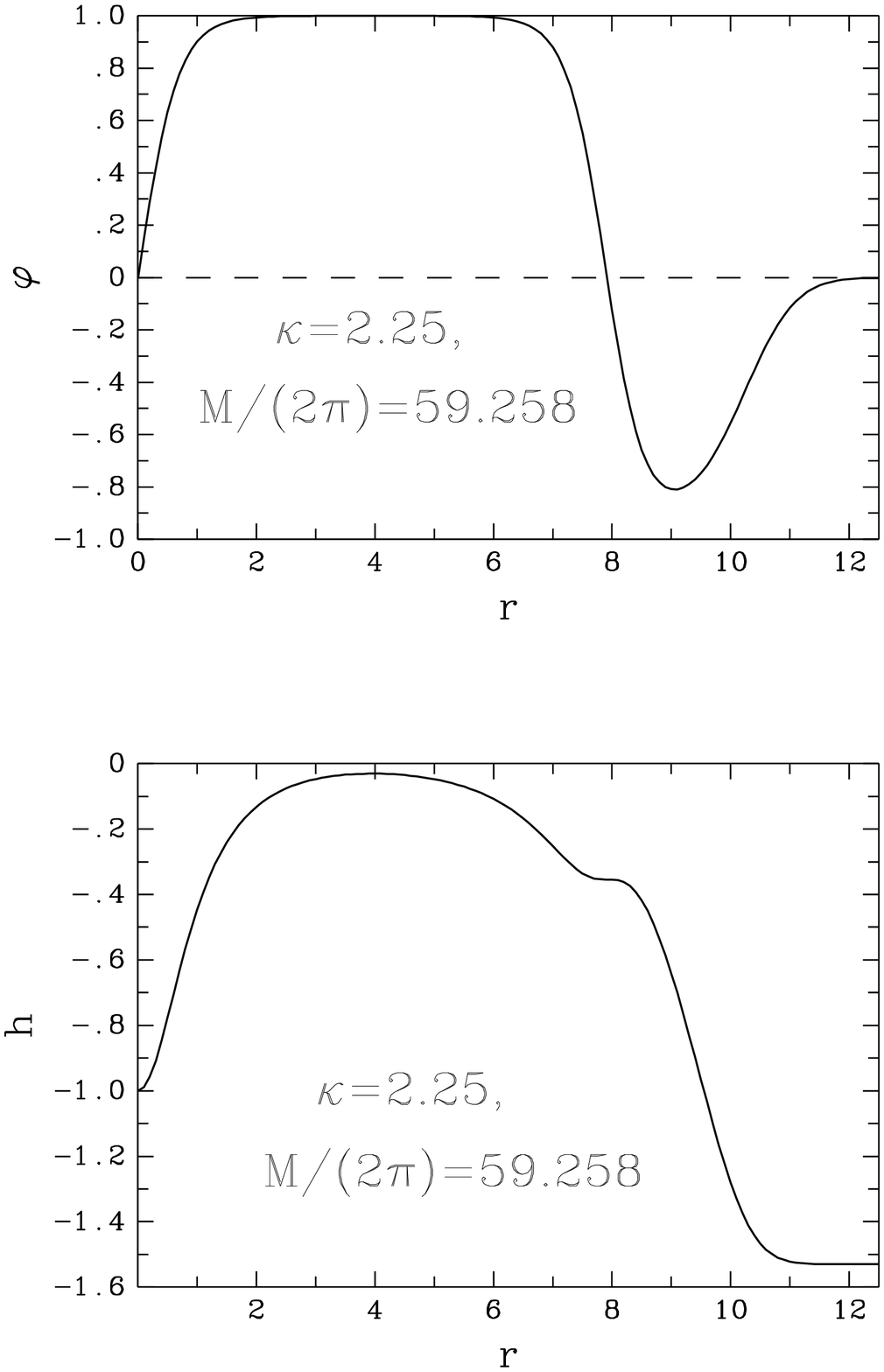,height=8cm} \hskip 2cm}
\pagebreak

\centerline{Figure 15:}
\centerline{\psfig{figure=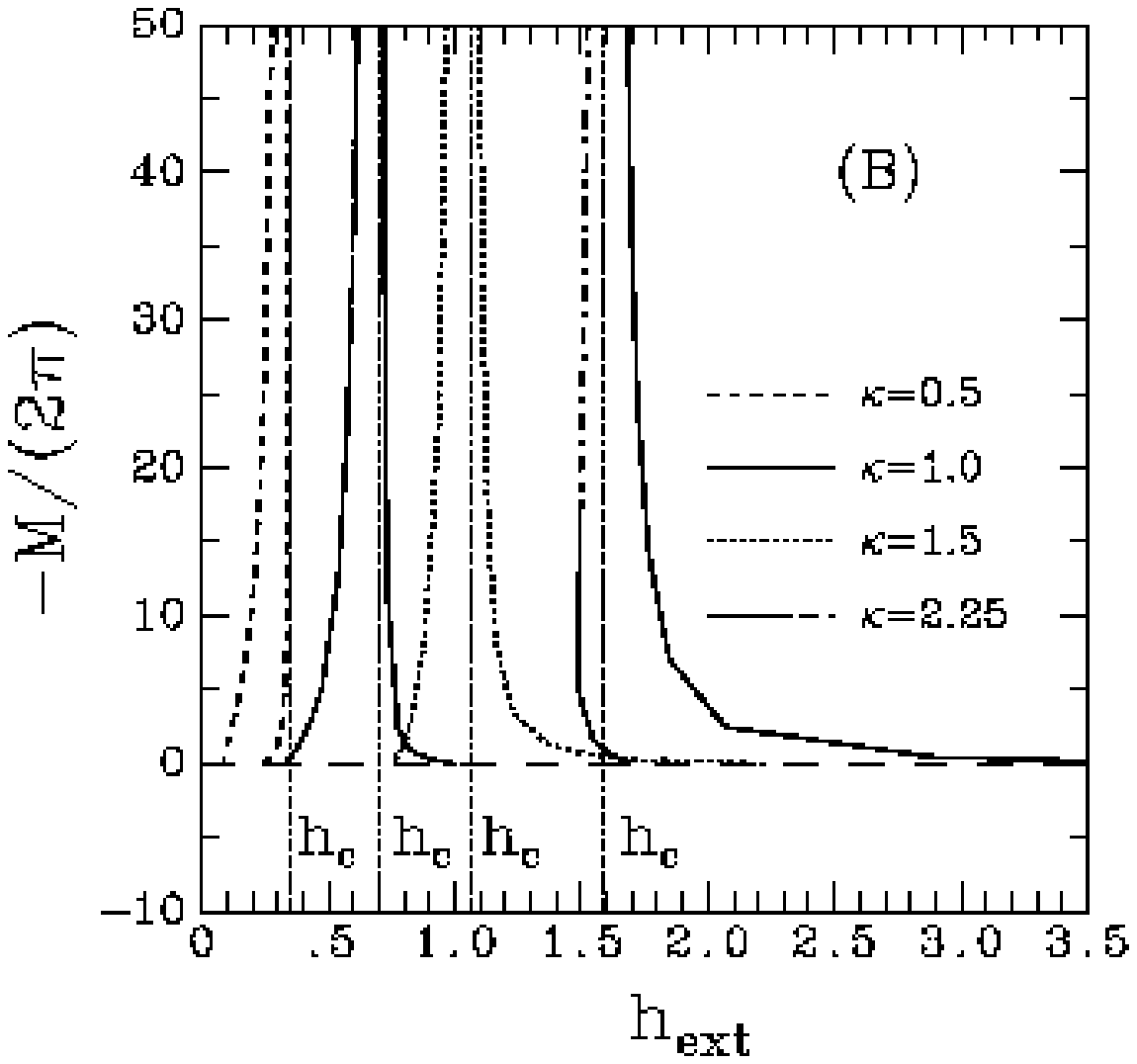,height=8cm} \hskip 2cm}
\pagebreak

\centerline{Figure 16:}
\centerline{\psfig{figure=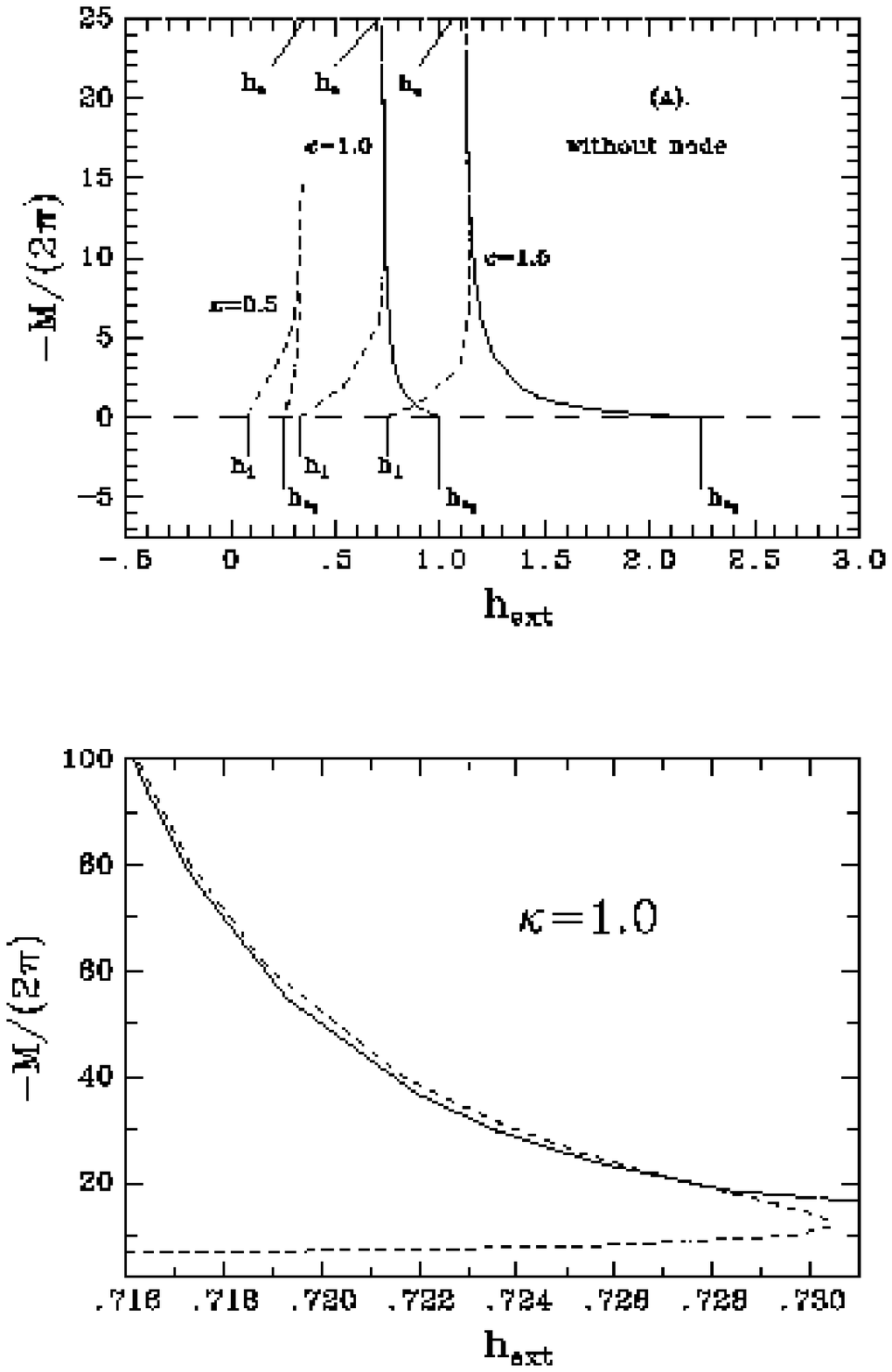,height=8cm} \hskip 2cm}
\pagebreak

\centerline{Figure 17:}
\vskip3cm
\centerline{\psfig{figure=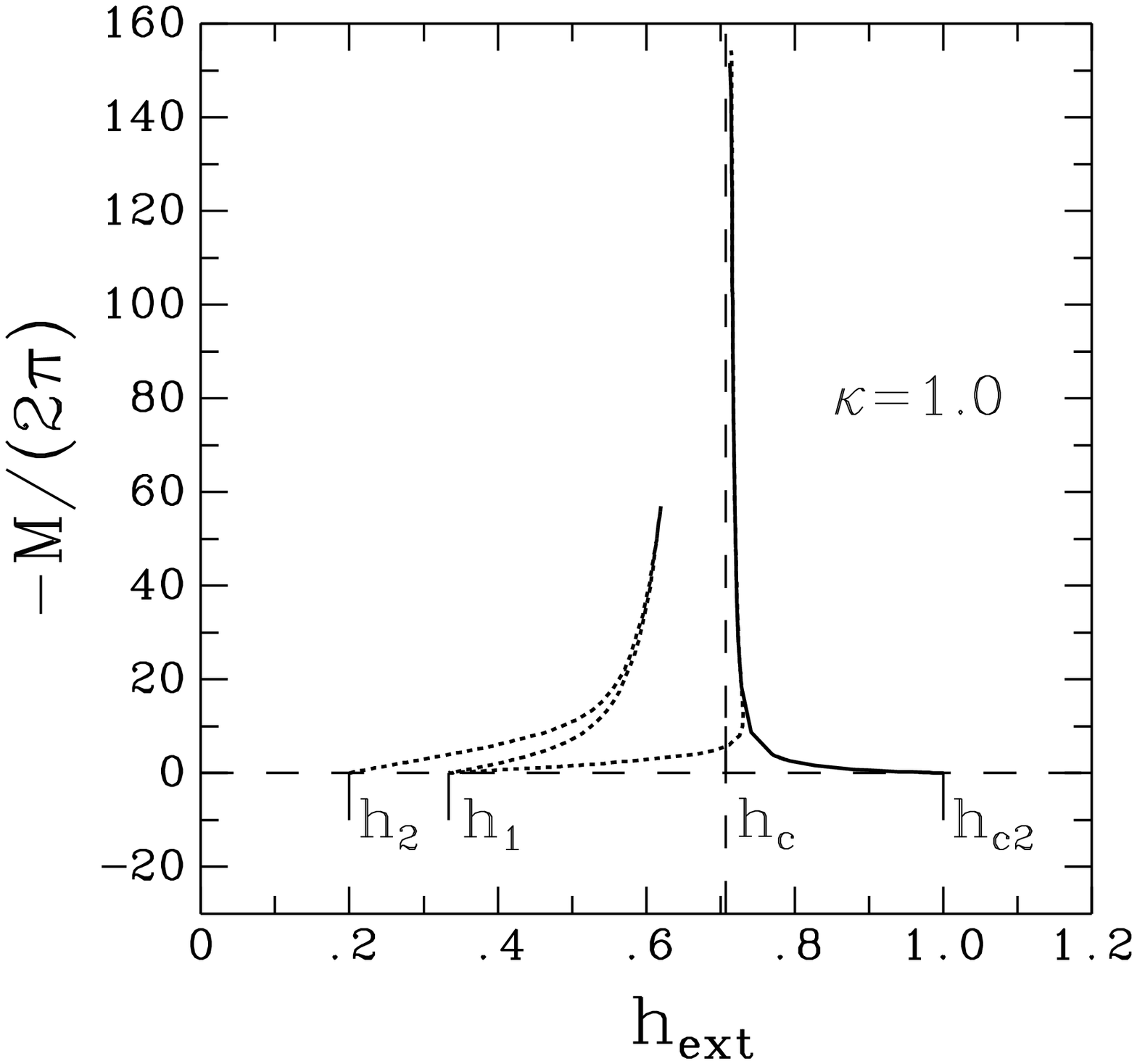,height=8cm} \hskip 2cm}
\pagebreak

\centerline{Figure 18:}
\vskip3cm
\centerline{\psfig{figure=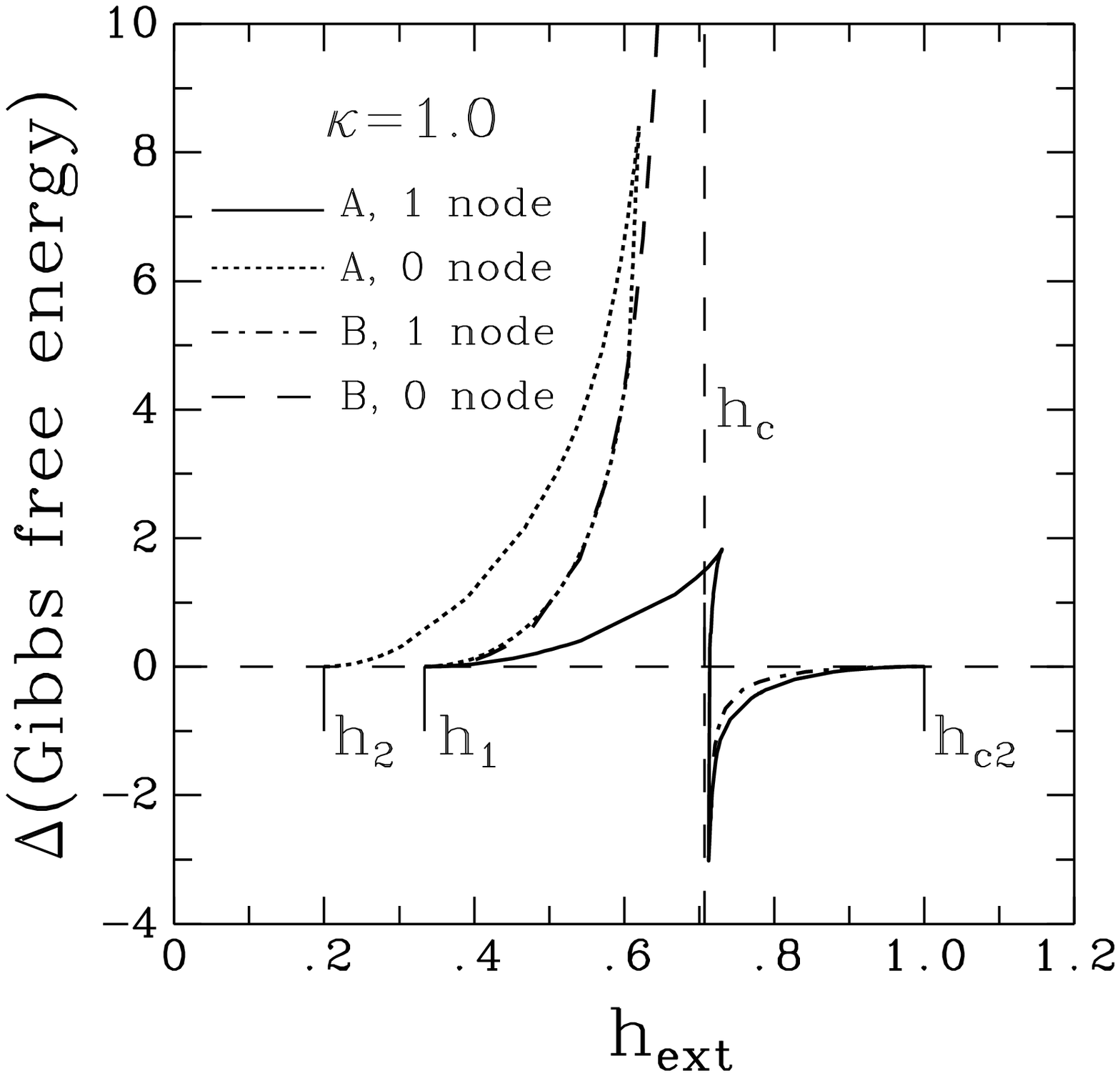,height=8cm} \hskip 2cm}
\pagebreak

\centerline{Figure 19:}
\centerline{\psfig{figure=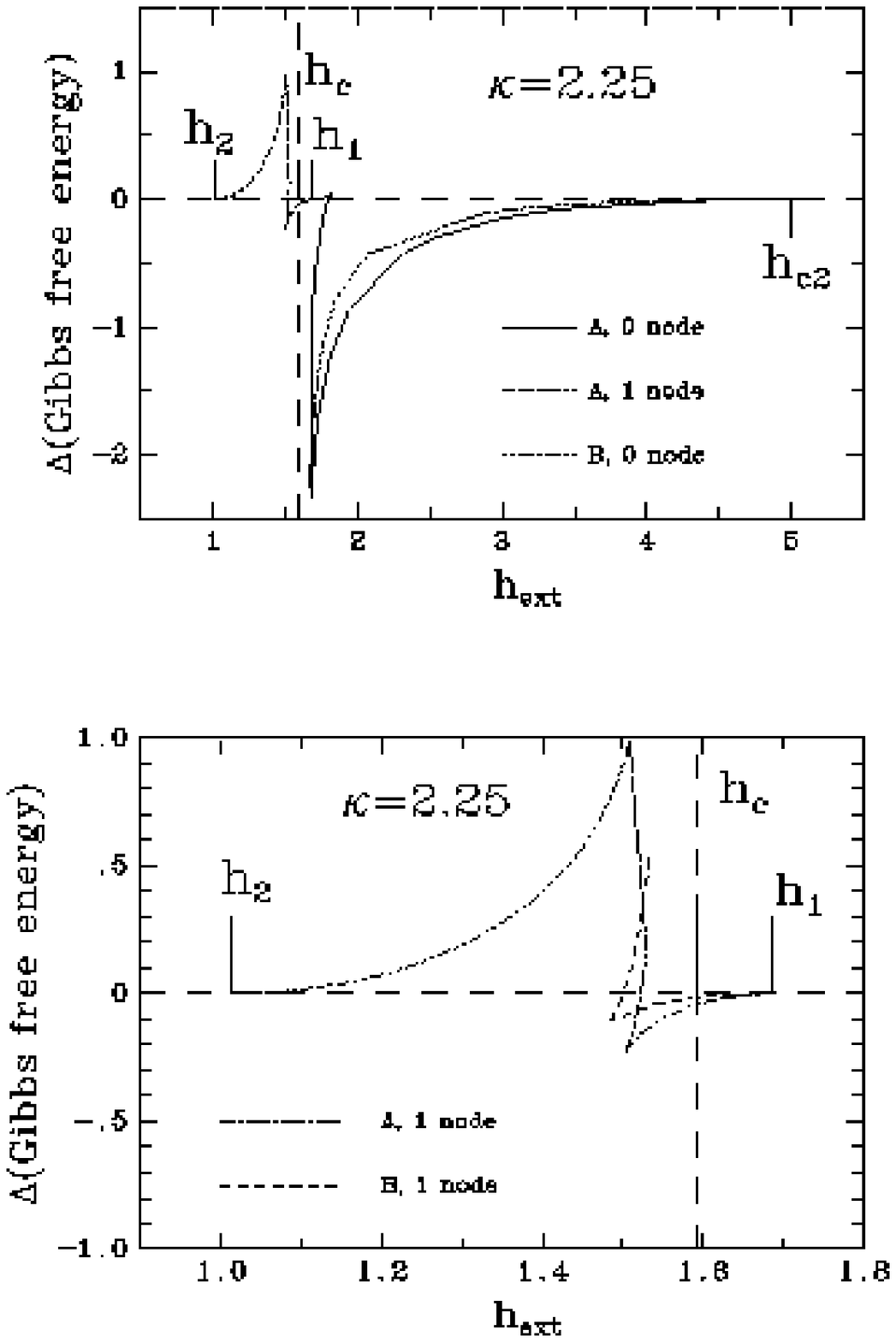,height=8cm} \hskip 2cm}
\pagebreak

\centerline{Figure 20:}
\vskip3cm
\centerline{\psfig{figure=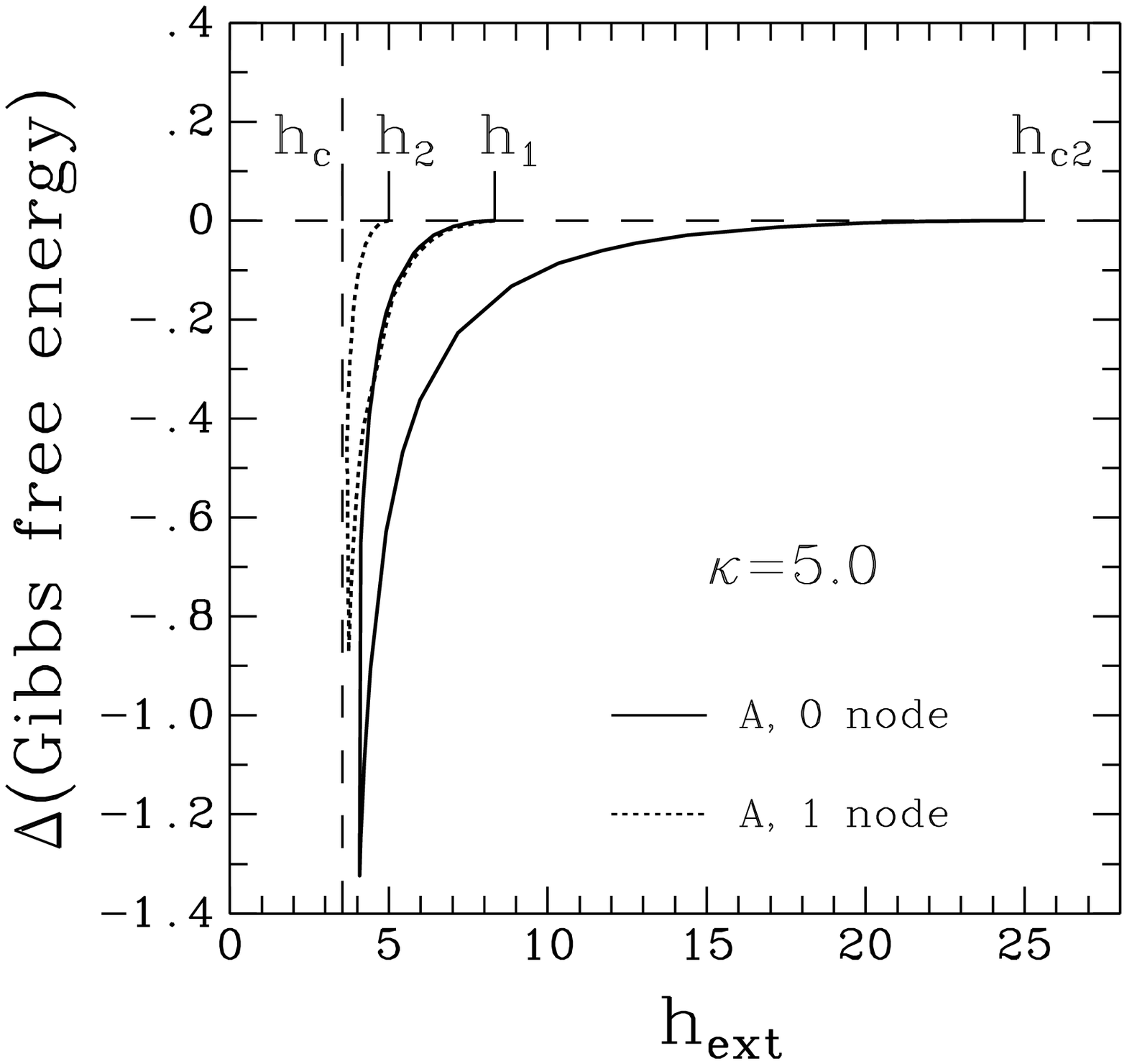,height=8cm} \hskip 2cm}
\pagebreak

\centerline{Figure 21:}
\vskip5cm
\centerline{\psfig{figure=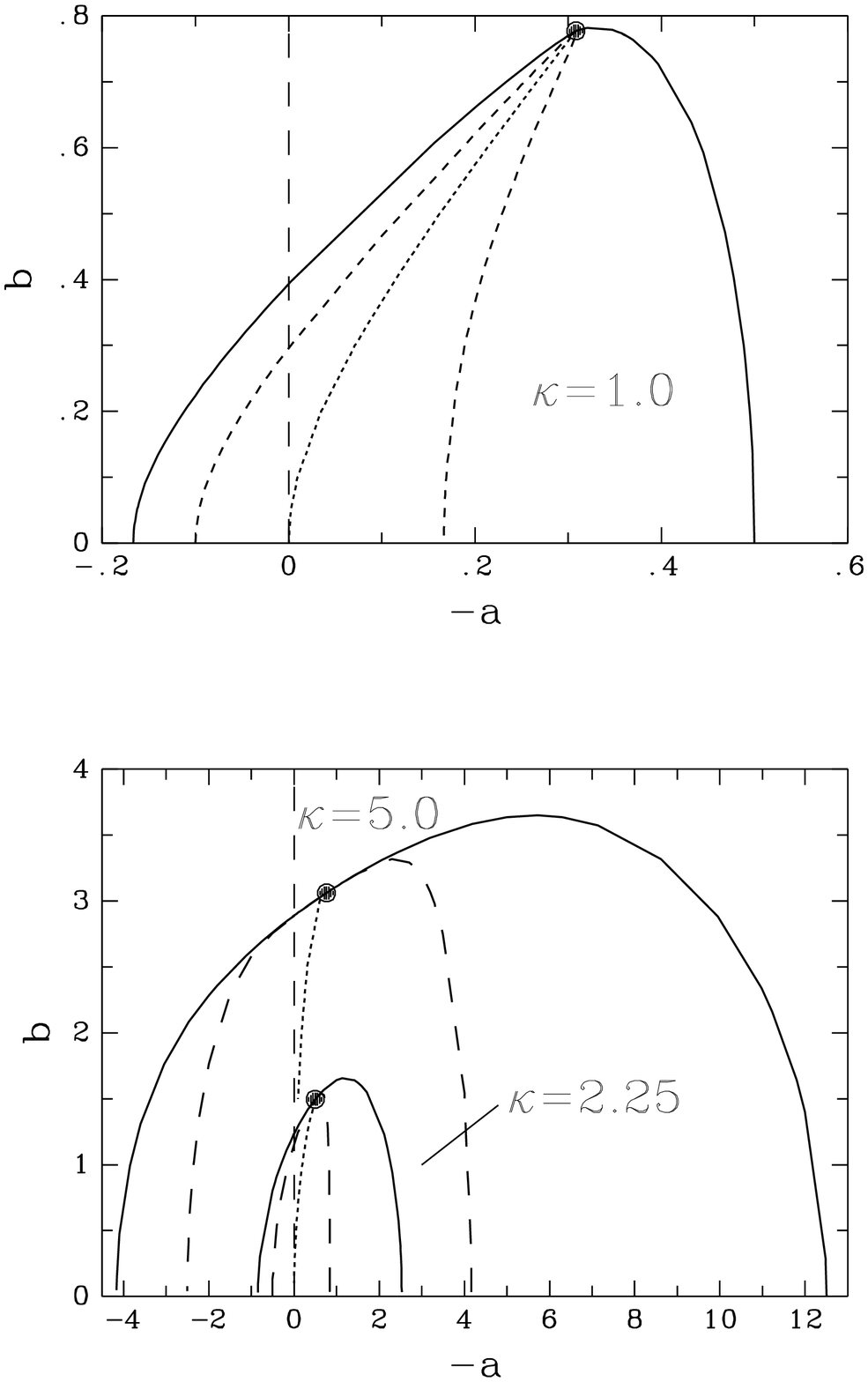,height=8cm} \hskip 2cm}
\pagebreak

\centerline{Figure 22:}
\vskip3cm
\centerline{\psfig{figure=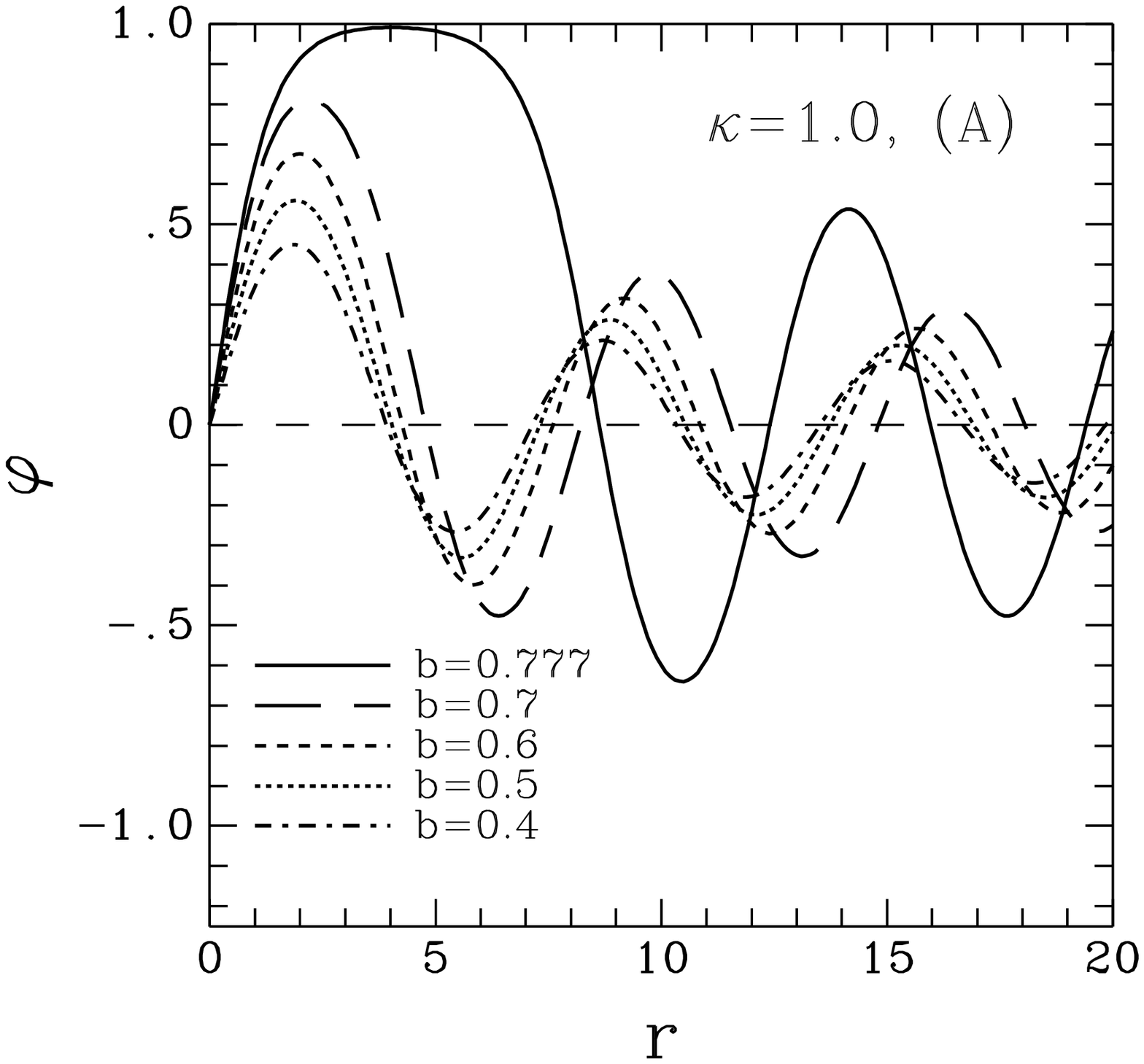,height=8cm} \hskip 2cm}
\pagebreak

\centerline{Figure 23:}
\vskip5cm
\centerline{\psfig{figure=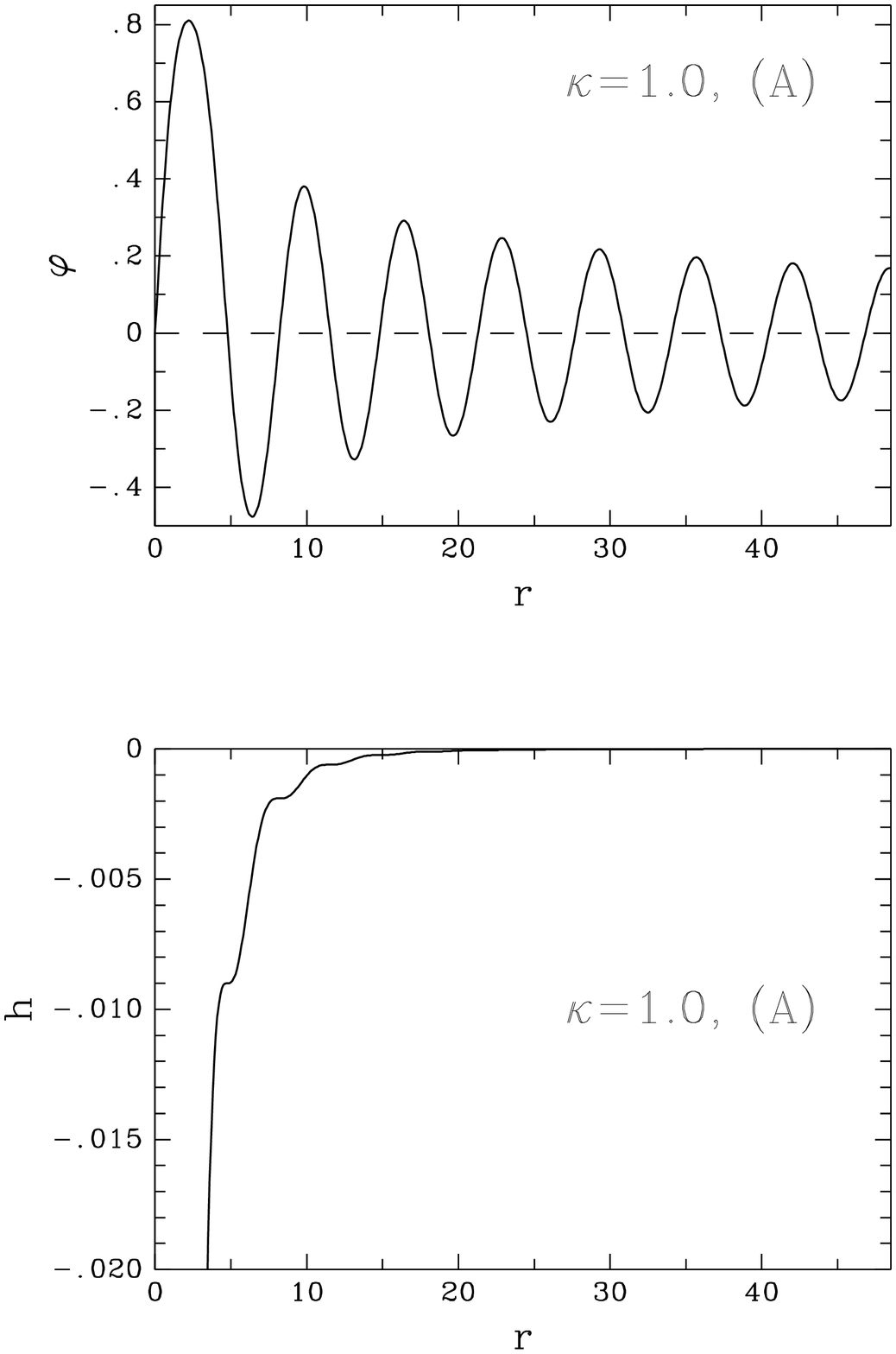,height=8cm} \hskip 2cm}
\pagebreak

\centerline{Figure 24:}
\vskip3cm
\centerline{\psfig{figure=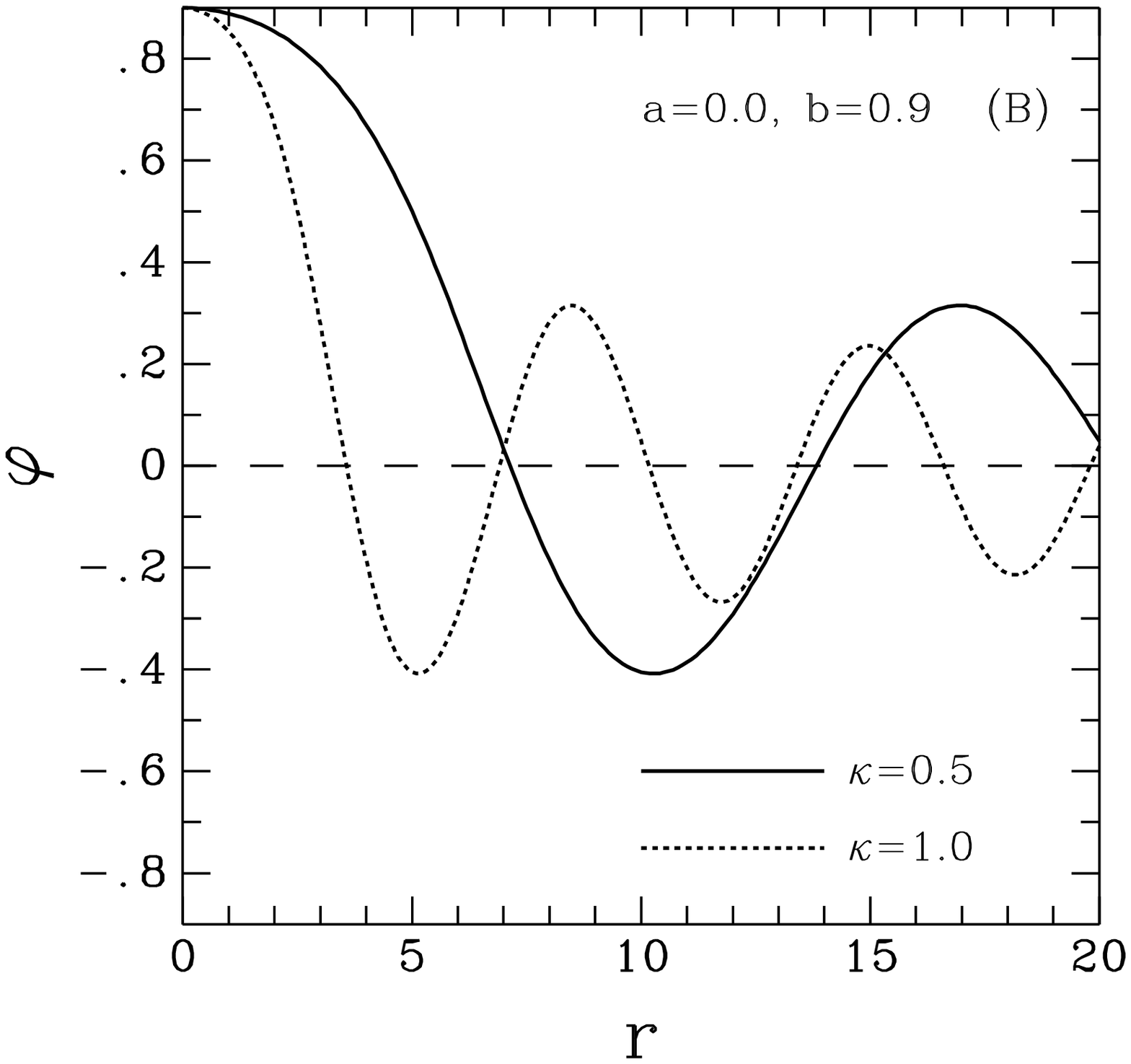,height=8cm} \hskip 2cm}
\pagebreak

\centerline{Figure 25:}
\vskip3cm
\centerline{\psfig{figure=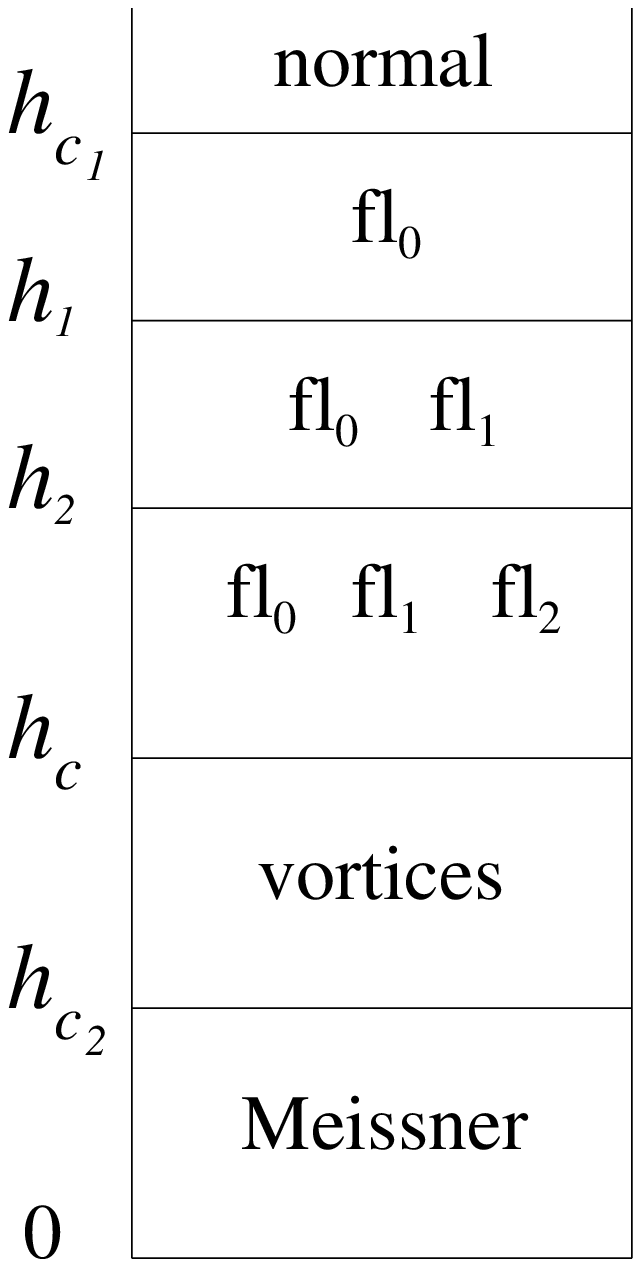,height=8cm}}
\pagebreak

\setcounter{table}{0}

\begin{table}
\caption[]{Vortex solutions for $N=1$.}
\begin{tabular}{lcccc}
$\kappa $ & $a$ & $b$ & ${\cal E}$[$\pi\mu^2$]&${H_{c_1}\over\sqrt{2}H_{c}}$\\
\tableline
2.50 & $-$0.52985763 & 1.63675335 & 0.27263936 & 0.34079920 \\
2.25 & $-$0.49897503 & 1.49452061 & 0.32221312 & 0.36248976 \\ 
2.00 & $-$0.46614139 & 1.35218556 & 0.38816478 & 0.38816478 \\
1.90 & $-$0.45240684 & 1.29519757 & 0.42091387 & 0.39986817 \\ 
1.80 & $-$0.43829401 & 1.23816023 & 0.45839933 & 0.41255939 \\ 
1.75 & $-$0.43108902 & 1.20961885 & 0.47922576 & 0.41932254 \\ 
1.70 & $-$0.42378083 & 1.18105958 & 0.50162817 & 0.42638394 \\ 
1.60 & $-$0.40884163 & 1.12387797 & 0.55190565 & 0.44152452 \\ 
1.50 & $-$0.39344716 & 1.06659333 & 0.61094895 & 0.45821171 \\
1.40 & $-$0.37756417 & 1.00917811 & 0.68104249 & 0.47672974 \\  
1.30 & $-$0.36115464 & 0.95159772 & 0.76529713 & 0.49744313 \\  
1.25 & $-$0.35273894 & 0.92273231 & 0.81402661 & 0.50876663 \\ 
1.20 & $-$0.34417457 & 0.89380826 & 0.86804706 & 0.52082823 \\  
1.10 & $-$0.32657239 & 0.83575334 & 0.99548949 & 0.54751921 \\  
1.00 & $-$0.30828665 & 0.77735953 & 1.1567616  & 0.5783808  \\  
0.95 & $-$0.29886475 & 0.74800633 & 1.25418337 & 0.59573710 \\  
0.90 & $-$0.28924269 & 0.71852956 & 1.36584219 & 0.61462898 \\  
0.85 & $-$0.27940823 & 0.68891163 & 1.49481540 & 0.63529654 \\  
0.80 & $-$0.26934755 & 0.65913181 & 1.64508815 & 0.65803526 \\  
0.75 & $-$0.25904488 & 0.62916556 & 1.82190232 & 0.68321337 \\  
0.70710678 & $-$0.25000000 & 0.60328785 & 1.9999998 & 0.70710678 \\ 
0.70 & $-$0.24848202 & 0.59898348 & 2.03227505 & 0.71129626 \\  
0.65 & $-$0.23763777 & 0.56855009 & 2.28578980 & 0.74288168 \\  
0.60 & $-$0.22648713 & 0.53782200 & 2.59583844 & 0.77875153 \\  
0.55 & $-$0.21500009 & 0.50674553 & 2.98163991 & 0.81995097 \\  
0.50 & $-$0.20314008 & 0.47525314 & 3.47164166 & 0.86791041 \\
0.25 & $-$0.13644748 & 0.30810000 & 10.65631899 & 1.33203987 \\
\end{tabular}
\label{table-vor1big}
\end{table}

\begin{table}
\caption[]{Vortex solutions for $N=2$.}
\begin{tabular}{dcccc}
$\kappa $ & $a$ & $b$ & ${\cal E}$[$\pi\mu^2$]&${H_{c_1}\over\sqrt{2}H_{c}}$\\
\tableline
2.00 & $-$0.52764322 & 1.01524213 & 0.88928974 & 0.44464487 \\
1.75 & $-$0.48000322 & 0.82862511 & 1.07872319 & 0.47194139 \\
1.70 & $-$0.47019357 & 0.79239817 & 1.12484629 & 0.47805967 \\
1.60 & $-$0.45027413 & 0.72409974 & 1.22772704 & 0.49109081 \\
1.50 & $-$0.42993524 & 0.65857003 & 1.34754217 & 0.50532831 \\ 
1.40 & $-$0.40915269 & 0.59580159 & 1.48856910 & 0.52099918 \\
1.30 & $-$0.38789871 & 0.53578681 & 1.65645675 & 0.53834844 \\
1.25 & $-$0.37708516 & 0.50680969 & 1.75284680 & 0.54776462 \\ 
1.20 & $-$0.36614139 & 0.47851821 & 1.85917290 & 0.55775187 \\
1.10 & $-$0.34384344 & 0.42398871 & 2.10795145 & 0.57968664 \\
1.00 & $-$0.32096055 & 0.37219117 & 2.41915478 & 0.60478869 \\
0.90 & $-$0.29743900 & 0.32311960 & 2.81759368 & 0.63395857 \\
0.80 & $-$0.27321196 & 0.27676880 & 3.34249568 & 0.66849913 \\
0.75 & $-$0.26080805 & 0.25461243 & 3.67126654 & 0.68836247 \\ 
0.70710678 & $-$0.25000000 & 0.23614630 & 3.99999979 & 0.70710674 \\
0.70 & $-$0.24819367 & 0.23313497 & 4.05933384 & 0.71038342 \\
0.60 & $-$0.22226965 & 0.19221625 & 5.08511208 & 0.76276681 \\
0.50 & $-$0.19527916 & 0.15401329 & 6.64856450 & 0.83107056 \\ 
\end{tabular}
\label{table-vor2big}
\end{table}

\begin{table}
\caption[]{Type B solutions for $\kappa =\sqrt{1/2}$.}
\begin{tabular}{ddccc}
$b$ & $a$ & $F/(2\pi)$ & $\Delta{\cal G}$ [$\pi \mu^2$] & $h_{\rm ext}$ \\
\tableline
$1-10^{-16}$ & 0.81463957 $10^{-19}$ &  $-$506.26448051 & 0.0 & 0.5 \\
$1-10^{-12}$ & 0.49998905 $10^{-12}$ &  $-$212.66407036 & 0.0 & 0.5 \\
$1-10^{-8}$  & 0.50000041 $10^{-8}$  &  $-$97.79618464 & 0.0 & 0.5 \\
$1-10^{-4}$ & 0.0000499975 & $-$26.47560174 & 0.0 & 0.5 \\
$1-10^{-3}$ & 0.00049975 & $-$15.55584217 & 0.0 & 0.5 \\
0.9 & 0.04750000 & $-$2.24829594 & 0.0 & 0.5 \\
0.8 & 0.09000000 & $-$1.23667262 & 0.0 & 0.5 \\
0.7 & 0.12750000 & $-$0.76485295 & 0.0 & 0.5 \\
0.6 & 0.16000000 & $-$0.48591998 & 0.0 & 0.5 \\
0.5 & 0.18750000 & $-$0.30398178 & 0.0 & 0.5 \\
0.4 & 0.21000000 & $-$0.18029540 & 0.0 & 0.5 \\
0.3 & 0.22750000 & $-$0.09603983 & 0.0 & 0.5 \\
0.2 & 0.24000000 & $-$0.04114477 & 0.0 & 0.5 \\
0.1 & 0.24750000 & $-$0.01006986 & 0.0 & 0.5 \\
0.0 & 0.25000000 & $-$0.00000000 & 0.0 & 0.5 \\
\end{tabular}
\label{tableBc}
\end{table}

\begin{table}
\caption[]{Type B solutions for $\kappa =0.5$.}
\begin{tabular}{ddccc}
$b$ & $a$ & $M/(2\pi)$ & $\Delta{\cal G}$ [$\pi \mu^2$] & $h_{\rm ext}$ \\
\tableline
$1-10^{-16}$ & 0.29397206 $10^{-29}$ &  $-$842.90290064 &  
11.741127  &  0.35182244 \\
$1-10^{-12}$ & 0.22785603 $10^{-17}$ &  $-$303.48173375 & 
7.0628467  &  0.35066627 \\
$1-10^{-8}$ & 0.11163709 $10^{-11}$ &  $-$140.17105915  &
4.9082144  &  0.34938890 \\
$1-10^{-4}$ & 0.57328821 $10^{-6}$ &  $-$38.35953571 &
2.5230830  &  0.34539301 \\ 
$1-10^{-2}$ & 0.00042699 &  $-$11.04294492 &   1.32277094 &   0.33805432 \\ 
0.9 &   0.01112676 &   $-$3.50056424 &     0.62184922 &     0.32363276 \\ 
0.8 &   0.02804562 &   $-$2.01439886 &     0.36480792 &     0.31151388 \\ 
0.7 &   0.04647666 &   $-$1.29869893 &     0.21255415 &     0.29980096 \\ 
0.6 &   0.06469237 &   $-$0.85736327 &     0.11638777 &     0.28851017 \\ 
0.5 &   0.08160630 &   $-$0.55534931 &     0.05742286 &     0.27797881 \\ 
0.4 &   0.09643545 &   $-$0.33958193 &     0.02412093 &     0.26860687 \\ 
0.3 &   0.10858262 &   $-$0.18551044 &     0.00781468 &     0.26079463 \\ 
0.2 &   0.11759100 &   $-$0.08099744 &     0.00157347 &     0.25490781 \\ 
0.1 &   0.12313071 &   $-$0.02005903 &     0.00009957 &     0.25124415 \\ 
0.0 &   0.12500000 &   $-$0.00000000 &     0.00000000 &     0.25000000 \\
\end{tabular}
\label{tableB05}
\end{table}

\begin{table}
\caption[]{Type B solutions for $\kappa =1.0$}
\begin{tabular}{ddccc}
$b$ & $a$ & $M/(2\pi)$ & $\Delta{\cal G}$ [$\pi \mu^2$] & $h_{\rm ext}$ \\
\tableline
$1-10^{-16}$ & 0.2626738 $10^{-12}$ & $-$319.83194287 &
$-$3.33629143 & 0.71216412 \\ 
$1-10^{-12}$ & 2.8028373 $10^{-9}$ & $-$150.42278724 & $-$2.30158296 &
0.71451434 \\ 
$1-10^{-8}$ & 0.00000179 & $-$69.24007330 & $-$1.58213465 & 0.71804013 \\ 
$1-10^{-4}$ & 0.00111758 & $-$18.77412288 & $-$0.84610904 & 0.72827673 \\
$1-10^{-3}$ & 0.00561929 & $-$10.98561767 & $-$0.65155679 & 0.73506133 \\
0.9 & 0.14665519 & $-$1.41282055 & $-$0.16978151 & 0.80012302 \\
0.8 & 0.23437248 & $-$0.72891143 & $-$0.08386589 & 0.84215100 \\
0.7 & 0.30301143 & $-$0.43023181 & $-$0.04233326 & 0.87880681 \\
0.6 & 0.35848496 & $-$0.26377020 & $-$0.02052442 & 0.91069735 \\
0.5 & 0.40331923 & $-$0.16055448 & $-$0.00913650 & 0.93782855 \\
0.4 & 0.43886076 & $-$0.09326563 & $-$0.00352478 & 0.96013136 \\
0.3 & 0.46590309 & $-$0.04892564 & $-$0.00106759 & 0.97754071 \\
0.2 & 0.48493325 & $-$0.02074118 & $-$0.00020468 & 0.99000792 \\
0.1 & 0.49624589 & $-$0.00504513 & $-$0.00001257 & 0.99750049 \\
0.0 & 0.50000000 & $-$0.00000000 & $-$0.00000000 & 1.00000000 \\
\end{tabular}
\label{tableB1}
\end{table}

\begin{table}
\caption[]{Type B solutions for $\kappa =1.5$}
\begin{tabular}{llccc}
$b$ & $a$ & $M/(2\pi)$ & $\Delta{\cal G}$ [$\pi \mu^2$] & $h_{\rm ext}$ \\
\tableline
$1-10^{-12}$ & 0.0000009048 & $-$120.08566366 & $-$2.90735164 & 1.08671648\\
$1-10^{-8}$ &  0.00009492 & $-$55.45237424 & $-$2.02045383 &  1.09897771 \\ 
$1-10^{-4}$ &  0.01076082 & $-$14.51995526 & $-$1.07824205 &  1.13651774 \\ 
$1-10^{-2}$ &  0.12487016 &  $-$3.50311345 & $-$0.51323746 &  1.22807544 \\ 
0.9 &     0.43216305 & $-$0.78071937 &    $-$0.15951625 &     1.47054176 \\ 
0.8 &     0.61870985 & $-$0.36990133 &    $-$0.07057877 &     1.65414678 \\ 
0.7 &     0.75504470 & $-$0.20814435 &    $-$0.03333402 &     1.80288132 \\ 
0.6 &     0.86159257 & $-$0.12380358 &    $-$0.01546007 &     1.92600329 \\ 
0.5 &     0.94608432 & $-$0.07384290 &    $-$0.00667256 &     2.02721532 \\ 
0.4 &     1.01229462 & $-$0.04230148 &    $-$0.00251856 &     2.10843492 \\ 
0.3 &     1.06230783 & $-$0.02198078 &    $-$0.00075128 &     2.17077409 \\ 
0.2 &     1.09734361 & $-$0.00926097 &    $-$0.00014259 &     2.21490851 \\ 
0.1 &     1.11811540 & $-$0.00224481 &    $-$0.00000871 &     2.24124437 \\ 
0.0 &     1.12500000 & $-$0.00000000 &    $-$0.00000000 &     2.25000000 \\
\end{tabular}
\label{tableB15}
\end{table}

\begin{table}
\caption[]{Type B solutions for $\kappa =2.25$}
\begin{tabular}{llccc}
$b$ & $a$ & $M/(2\pi)$ & $\Delta{\cal G}$ [$\pi \mu^2$] & $h_{\rm ext}$ \\
\tableline
$1-10^{-8}$ & 0.00039686187 & $-$66.11152808 & $-$2.2499749 &  1.67050641 \\ 
$1-10^{-4}$ & 0.039632414   & $-$14.18949823 & $-$1.1073683 &  1.76576571 \\ 
$1-10^{-2}$ & 0.38362876    &  $-$2.44323885 & $-$0.42859875 & 2.06919235 \\ 
0.9 &  1.1088587  &  $-$0.40297816 & $-$0.099959198 &    2.89279937 \\ 
0.8 &  1.5013300  &  $-$0.17747571 & $-$0.040526105 &    3.44436324 \\ 
0.7 &  1.7810811  &  $-$0.09671914 & $-$0.018347512 &    3.86542221 \\ 
0.6 &  1.9978592  &  $-$0.05652767 & $-$0.0083010393 &   4.20301812 \\ 
0.5 &  2.1691771  &  $-$0.03335736 & $-$0.0035262489 &   4.47522932 \\ 
0.4 &  2.3032325  &  $-$0.01897809 & $-$0.0013169381 &   4.69097359 \\ 
0.3 &  2.4044317  &  $-$0.00981733 & $-$0.00039005489 &  4.85521398 \\ 
0.2 &  2.4753075  &  $-$0.00412457 & $-$0.000073694017 & 4.97087202 \\ 
0.1 &  2.5173243  &  $-$0.00099821 & $-$0.0000044888587 &5.03966430 \\ 
0.0 &  2.5312500  &     0.00000000 &    0.00000000 &     5.06250000 \\
\end{tabular}
\label{tableB225}
\end{table}

\begin{table}
\caption[]{Type B solutions with one node for $\kappa =0.5$.}
\begin{tabular}{llccc}
$b$ & $a$ & $M/(2\pi)$ & $\Delta{\cal G}$ [$\pi \mu^2$] & $h_{\rm ext}$ \\
\tableline
$1-10^{-12}$ & 0.20777588 $10^{-19}$ & $-$407.83814571 & 76.685566 & 
0.32415859 \\
$1-10^{-8}$ & 0.16968270 $10^{-13}$ & $-$207.47249709 & 50.180428 & 
0.31279238 \\ 
$1-10^{-4}$ & 1.98235 $10^{-8}$ & $-$70.92516787 & 23.94263827 & 0.28591112 \\ 
$1-10^{-2}$ & 0.00003060 & $-$27.38806803 & 11.06437630 &     0.24968540 \\ 
0.9 &     0.00145092 &  $-$12.10387413 &     4.79404931 &     0.20705162 \\ 
0.8 &     0.00470204 &   $-$8.24390469 &     2.91014633 &     0.18357733 \\ 
0.7 &     0.00925886 &   $-$6.05828898 &     1.82769449 &     0.16452267 \\ 
0.6 &     0.01475161 &   $-$4.49834233 &     1.10800369 &     0.14740092 \\ 
0.5 &     0.02078605 &   $-$3.26222813 &     0.61837208 &     0.13157619 \\ 
0.4 &     0.02691416 &   $-$2.23078265 &     0.29897268 &     0.11700553 \\ 
0.3 &     0.03262722 &   $-$1.36003756 &     0.11262889 &     0.10401313 \\ 
0.2 &     0.03736276 &   $-$0.65666061 &     0.02623761 &     0.09329940 \\ 
0.1 &     0.04053971 &   $-$0.17538241 &     0.00185854 &     0.08598007 \\ 
0.0 &     0.04144444 &   $-$0.00000000 &     0.00000000 &     0.08333333 \\ 
\end{tabular}
\label{tableBn05}
\end{table}

\begin{table}
\caption[]{Type B solutions with one node for $\kappa =1.0$.}
\begin{tabular}{llccc}
$b$ & $a$ & $M/(2\pi)$ & $\Delta{\cal G}$ [$\pi \mu^2$] & $h_{\rm ext}$ \\
\tableline
$1-10^{-16}$ & 0.18816282 $10^{-13}$& $-$399.13713984 & 22.180605 &
0.67281714 \\
$1-10^{-12}$ & 0.26227586 $10^{-9}$& $-$201.85338653 & 14.548888 &
0.65939048 \\ 
$1-10^{-8}$  & 0.00000022 &  $-$102.17627279 &  9.27215886 &   0.64105359 \\ 
$1-10^{-4}$  & 0.00021123 &   $-$34.38826735 &  4.1088762  &   0.59787403 \\
$1-10^{-3}$  & 0.00123249 &   $-$22.55431617 &  2.86626778 &   0.57561429 \\
$1-10^{-2}$  & 0.00740425 &   $-$12.84295319 &  1.6778879  &   0.54090402 \\
0.9 &     0.04397343 &  $-$5.29912852 &   0.61000069 &     0.47769745 \\ 
0.8 &     0.07272152 &  $-$3.41688303 &   0.33196266 &     0.44532585 \\ 
0.7 &     0.09578531 &  $-$2.37610307 &   0.18841580 &     0.42035415 \\ 
0.6 &     0.11488290 &  $-$1.66122221 &   0.10272050 &     0.39901535 \\ 
0.5 &     0.13071774 &  $-$1.12628291 &   0.05103931 &     0.38039878 \\ 
0.4 &     0.14359881 &  $-$0.71443290 &   0.02171670 &     0.36441468 \\ 
0.3 &     0.15364041 &  $-$0.40169570 &   0.00713976 &     0.35131626 \\ 
0.2 &     0.16085504 &  $-$0.17911859 &   0.00145678 &     0.34150658 \\ 
0.1 &     0.16520978 &  $-$0.04493592 &   0.00009304 &     0.33540631 \\ 
0.0 &     0.16666666 &  $-$0.00000000 &   0.00000000 &     0.33333333 \\ 
\end{tabular}
\label{tableBn1}
\end{table}

\begin{table}
\caption[]{Type B solutions with one node for $\kappa =1.5$.}
\begin{tabular}{llccc}
$b$ & $a$ & $M/(2\pi)$ & $\Delta{\cal G}$ [$\pi \mu^2$] & $h_{\rm ext}$ \\
\tableline
$1-10^{-16}$ & 0.19551488 $10^{-8}$ & $-$253.65744393 & 6.8992671 & 
1.01900652 \\
$1-10^{-12}$ & 0.52845869 $10^{-6}$ & $-$137.93602705 & 4.5873181 & 
1.00509209 \\
$1-10^{-8}$ &  0.48741567 $10^{-4}$ &  $-$69.41302070 & 2.7590523 & 
0.98413629 \\ 
$1-10^{-4}$ &  0.48556580 $10^{-2}$ &  $-$22.89693286 & 1.0335935 & 
0.93620705 \\
$1-10^{-2}$ &  0.05185846 & $-$8.13129020 & 0.31314228 &     0.87891501 \\ 
0.9 &     0.16380734 &  $-$3.07893180 &     0.08550524 &     0.83074850 \\ 
0.8 &     0.22458591 &  $-$1.88505320 &     0.04289325 &     0.81115382 \\ 
0.7 &     0.26661104 &  $-$1.25602039 &     0.02314474 &     0.79687633 \\ 
0.6 &     0.29844532 &  $-$0.84488327 &     0.01207324 &     0.78493617 \\ 
0.5 &     0.32324161 &  $-$0.55284397 &     0.00574979 &     0.77469908 \\ 
0.4 &     0.34247940 &  $-$0.33962509 &     0.00234955 &     0.76608502 \\ 
0.3 &     0.35693478 &  $-$0.18577883 &     0.00074493 &     0.75918366 \\ 
0.2 &     0.36703569 &  $-$0.08108642 &     0.00014762 &     0.75412788 \\ 
0.1 &     0.37301794 &  $-$0.02006596 &     0.00000925 &     0.75103921 \\ 
0.0 &     0.37500000 &  $-$0.00000000 &     0.00000000 &     0.75000000 \\
\end{tabular}
\label{tableBn15}
\end{table}

\begin{table}
\caption[]{Type B solutions with one node for $\kappa =2.25$.}
\begin{tabular}{llccc}
$b$ & $a$ & $M/(2\pi)$ & $\Delta{\cal G}$ [$\pi \mu^2$] & $h_{\rm ext}$ \\
\tableline
$1-10^{-8}$ & 0.0003961775 & $-$70.82155158 & 0.52680199  &  1.53403702 \\ 
$1-10^{-4}$ & 0.033788231  & $-$17.35196356 & $-$0.043408899 &  1.49548933 \\ 
$1-10^{-2}$ & 0.21543634  &  $-$5.05538676 & $-$0.10152137 &    1.48666351 \\ 
0.9 &  0.47568332 & $-$1.67363013 & $-$0.037872528  &  1.54467090 \\ 
0.8 &  0.58990955 & $-$0.96900019 & $-$0.017383337  &  1.58483248 \\ 
0.7 &  0.66407250 & $-$0.62129733 & $-$0.0083894957 &  1.61373320 \\ 
0.6 &  0.71839505 & $-$0.40560746 & $-$0.0039511917 &  1.63573439 \\ 
0.5 &  0.75977286 & $-$0.25909793 & $-$0.0017251285 &  1.65275343 \\ 
0.4 &  0.79135512 & $-$0.15613863 & $-$0.0006568765 &  1.66581546 \\ 
0.3 &  0.81479746 & $-$0.08415334 & $-$0.0001972193 &  1.67552249 \\ 
0.2 &  0.83103287 & $-$0.03634436 & $-$0.0000376004 &  1.68224194 \\ 
0.1 &  0.84059205 & $-$0.00893709 & $-$0.23022948 $10^{-5}$ & 1.68619486 \\ 
0.0 &  0.84375000 &    0.00000000 & 0.00000000 & 1.68750000 \\
\end{tabular}
\label{tableBn225}
\end{table}

\begin{table}
\caption[]{Flux-tube solutions for $\kappa=0.50$}
\begin{tabular}{llccc}
$a$ & $b$ & $M/(2\pi)$ & $\Delta{\cal G}$ [$\pi \mu^2$] & $h_{\rm ext}$ \\
\tableline
0.04153310 &   0.00482634 &  $-$0.01000000 &   0.00001160 &   0.08362321 \\ 
0.03881390 &   0.02291673 &  $-$0.20000000 &   0.00485667 &   0.08933301 \\ 
0.03377073 &   0.03987468 &  $-$0.49999998 &   0.03251648 &   0.09914171 \\ 
0.03187151 &   0.04510360 &  $-$0.59999997 &   0.04786481 &   0.10262833 \\ 
0.02547804 &   0.06080732 &  $-$0.89999996 &   0.11465226 &   0.11372799 \\ 
0.02310788 &   0.06616722 &  $-$1.00000005 &   0.14434866 &   0.11763453 \\ 
0.00943295 &   0.09464338 &  $-$1.49999992 &   0.35425328 &   0.13855274 \\ 
$-$0.00711859 &   0.12627010 &  $-$2.00000011 &   0.67207379 &   0.16122362 \\
$-$0.02593361 &   0.16051533 &  $-$2.50000020 &   1.09285215 &   0.18459837 \\ 
$-$0.06611321 &   0.23111422 &  $-$3.50000033 &   2.14659123 &   0.22868403 \\ 
$-$0.12500000 &   0.33318466 &  $-$5.25759258 &   4.0155314  &   0.28313110 \\
$-$0.14000000 &   0.35942324 &  $-$5.89146805 &   4.5611306  &   0.29539602 \\ 
$-$0.17000000 &   0.41279518 &  $-$7.87229590 &   5.7877739  &   0.31812023 \\ 
$-$0.18000000 &   0.43102350 &  $-$9.05838675 &   6.2658005  &   0.32521571 \\ 
$-$0.19000000 &   0.44962804 & $-$11.08773185 &   6.8232515  &   0.33221084 \\ 
0.20029645 & 0.46875000 & $-$14.60379550 &  2.2039074  &   0.33977157 \\ 
0.19100000 & 0.44513701 &  $-$7.67322461 &  1.4931287  &   0.33119462 \\ 
0.17800000 & 0.40691601 &  $-$4.65587992 &  0.94553126 &   0.31967926 \\ 
0.17500000 & 0.39713515 &  $-$4.20761305 &  0.84563339 &   0.31685806 \\ 
0.16000000 & 0.34108654 &  $-$2.52520702 &  0.43874152 &   0.30136833 \\ 
0.15000000 & 0.29422019 &  $-$1.69236974 &  0.23906195 &   0.28941054 \\ 
0.14000000 & 0.23346883 &  $-$0.97543913 &  0.09420080 &   0.27567022 \\ 
0.13250000 & 0.16870544 &  $-$0.48035611 &  0.02572468 &   0.26378724 \\ 
0.12750000 & 0.09906230 &  $-$0.15975047 &  0.00307150 &   0.25485285 \\ 
0.12500000 & 0.00100000 &  $-$0.00001600 &  0.31997933 $10^{-10}$& 0.25000000\\
\end{tabular}
\label{ft05}
\end{table}

\begin{table}
\caption[]{Flux-tube solutions for $\kappa=1.00$}
\begin{tabular}{llccc}
$a$ & $b$ & $M/(2\pi)$ & $\Delta{\cal G}$ [$\pi \mu^2$] & $h_{\rm ext}$ \\
\tableline
0.16666065 & 0.00192455 & $-$0.00010000 & 0.10185308 $10^{-8}$ & 0.33334352 \\
0.16048691 &   0.06238119 &  $-$0.10000000 &   0.00103043 &   0.34360827 \\ 
0.09149278 &   0.24083303 &  $-$0.99999998 &   0.10874652 &   0.44085382 \\ 
0.00770227 &   0.38158876 &  $-$1.90000005 &   0.37497658 &   0.53338571 \\ 
$-$0.00172647 &  0.39592250 &   $-$2.00000003 &   0.41094947 &  0.54261067 \\
$-$0.23437054 &  0.70095040 &   $-$5.99999872 &   1.54794907 &  0.71152198 \\
$-$0.28685815 &  0.75749117 &  $-$10.00000276 &   1.81154073 &  0.72966737 \\
$-$0.29562992 &  0.76607871 &  $-$12.00000628 &   1.82881244 &  0.73048079 \\
$-$0.30000000 &  0.77018431 &  $-$13.75256885 &   1.82095047 &  0.73018163 \\
$-$0.30326892 &  0.77314720 &  $-$15.99995420 &   1.79510039 &  0.72931654 \\
$-$0.30818466 &  0.77728463 &  $-$40.00000208 &   1.39704713 &  0.72168651 \\ 
$-$0.30827686 &  0.77735251 &  $-$60.00005351 &   1.12951015 &  0.71895537 \\ 
$-$0.30828641 &  0.77735936 &  $-$99.99974294 &   0.70930334 &  0.71624176 \\ 
$-$0.3082866485 &0.77735952553 & $-$154.22042007 & 0.28830787 & 0.71452845 \\ 
0.3082866575 &0.7773595318839 & $-$151.56514834 &$-$3.02337416 &0.71300061\\
0.308286675 & 0.77735954 &  $-$129.08124738 &  $-$2.41023205 &  0.71510625\\
0.30828685 &  0.77735967 &  $-$100.37162393 &  $-$2.17454851 &  0.71614669\\
0.30828800 &  0.77736047 &   $-$78.04813009 &  $-$1.96478896 &  0.71733231\\
0.30840000 &  0.77743589 &   $-$36.78370133 &  $-$1.46832294 &  0.72196383\\
0.32000000 &  0.78196826 &  $-$8.71898303 &  $-$0.81916267 &  0.74119260 \\ 
0.39637726 &  0.72724186 &  $-$1.59999998 &  $-$0.20076202 &  0.82705431 \\
0.44486258 &  0.59258754 &  $-$0.60000000 &  $-$0.05241147 &  0.89933482 \\ 
0.48822938 &  0.29977039 &  $-$0.10000000 &  $-$0.00225057 &  0.97688934 \\ 
0.49875776 &  0.09944583 &  $-$0.01000000 &  $-$0.00002473 &  0.99752029 \\ 
0.49998750 &  0.00999947 &  $-$0.00010000 &  $-$0.000000002499 &  0.99997500\\
\end{tabular}
\label{ft1big}
\end{table}

\begin{table}
\caption[]{Flux-tube solutions for $\kappa=1.50$}
\begin{tabular}{llccc}
$a$ & $b$ & $M/(2\pi)$ & $\Delta{\cal G}$ [$\pi \mu^2$] & $h_{\rm ext}$ \\
\tableline
0.37338330 & 0.04336260 & $-$0.01000000 & 0.00000786 & 0.75176971 \\ 
0.35865108 & 0.13882713 & $-$0.10000000 & 0.00077989 & 0.76758881 \\ 
0.29017559 & 0.32503549 & $-$0.50000000 & 0.01852335 & 0.83476960 \\ 
0.20216876 & 0.47718137 & $-$1.00000001 & 0.06744250 & 0.90884380 \\ 
$-$0.08153028 & 0.81864927 &  $-$2.99999992 & 0.34444650 & 1.07986153 \\ 
$-$0.32032191 & 1.02063593 &  $-$8.09999184 & 0.58261046 & 1.14350498 \\ 
$-$0.37526158 & 1.05617319 & $-$15.00016168 & 0.46855172 & 1.13288052 \\ 
$-$0.38396193 & 1.06126211 & $-$19.00079350 & 0.36778303 & 1.12618616 \\ 
$-$0.38882706 & 1.06402751 & $-$24.00200915 & 0.24189747 & 1.11955789 \\ 
$-$0.39033430 & 1.06487144 & $-$26.99363351 & 0.16315885 & 1.11607308 \\ 
0.39368037 & 1.06672106 & $-$43.31969516 & $-$5.99946594 & 1.00392926 \\ 
0.39942951 & 1.06981657 & $-$20.00000303 & $-$1.63731205 & 1.12470204 \\ 
0.40420067 & 1.07231309 & $-$16.00005584 & $-$1.50098647 & 1.13328380 \\ 
0.42313776 & 1.08161666 & $-$10.00000397 & $-$1.22846970 & 1.15769607 \\ 
0.43745867 & 1.08803048 &  $-$8.00000030 & $-$1.10316720 & 1.17348368 \\ 
0.46163405 & 1.09765879 &  $-$6.00000061 & $-$0.94558871 & 1.19913931 \\ 
0.50659366 & 1.11151317 &  $-$4.00000004 & $-$0.73520293 & 1.24765087 \\ 
0.60827205 & 1.12205983 &  $-$2.00000014 & $-$0.42922215 & 1.36989865 \\ 
0.65684841 & 1.11579485 &  $-$1.50000002 & $-$0.32846827 & 1.43530562 \\ 
0.72822112 & 1.09126036 &  $-$1.00000002 & $-$0.21488613 & 1.53932635 \\ 
0.84790088 & 1.00003242 &  $-$0.50000000 & $-$0.09110741 & 1.73274206 \\ 
1.04053433 & 0.62739169 &  $-$0.10000000 & $-$0.00695891 & 2.08404295 \\ 
1.11531973 & 0.22190830 &  $-$0.01000000 & $-$0.00008535 & 2.23067648 \\ 
1.12401729 & 0.07105192 &  $-$0.00100000 & $-$0.00000087 & 2.24803495 \\ 
1.12490158 & 0.02249687 &  $-$0.00010000 & $-$0.00000001 & 2.24980316 \\ 
\end{tabular}
\label{ft15}
\end{table}

\begin{table}
\caption[]{Flux-tube solutions for $\kappa=2.25$}
\begin{tabular}{llccc}
$a$ & $b$ & $M/(2\pi)$ & $\Delta{\cal G}$ [$\pi \mu^2$] & $h_{\rm ext}$ \\
\tableline
0.84325497 & 0.03080636 & $-$0.00100000 & 0.26603798 $10^{-7}$ & 1.68763471 \\ 
0.79515994 & 0.30501597 & $-$0.10000000 & 0.25079340 $10^{-3}$ & 1.70038988 \\ 
0.43741045 & 0.87389870 & $-$1.00000000 & 0.014681371 & 1.77541273 \\ 
0.00233172 & 1.23081820 & $-$3.00000000 & 0.042078389 & 1.81785677 \\ 
0.00000000 & 1.23234317 & $-$3.01667010 & 0.042145730 & 1.81791344 \\ 
$-$0.01143546 & 1.23977546 &  $-$3.10000000 &    0.04244274 & 1.81815942 \\ 
$-$0.11487304 & 1.30363200 &  $-$4.00000000 &    0.04174682 & 1.81776491 \\ 
$-$0.33764968 & 1.42257539 &  $-$8.00000000 & $-$0.01708379 & 1.79367785 \\
$-$0.38602425 & 1.44534468 & $-$10.00000001 & $-$0.06326788 & 1.78066590 \\
$-$0.49726408 & 1.49381678 & $-$50.03000994 & $-$0.97225009 & 1.68350270 \\ 
$-$0.49780000 & 1.49403737 & $-$55.15382089 & $-$1.0370393  & 1.68033109 \\
0.50045166    & 1.49512705 & $-$49.99032149 & $-$2.3225981  & 1.68233836 \\ 
0.52081765    & 1.50339534 & $-$20.00000037 & $-$1.6410071  & 1.73773679 \\ 
0.58020017    & 1.52647271 & $-$10.00000000 & $-$1.2232806  & 1.81341932 \\ 
0.70035444    & 1.56827220 & $-$5.00000000  & $-$0.85686163 & 1.94614089 \\ 
1.00000000    & 1.63985366 & $-$1.74625522  & $-$0.42190989 & 2.32313287 \\ 
1.13578728    & 1.65419047 & $-$1.20000000  & $-$0.30921646 & 2.52009576 \\ 
1.40000000    & 1.64269550 & $-$0.62208273  & $-$0.16450278 & 2.94138042 \\ 
1.70971180    & 1.54816802 & $-$0.30000000  & $-$0.07041132 & 3.48319881 \\ 
2.12946840    & 1.21592253 & $-$0.09568099  & $-$0.01331646 & 4.27157414 \\ 
2.43883772    & 0.62758267 & $-$0.01711513  & $-$0.00060838 & 4.87826672 \\
2.45577270    & 0.56931287 & $-$0.01380428  & $-$0.00040280 & 4.91193731 \\
2.50000000 & 0.36990731 & $-$0.00553393 &$-$6.7727614 $10^{-5}$ & 5.00006609 \\
2.52100000 & 0.21281951 & $-$0.00178806 &$-$7.2206406 $10^{-6}$ & 5.04200706 \\
2.53100000 & 0.03330876 & $-$0.00004330 &$-$4.2770818 $10^{-9}$ & 5.06200000 \\
\end{tabular}
\label{ft225big}
\end{table}

\begin{table}
\caption[]{Flux-tube solutions for $\kappa=5.00$}
\begin{tabular}{llccc}
$a$ & $b$ & $M/(2\pi)$ & $\Delta{\cal G}$ [$\pi \mu^2$] & $h_{\rm ext}$ \\
\tableline
4.16596082 & 0.04810398 & $-$0.00010000 & $-$3.4267426 $10^{-9}$ & 8.33247708\\
4.09778819 & 0.47269111 & $-$0.01000000 & $-$3.3077416 $10^{-5}$ & 8.24994494\\
3.59756383    & 1.30756971 &  $-$0.10000000 & $-$0.00250402 & 7.65560143 \\
2.46720689    & 2.08399110 &  $-$0.50000000 & $-$0.02888276 & 6.41076481 \\
1.12425543    & 2.58452071 &  $-$2.00000000 & $-$0.13149506 & 5.19549809 \\
0.28823276    & 2.81635033 &  $-$4.99999563 & $-$0.28008789 & 4.61185532 \\ 
0.00000000    & 2.88809220 &  $-$7.25261516 & $-$0.36536655 & 4.43486702 \\ 
$-$0.06903770 & 2.90474500 &  $-$8.00000145 & $-$0.39067762 & 4.39333130 \\
$-$0.10000000 & 2.91214944 &  $-$8.37263989 & $-$0.40286046 & 4.37472398 \\
$-$0.21388850 & 2.93905032 &  $-$9.99944370 & $-$0.45318461 & 4.30597683 \\ 
$-$0.40000000 & 2.98190822 & $-$14.06339412 & $-$0.56348072 & 4.18974698 \\ 
$-$0.50000000 & 3.00438855 & $-$17.68306279 & $-$0.64910298 & 4.12187923 \\
0.90370250  & 3.09142010 & $-$25.02259657 & $-$1.3240437  &    4.07299171 \\
0.95902524  & 3.10289605 & $-$20.00180865 & $-$1.2461821  &    4.11784001 \\
1.05510050  & 3.12257405 & $-$14.99986439 & $-$1.0959379  &    4.22641672 \\ 
1.23355635  & 3.15828834 & $-$10.00002257 & $-$0.90417624 &    4.42267078 \\ 
1.65490767  & 3.23837892 &  $-$5.00000000 & $-$0.62843309 &    4.91311818 \\ 
3.19773261  & 3.48019659 &  $-$1.00000000 & $-$0.22660722 &    7.16833774 \\
4.18000000  & 3.58538926 &  $-$0.48951133 & $-$0.13181569 &    8.85545777 \\
5.00000000  & 3.63660608 &  $-$0.29690981 & $-$0.08628561 &   10.34339701 \\ 
7.14003165  & 3.57240809 &  $-$0.10000000 & $-$0.02872954 &   14.40566028 \\
8.61436325  & 3.31817152 &  $-$0.05000000 & $-$0.01209595 &   17.28374629 \\
9.94894093  & 2.88171879 &  $-$0.02500000 & $-$0.00440641 &   19.91846903 \\
12.00000000 & 1.39965274 &  $-$0.00349371 & $-$0.00013615 &   24.00065683 \\
12.48466309 & 0.25004504 &  $-$0.00010000 & $-$1.2234457 $10^{-7}$ & 
24.96932678 \\
\end{tabular}
\label{ft5}
\end{table}

\begin{table}
\caption[]{Flux-tube solutions with one node for $\kappa=1.00$}
\begin{tabular}{llccc}
$a$ & $b$ & $M/(2\pi)$ & $\Delta{\cal G}$ [$\pi \mu^2$] & $h_{\rm ext}$ \\
\tableline
0.10000000 & 0.00000000 &    0.00000000 & 0.00000000 & 0.20000000 \\ 
0.09500000 & 0.05752439 & $-$0.19829029 & 0.00122474 & 0.20615685 \\ 
0.09000000 & 0.08208260 & $-$0.38793945 & 0.00474348 & 0.21215358 \\ 
0.05000000 & 0.19570689 & $-$1.69647232 & 0.09688239 & 0.25596041 \\ 
0.02000000 & 0.25807049 & $-$2.53551297 & 0.22211571 & 0.28553094 \\ 
0.00000000 & 0.29595722 & $-$3.05989492 & 0.32649272 & 0.30418516 \\ 
$-$0.10000000 & 0.46591028 &  $-$5.59633932 & 1.05752348 & 0.38929139 \\ 
$-$0.20000000 & 0.62101985 &  $-$8.76927872 & 2.16007539 & 0.46739533 \\ 
$-$0.25000000 & 0.69538079 & $-$11.44975608 & 2.96501895 & 0.50763559 \\ 
$-$0.28500000 & 0.74590688 & $-$15.43754890 & 3.86498975 & 0.54164162 \\ 
$-$0.30500000 & 0.77336546 & $-$25.16144125 & 5.31339725 & 0.57868048 \\ 
$-$0.30800000 & 0.77706223 & $-$40.62038139 & 6.96751752 & 0.60456234 \\ 
$-$0.30820000 & 0.77727643 & $-$49.74797375 & 7.80093365 & 0.61383054 \\ 
$-$0.30825000 & 0.77732618 & $-$56.97949955 & 8.41376235 & 0.61958525 \\ 
0.30840000 &  0.77735374 & $-$39.22317054 & 4.61063069 &  0.60533567 \\ 
0.30864000 &  0.77650569 & $-$25.20251927 & 3.26378469 &  0.58397801 \\ 
0.308645625 & 0.77627525 & $-$24.24790512 & 3.16433654 &  0.58196662 \\
0.30800000 &  0.77141194 & $-$17.49003861 & 2.41915630 &  0.56390083 \\ 
0.30000000 &  0.74194860 & $-$10.80591442 & 1.55832296 &  0.53261908 \\ 
0.25000000 &  0.57825777 &  $-$4.54518747 & 0.51406021 &  0.45932507 \\ 
0.20000000 &  0.36647786 &  $-$1.73955332 & 0.09674361 &  0.39071118 \\ 
0.18000000 &  0.23311025 &  $-$0.71579493 & 0.01761512 &  0.35821647 \\ 
0.17000000 &  0.11709493 &  $-$0.18374668 & 0.00119903 &  0.33987571 \\ 
0.16700000 &  0.03709171 &  $-$0.01855901 & 0.00001235 &  0.33399871 \\ 
0.16670000 &  0.01173153 &  $-$0.00185797 & 0.00000012 &  0.33339999 \\ 
\end{tabular}
\label{ftn1}
\end{table}

\begin{table}
\caption[]{Flux-tube-solutions with one node for $\kappa=1.50$}
\begin{tabular}{llccc}
$a$ & $b$ & $M/(2\pi)$ & $\Delta{\cal G}$ [$\pi \mu^2$] & $h_{\rm ext}$ \\
\tableline
0.22500000 & 0.00000000 &    0.00000000 & 0.00000000 & 0.45000000 \\
0.22000000 & 0.08313281 & $-$0.08422212 & 0.00019629 & 0.45524174 \\ 
0.21000000 & 0.14475946 & $-$0.24910164 & 0.00172214 & 0.46553805 \\ 
0.20000000 & 0.18785321 & $-$0.40973461 & 0.00466970 & 0.47560319 \\ 
0.19000000 & 0.22339176 & $-$0.56665414 & 0.00894568 & 0.48545641 \\ 
0.18000000 & 0.25454495 & $-$0.72032371 & 0.01446994 & 0.49511432 \\ 
0.17000000 & 0.28275365 & $-$0.87115056 & 0.02117316 & 0.50459133 \\ 
0.16000000 & 0.30881502 & $-$1.01949636 & 0.02899498 & 0.51390008 \\ 
0.15000000 & 0.33322364 & $-$1.16568537 & 0.03788266 & 0.52305168 \\ 
0.14000000 & 0.35631066 & $-$1.31001117 & 0.04778985 & 0.53205603 \\ 
0.13000000 & 0.37831017 & $-$1.45274190 & 0.05867577 & 0.54092196 \\ 
0.12000000 & 0.39939456 & $-$1.59412474 & 0.07050440 & 0.54965737 \\ 
0.11000000 & 0.41969495 & $-$1.73438975 & 0.08324393 & 0.55826941 \\ 
0.10000000 & 0.43931360 & $-$1.87375249 & 0.09686622 & 0.56676455 \\ 
0.09000000 & 0.45833199 & $-$2.01241726 & 0.11134644 & 0.57514868 \\ 
0.08000000 & 0.47681616 & $-$2.15057901 & 0.12666273 & 0.58342721 \\ 
0.07000000 & 0.49482045 & $-$2.28842532 & 0.14279588 & 0.59160507 \\ 
0.06000000 & 0.51239011 & $-$2.42613867 & 0.15972916 & 0.59968684 \\ 
0.05000000 & 0.52956326 & $-$2.56389763 & 0.17744808 & 0.60767672 \\ 
0.04000000 & 0.54637228 & $-$2.70187844 & 0.19594026 & 0.61557866 \\ 
0.03000000 & 0.56284492 & $-$2.84025694 & 0.21519532 & 0.62339630 \\ 
0.02000000 & 0.57900511 & $-$2.97920954 & 0.23520479 & 0.63113309 \\ 
0.01000000 & 0.59487361 & $-$3.11891442 & 0.25596195 & 0.63879225 \\ 
\end{tabular}
\label{ftn15}
\end{table}

\begin{table}
\caption[]{Flux-tube solutions with one node for $\kappa=2.25$}
\begin{tabular}{llccc}
$a$ & $b$ & $M/(2\pi)$ & $\Delta{\cal G}$ [$\pi \mu^2$] & $h_{\rm ext}$ \\
\tableline
0.50620000 & 0.01167690 & $-$0.00033251 & 0.22768052 $10^{-8}$ & 1.01253467 \\ 
0.50500000 & 0.05838531 & $-$0.00831132 & 0.00000142 & 1.01336606 \\ 
0.50000000 & 0.13056159 & $-$0.04152946 & 0.00003540 & 1.01681860 \\ 
0.10000000 & 1.03984381 & $-$2.85901191 & 0.12071005 & 1.24799509 \\ 
0.00000000 & 1.15113625 & $-$3.75047509 & 0.18242766 & 1.29551424 \\ 
$-$0.10000000 & 1.24623535 &  $-$4.83323620 & 0.25732584 & 1.33996121 \\ 
$-$0.30000000 & 1.39490448 &  $-$8.40318054 & 0.46201774 & 1.42158269 \\ 
$-$0.45000000 & 1.47366214 & $-$17.70122111 & 0.75503469 & 1.48389431 \\ 
$-$0.48500000 & 1.48872942 & $-$28.96559305 & 0.95568218 & 1.50645779 \\ 
$-$0.48750000 & 1.48977303 & $-$30.99115126 & 0.98537998 & 1.50896690 \\ 
0.50000000 & 1.49494165 & $-$59.25824667 & 0.14568520 & 1.53019652 \\ 
0.50750000 & 1.49800748 & $-$32.01239549 & $-$0.12085067 & 1.51498887 \\ 
0.52000000 & 1.50304576 & $-$22.66014509 & $-$0.19695500 & 1.50793798 \\ 
0.55000000 & 1.51459409 & $-$14.74916477 & $-$0.22530268 & 1.50431979 \\ 
0.57500000 & 1.52330439 & $-$11.58013568 & $-$0.21460986 & 1.50645849 \\ 
0.60000000 & 1.53070812 &  $-$9.45207563 & $-$0.19508325 & 1.51122284 \\ 
0.65000000 & 1.53870849 &  $-$6.58697807 & $-$0.14762030 & 1.52662205 \\ 
0.70000000 & 1.52843226 &  $-$4.58431344 & $-$0.09886668 & 1.54911417 \\ 
0.75000000 & 1.47278637 &  $-$2.95284051 & $-$0.05379746 & 1.57988712 \\ 
0.80000000 & 1.27208198 &  $-$1.43082804 & $-$0.01658306 & 1.62390581 \\ 
0.82500000 & 0.97248090 &  $-$0.63822309 & $-$0.00388757 & 1.65544607 \\ 
0.83500000 & 0.71480687 &  $-$0.30403051 & $-$0.00095361 & 1.67131472 \\ 
0.84000000 & 0.48692825 &  $-$0.13178990 & $-$0.00018702 & 1.68025546 \\ 
0.84370000 & 0.05799446 &  $-$0.00177295 & $-$3.50067 $10^{-8}$ & 1.68740005 \\
0.84373500 & 0.03177454 &  $-$0.00053194 & 0.00000000 &   1.68747000 \\ 
\end{tabular}
\label{ftn225}
\end{table}

\begin{table}
\caption[]{Flux-tube solutions with one node for $\kappa=5.00$}
\begin{tabular}{llccc}
$a$ & $b$ & $M/(2\pi)$ & $\Delta{\cal G}$ [$\pi \mu^2$] & $h_{\rm ext}$ \\
\tableline
4.16660000 & 0.03194924 & $-$0.00002206 &$-$1.1760422 $10^{-10}$& 8.33320000\\ 
4.00000000 & 1.54205972 & $-$0.05963664 &   $-$0.00076581 &    8.00283544 \\
3.50000000 & 2.74740880 & $-$0.30980879 &   $-$0.01394038 &    7.05068132 \\
2.90000000 & 3.23595155 & $-$0.85308464 &   $-$0.05940588 &    6.00933253 \\
2.30000000 & 3.31897602 & $-$1.97604538 &   $-$0.15417819 &    5.11558078 \\
1.30000000 & 3.17127693 & $-$9.15487152 &   $-$0.52003242 &    4.03145976 \\
1.20000000 & 3.15164667 & $-$11.20374194 &  $-$0.58424333 &    3.95212203 \\ 
1.10000000 & 3.13166024 & $-$14.07329927 &  $-$0.65910757 &    3.87747347 \\
1.00000000 & 3.11132720 & $-$18.46102014 &  $-$0.75012231 &    3.80669855 \\
0.90000000 & 3.09064825 & $-$26.43953266 &  $-$0.87154430 &    3.73762810 \\
$-$0.69000000 & 3.04608002& $-$37.39455337 &  $-$0.44849527 & 3.68172157 \\
$-$0.60000000 & 3.02649613& $-$25.05396233 &  $-$0.36250996 & 3.71708280 \\ 
$-$0.50000000 & 3.00438828& $-$18.71862741 &  $-$0.30676287 & 3.74928043 \\
0.00000000 & 2.88783642 &   $-$7.65473040  &  $-$0.16951423 & 3.89400437 \\
0.50000000 & 2.75540523 &   $-$3.98870419  &  $-$0.09554523 & 4.06148335 \\
1.00000000 & 2.57972937 &   $-$2.15981734  &  $-$0.04797991 & 4.26240268 \\
1.50000000 & 2.29453336 &   $-$1.10535940  &  $-$0.01910902 & 4.49078655 \\
2.00000000 & 1.76681579 &   $-$0.44288754  &  $-$0.00431046 & 4.73836093 \\
2.25000000 & 1.30125018 &   $-$0.20067548  &  $-$0.00102708 & 4.86761408 \\
2.35000000 & 1.02401580 &   $-$0.11597628  &  $-$0.00036295 & 4.92020750 \\
2.45000000 & 0.60047219 &   $-$0.03727405  &  $-$0.00003960 & 4.97328463 \\
2.47500000 & 0.42623100 &   $-$0.01847061  &  $-$0.00000986 & 4.98662776 \\
2.49500000 & 0.19120008 &   $-$0.00366787  &  $-$0.00000039 & 4.99732323 \\ 
2.49750000 & 0.13525045 &$-$0.00183231 & $-$9.8159338 $10^{-8}$ & 4.99866147 \\
2.49975000 & 0.04278467 &$-$0.00018309 & $-$9.8120083 $10^{-10}$& 4.99986614 \\
\end{tabular}
\label{ftn5}
\end{table}

\begin{table}
\caption[]{Oscillating solutions for $\kappa =1.0$ and flux $F/(2\pi)=-1$.}
\begin{tabular}{ldcc}
$a$ & $b$ & ${\cal E}_{mag}$ [$\pi \mu^2$] & $R$ (1st max) \\
\tableline
$-$0.30828665 & 0.77735953 & 0.24523135 & $\infty$   \\
$-$0.30809193504244 & 0.77700000 & 0.24493325 & 4.02791712 \\
$-$0.30700707 & 0.77500000 & 0.24328646 & 3.36848331 \\
$-$0.26629161 & 0.70000000 & 0.18786525 & 2.21217601 \\
$-$0.21314184 & 0.60000000 & 0.12758654 & 1.99170090 \\
$-$0.16217784 & 0.50000000 & 0.08008222 & 1.89958297 \\
$-$0.11438161 & 0.40000000 & 0.04429725 & 1.85499121 \\
$-$0.07120571 & 0.30000000 & 0.01975218 & 1.83620502 \\
$-$0.03500354 & 0.20000000 & 0.00579848 & 1.83312370 \\
$-$0.0095547963444034 & 0.10000000 & 0.00058777 & 1.83776768 \\
$-$0.0024645497256301 & 0.05000000 & 0.00005105 & 1.84246831 \\
\end{tabular}
\label{table-osz}
\end{table}

\begin{table}
\caption[]{Type B oscillating solutions for $\kappa=1.0$.
$R$ gives positions of extrema.} 
\begin{tabular}{c|cc|cc}
& $R$ (minima)& ${\cal G}$[$\pi\mu^2$] &$R$ (maxima)& ${\cal G}$[$\pi\mu^2$]\\
\tableline
          & 3.84  & $-$0.00002484 & 7.02  & $-$0.00002978 \\
$\{a=0.0$ & 10.18 & $-$0.00003270 & 13.33 & $-$0.00003479 \\
$b=0.1\}$ & 73.1  & $-$0.00004785 & 101.3 & $-$0.00005033 \\
          & 98.2  & $-$0.00005009 & 208.1 & $-$0.00005583 \\
\tableline
          & 5.13  & $-$0.54496245 & 8.48  &  $-$0.62000305 \\
$\{a=0.0$ & 11.74 & $-$0.66506689 & 14.97 &  $-$0.69788474 \\
$b=0.9\}$ & 18.18 & $-$0.72371867 & 97.3  &  $-$0.94062109 \\
          & 295.5 & $-$1.08209885 & 198.0 &  $-$1.03113511 \\
\end{tabular}
\label{table-osB1}
\end{table}

\end{document}